\documentclass[%
 aip,
 jmp,%
 amsmath,amssymb,
 reprint,%
]{revtex4-2}

\usepackage{graphicx}
\usepackage{dcolumn}
\usepackage{bm}
 \usepackage{geometry}

\newgeometry{vmargin={25mm}, hmargin={20mm,25mm}}   

\graphicspath{{figures_main/}}

\begin{document}

\preprint{AIP/123-QED}

\title[]{Mechanics of two filaments in tight contact: \\ The orthogonal clasp}
\thanks{P. Grandgeorge and C. Baek contributed equally.}

\author{Paul Grandgeorge}
 \affiliation{
 Flexible Structures Laboratory, Institute of Mechanical Engineering, École Polytechnique Fédérale de Lausanne (EPFL), Lausanne, Switzerland
 }
 
\author{Changyeob Baek}%
\affiliation{ 
Department of Mechanical Engineering, Massachusetts Institute of Technology, Cambridge, USA
}%

\author{Harmeet Singh}
\affiliation{%
Laboratory for Computation and Visualization in Mathematics and Mechanics, Institute of Mathematics, École Polytechnique Fédérale de Lausanne (EPFL), Switzerland
}%

\author{Paul Johanns}
 \affiliation{
 Flexible Structures Laboratory, Institute of Mechanical Engineering, École Polytechnique Fédérale de Lausanne (EPFL), Lausanne, Switzerland
 }

 \author{Tomohiko G. Sano}
 \affiliation{
 Flexible Structures Laboratory, Institute of Mechanical Engineering, École Polytechnique Fédérale de Lausanne (EPFL), Lausanne, Switzerland
 }
 
 \author{Alastair Flynn}
\affiliation{%
Laboratory for Computation and Visualization in Mathematics and Mechanics, Institute of Mathematics, École Polytechnique Fédérale de Lausanne (EPFL), Switzerland
}%

\author{John H. Maddocks}
 \email{john.maddocks@epfl.ch}
\affiliation{%
Laboratory for Computation and Visualization in Mathematics and Mechanics, Institute of Mathematics, École Polytechnique Fédérale de Lausanne (EPFL), Switzerland
}%
\author{Pedro M. Reis}
 \email{pedro.reis@epfl.ch}
\affiliation{
 Flexible Structures Laboratory, Institute of Mechanical Engineering, École Polytechnique Fédérale de Lausanne (EPFL), Lausanne, Switzerland
 }%

\date{\today}

\begin{abstract}
Networks of flexible filaments often involve regions of tight contact. Predictively understanding the equilibrium configurations of these systems is challenging due to intricate couplings between topology, geometry, large nonlinear deformations, and friction. Here, we perform an in-depth study of a simple yet canonical problem that captures the essence of contact between filaments. In the {\em orthogonal clasp}, two filaments are brought into contact, with each centerline lying in one of a pair of orthogonal planes. Our data from X-ray tomography ($\mu$CT) and mechanical testing experiments are in excellent agreement with the finite element method (FEM) simulations. Despite the apparent simplicity of the physical system, the data exhibits strikingly unintuitive behavior, even when the contact is frictionless.
Specifically, we observe a curvilinear diamond-shaped ridge in the contact pressure field between the two filaments, sometimes with an inner gap.
When a relative displacement is imposed between the filaments, friction is activated, and a highly asymmetric pressure field develops.
These findings contrast to the classic capstan analysis of a single filament wrapped around a rigid body. Both the $\mu$CT and the FEM data indicate that the cross-sections of the filaments can deform significantly. Nonetheless, an idealized geometrical theory assuming undeformable tube cross-sections and neglecting elasticity rationalizes our observations qualitatively and highlights the central role of the small but finite tube radius of the filaments. We believe that our orthogonal clasp analysis provides a building block for future modeling efforts in frictional contact mechanics of more complex filamentary structures.
\end{abstract}

\maketitle

Flexible filamentary structures have been handcrafted and employed by humans since prehistoric times for fastening, lifting, hunting, weaving, sailing, and climbing~\cite{Hardy2020}. The associated engineering of ropes and fabrics has evolved substantially~\cite{Buckner2020}, reflecting the need to predict and enhance their mechanical performance (\textit{e.g.}, flexibility, strength, durability). Towards rationalizing the behavior of touching filaments, pioneering contributions on the mechanics of one-dimensional structures (\textit{e.g.}, the Euler elastica~\cite{euler1952methodus, landau1986lifshitz} and Kirchhoff theory of rods~\cite{Kirchhoff1859,Kirchhoff1876}) have been gradually augmented to describe more complex assemblies of filaments, including frictional elastica~\cite{Sano2017Elastica}, plant tendrils~\cite{GorielyNeukirch2006Plant}, knitted~\cite{Poincloux2018} and woven~\cite{ayres2018beyond, vekhter2019weaving} fabrics, gridshells~\cite{baek2018form, baek2019rigidity}, networks~\cite{Yamaguchi2020Network}, filament and wire bundles~\cite{Stoop2008Wire,grason2009braided, Ward:2015fc,panaitescu2017measuring, panaitescu2018persistence,Warren2018}, and knots~\cite{Maddocks1987a, Audoly2007, Jawed2015,Patil2020}. However, notwithstanding centuries of advances in the mechanics of filamentary networks across length scales, the descriptive understanding of tight filament-filament interactions remains intuitive and empirical at best. In these systems, the intricate coupling between the highly nonlinear fiber deformations, their contact geometry, and mechanics, limits the applicability of conventional one-dimensional centerline-based models such as the Kirchhoff or other rod-based frameworks. 

\begin{figure}[ht!]
\centering
\includegraphics[width=10cm]{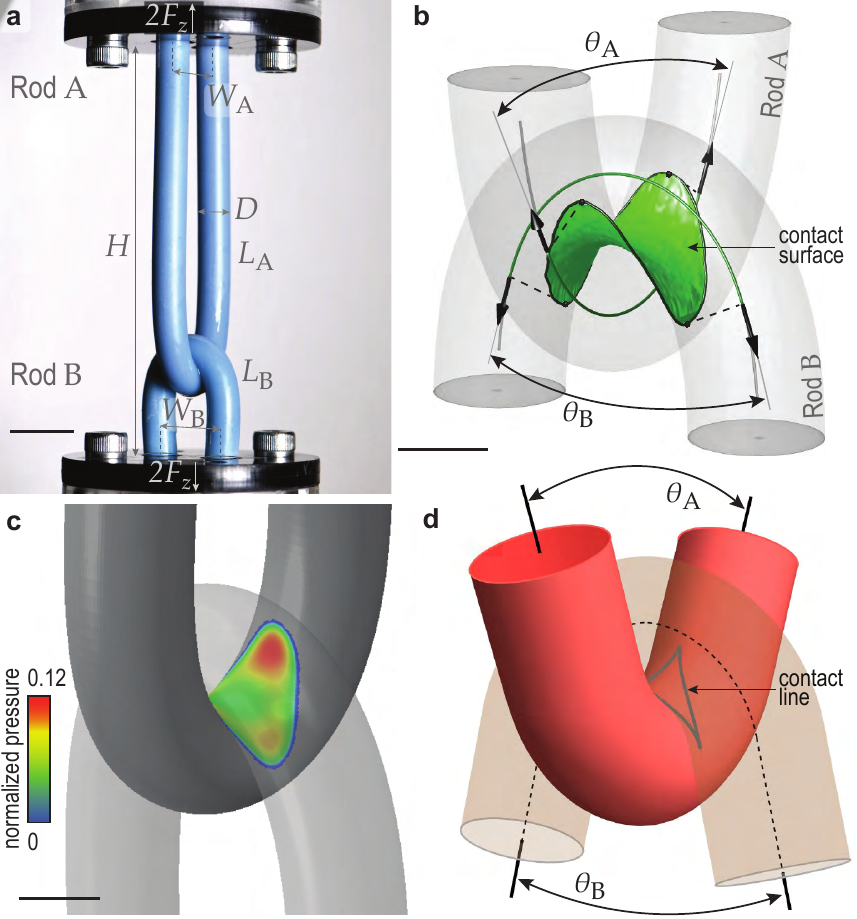}
\caption{\textbf{The orthogonal clasp: two rods in physical contact.} 
\textbf{a},~Photograph of an \textit{elastic orthogonal clasp}: two homogeneous rods (A and B) lying in orthogonal planes are brought into contact.
Both rods have a rest diameter~$D$, respective rest lengths~$L_\text{A}$ and~$L_\text{B}$ and their extremities are fixed at distances~$W_\text{A}$ and~$W_\text{B}$ apart.
The loading is performed by varying the wall-to-wall distance~$H$. Scale bar:~20\,mm.
\textbf{b},~Tomographic rendering of the elastic clasp in the vicinity of the contact surface obtained from $\mu$CT. The physical centerlines of the rods are represented as thin dark gray tubes. The local opening angles of the two rods, $\theta_\mathrm{A}$ and $\theta_\mathrm{B}$, are defined as the angle between the two tangents (black arrows) to the respective centerlines at the extremities of the (green) portions of the centerline attributed to the contact (or green) surface. Scale bar:~5\,mm.
\textbf{c},~Corresponding FEM computation of the elastic orthogonal clasp for the same parameters as in \textbf{b}. The colormap represents the contact pressure distribution (normalized by $EA/D^2$) between the two rods. Scale bar: 5\,mm. 
\textbf{d},~Equilibrium configuration of an ideal orthogonal clasp with two perfectly flexible tubes in contact. The tubes have circular cross-sections orthogonal to the centerline. The idealized contact set is a closed, curvilinear diamond-shaped, \textit{line}, meaning that there is a gap region between the tips of the two tubular surfaces.
All configurations in \textbf{a-c} were obtained for the same physical parameters: $D=8.5$~mm, $L_\mathrm{A}=170$~mm, $W_\mathrm{A}=17$~mm, $L_\mathbf{B}=63.7$~mm, $W_\mathrm{B}=17$ and~$H=103.2$~mm. The ideal clasp in~\textbf{d} was computed for the opening angles, $\theta_\mathrm{A}$ and $\theta_\mathrm{B}$, measured on the elastic clasp of~\textbf{a-c} ($\theta_\mathrm{A}=58^\mathrm{o}$ and~$\theta_\mathrm{B}=33^\mathrm{o}$).}
\label{fig:Fig1}
\end{figure}

Here, we study a deceptively simple, yet we believe canonical, system comprising the mechanical contact between two elastic rods, 
whose respective centerlines lie in one of two orthogonal planes (Fig.~\ref{fig:Fig1}\textbf{a}); a problem that we refer to as the {\em elastic orthogonal clasp}. Using precision X-ray  tomography ($\mu$CT; Fig.~\ref{fig:Fig1}\textbf{b}) and finite element simulations (FEM; Fig.~\ref{fig:Fig1}\textbf{c}), we first study the contact equilibria of a  physical elastic clasp under quasi-static conditions where friction can reasonably be neglected. Throughout, we find excellent quantitative agreement between experiment and FEM. For example, both predict that the two tubular rod surfaces touch in a saddle-shaped {\em patch}. 
The FEM simulations additionally give access to the pressure distribution in the contact region between the rods. For a wide range of loading regimes the pressure distribution is strongly heterogeneous and highly localized along ridgelines that link four isolated peaks, forming a two-fold symmetric, curvilinear diamond pattern. This  surprising localization can be explained qualitatively using a version of a $1D$, primarily geometrical theory, called the {\em ideal orthogonal clasp}. This 1D theory exhibits contact {\em lines} (cf.\ Fig.~\ref{fig:Fig1}\textbf{d}), which form a remarkably accurate skeleton of the ridgelines of the FEM pressure field computed for the elastic orthogonal clasp. The accuracy of this approximation is all the more remarkable given that the ideal clasp model assumes undeformable tube cross-sections, while both experiment and FEM simulation reveal significant deformations of cross-sections in the elastic clasp. 
Finally, we investigate the effect of friction by carrying out capstan-inspired experiments, where one of the elastic rods in the clasp is made to slide against the other. We find that the local contact mechanics strongly influences the tension drop along the sliding rod, and that the contact pressure field now only exhibits three peaks. We provide a qualitative explanation of these observations by considering the analogous system of a V-belt capstan problem built from the ideal orthogonal clasp case. Overall, our findings demonstrate the central role played by the small, but non-vanishing, tube radius in the underlying complex geometry of two contacting filaments in dictating their mechanical response.

\section*{Contact surface of the elastic orthogonal clasp}

Fig.~\ref{fig:Fig1}\textbf{a} presents the experimental setup that we designed to study elastic orthogonal clasps systematically. We clamp two homogeneous elastic rods, rod~A and rod~B, to two rigid walls, the distance between which ($H$) is varied to bring the rods into contact. The two rods have equal rest diameters~$D$, respective rest lengths~$L_\text{A}$ and~$L_\text{B}$, and their extremities are clamped at distances~$W_\text{A}$ and~$W_\text{B}$ apart. As the imposed wall-to-wall distance~$H$ is varied, the overall geometry of the elastic clasp changes. Each wall applies a total vertical force of $2F_z$ to the elastic clasp ($F_z$ per extremity). We define the normalized vertical force $f_z=F_z/EA$, where $E$ is Young's modulus of the material and~$A=\pi D^2/4$ is the unloaded rod cross-section area. The experiments enable us to investigate the contact geometry between the touching rods by analyzing the $\mu$CT volumetric images of tomographically scanned elastic clasps. To enable such a 3D image analysis, we customized a rod fabrication protocol (detailed in the \textit{SI text}, Sec.~1.A) to produce (coaxial) composite rods. These rod designs, together with X-ray tomography, enabled the precise determination of the material centerlines of each rod, along with the contact-surface geometry between the rods. 

In Fig.~\ref{fig:Fig1}\textbf{b}, we show a representative example of a rendered three-dimensional $\mu$CT image of an elastic clasp along with rod centerlines and contact surface (see \textit{Methods}, \textit{SI text}, Sec.~1.B and Sec.~1.C). The saddle-shaped contact surface, physically hidden in between the two rods, is the primary object of our study. 
In parallel to our experimental investigation, we conducted full 3D simulations using the finite element method (FEM) to extract quantities that are not readily available from the experiment, with particular focus on the contact pressure between the rods (colorbar in Fig.~\ref{fig:Fig1}\textbf{c}). In the \textit{SI text}, Sec.~3, we detail our validation procedure of the FEM numerics using the force-displacement curves, $F_z(H)$, for the clasp configuration presented Fig.~\ref{fig:Fig1}\textbf{a}.
Beyond the precise localization of the saddle-shaped contact surface, our computational FEM procedure reveals a highly heterogeneous pressure field between the two rods, which we will analyze in detail. One of the central goals of our study is to rationalize the distribution of this pressure field using the geometrical theory of idealized orthogonal clasps.

\section*{The geometrical theory of idealized orthogonal clasps}
The geometrical problem of predicting contact sets between tightly interwound filaments subject to a non-penetration constraint has been addressed previously in the context of ideal knot shapes, for example, Refs.~\cite{Gonzalez1999, Carlen2005}, where it was observed that it is common for double-contact lines to arise.
Ideal shapes involve filaments that, by assumption, have undeformable, circular, orthogonal cross-sections of finite radius. Typically ideal knot problems are formulated as a purely geometrical problem with no mechanics, which corresponds to their centerlines being considered as inextensible yet perfectly flexible so that they can support no bending moment. 
Starostin~\cite{Starostin2003} first considered the particular case of the ideal shape of an orthogonal clasp, but only for the four-fold symmetric case, where the two components are congruent. 
In our notation, Starostin assumed that the two opening angles (defined in\  Fig.~\ref{fig:Fig1}\textbf{d}) were equal $\theta_\text{A}=\theta_\text{B}$. In the \textit{Methods} and the \textit{SI text}, Sec.~4.D, we present a self-contained variant of Starostin's analysis that allows for an explicit computation of the contact set (\textit{e.g.}, Fig.~\ref{fig:Fig1}\textbf{d}), including the cases $\theta_\text{A}\neq \theta_\text{B}$, via the numerical solution of a set of ordinary differential equations. 
For each $0\le \theta_\text{A}, \theta_\text{B} <\pi$, the contact set is a closed, curvilinear diamond-shaped, \textit{line},  which stands in contrast to the surface-patch contact set observed in both the experimental $\mu$CT and numerical FEM data for the elastic orthogonal clasp (where cross-sections are deformable); see Fig.~\ref{fig:Fig1}\textbf{b} and~\textbf{c}).  
A significant part of our study seeks to explain the connection between the contact patches observed for elastic orthogonal clasps and the contact lines predicted by the ideal orthogonal clasp model.

Double-contact lines arise in ideal knot shapes because at each arc-length along the tube centerline, the associated circular arc in the tube surface has {\em two} contact points with other surface arcs, each of which corresponds to another two distinct arc-lengths. This double-contact feature is found in the present case of the ideal orthogonal clasp, where the phenomenon is perhaps less surprising due to the presence of two planes of reflection symmetry. Moreover, due to the reflection symmetry, the set of all contact chords between two touching centerline points decomposes into a one-parameter family of closed equilateral tetrahedra. It is the one-parameter family of mid-points of these four tetrahedral edges that traces out the curvilinear diamond-shaped contact line. The corners of the diamond arise when the family of open tetrahedra collapses to two double-covered straight line segments. In turn, this coalescent limit arises at first and last contact points along the curve centerlines (or touch-down and lift-off points, cf.\ \textit{SI text} Sec.~4.D for more detail). The diamond-shape contact line surrounds the two tips of the clasp equilibria, and consequently, there is an enclosed gap, or physical space, separating the two tubes close to their tips.

Heuristically, the connection between the contact lines of the ideal orthogonal clasp and the contact patches of the elastic orthogonal clasp can be explained, in both experiment and FEM, by cross-section deformation.
Consequently, the idealized contact lines are, in reality, `fattened' to become surface patches. 
In general, the deformation of the cross-sections can be sufficiently large that the gap between the tips of the ideal orthogonal clasp closes in the corresponding elastic orthogonal clasp. 
Nevertheless, in most cases, we do observe an overall resemblance between the shape of the ideal contact line and the boundary of the elastic clasp surface patch, as in Fig.~\ref{fig:Fig1}\textbf{b}-\textbf{d}.
And, as presented in Fig.~\ref{fig:Fig3}\textbf{b2}, we observed one elastic clasp configuration in which the tip gap does persist.
Furthermore, we will show below that in many cases, the pressure distribution in our FEM simulations of the elastic orthogonal clasp are highly concentrated on the diamond-shaped contact lines in the corresponding ideal orthogonal clasp and that these high pressures arise toward the boundary of the contact patch, with comparatively low contact pressures close to the central tip regions.

\section*{Opening angles of the elastic clasp} 

Next, we perform a quantitative comparison between the equilibria of the ideal and elastic orthogonal clasp configurations. To do so, we use the local opening angles $\theta_\text{A}$ and $\theta_\text{B}$, defined as the angle described by the tangents to centerlines of either the tube (for the ideal case) or rod (for the elastic case) at the touch-down and lift-off contact points (see Fig.~\ref{fig:Fig1}\textbf{b} and~\ref{fig:Fig1}\textbf{d}). Since these opening angles appear in both cases, we take them as the common denominator to identify corresponding equilibria in ideal and elastic orthogonal clasps. In the ideal orthogonal clasp,  these two opening angles are the only model input parameters; the tube diameter and the magnitude of the tensions applied at the tube ends are scale factors. However, in the elastic orthogonal clasp, the two opening angles are observables rather than inputs. The inputs are the applied vertical load (or the vertical imposed displacement) and the displacement boundary conditions enforced at the ends of both rods. Still, for any elastic orthogonal clasp equilibrium configuration, the opening angles can be measured experimentally from the $\mu$CT  data or computed as part of the FEM simulation. As we will discuss later, the opening angles are also of key importance when describing the frictional mechanics of touching filaments, as they reflect the extent of the contact region.

\begin{figure}[b!]
\centering
\includegraphics[width=12cm]{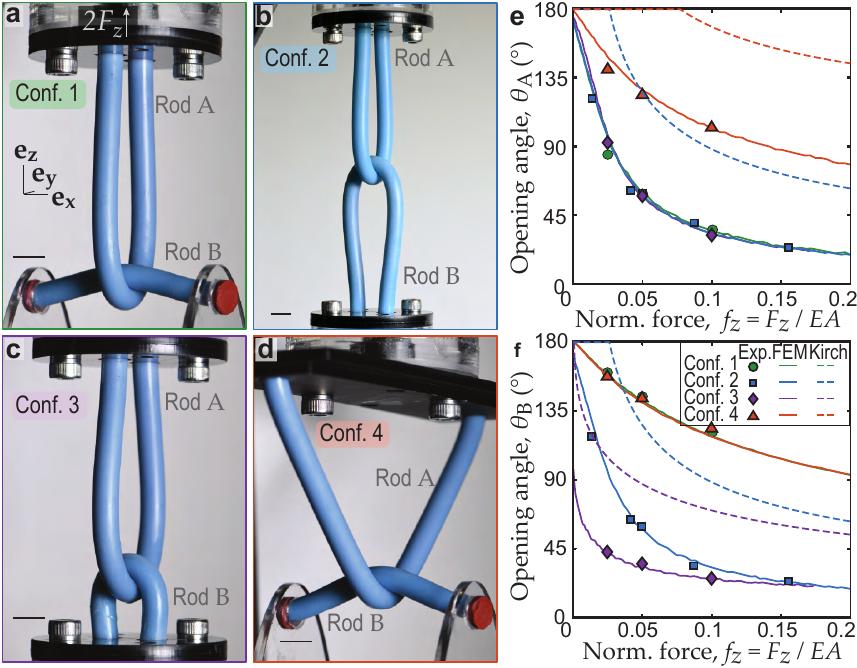}
\caption{
\textbf{Opening angles of the elastic orthogonal clasp, as a function of applied load, for four specific configurations of displacement boundary conditions.}
\textbf{a},~The elastic rods of configuration~1 have the following normalized geometric attributes: rod~A, $l_\text{A}=L_\text{A}/D=20$, $w_\text{A}=W_\text{A}/D=2$; rod~B, $l_\text{B}=L_\text{B}/D=7.5$ and $w_\text{B}=W_\text{B}/D=7.5$ (rod~B is straight when no load is applied).
\textbf{b},~Configuration~2: rod~A, $l_\text{A}=20$, $w_\text{A}=2$; rod~B, $l_\text{B}=20$, $w_\text{B}=2$.
\textbf{c},~Configuration~3: rod~A, $l_\text{A}=20$, $w_\text{A}=2$; rod~B, $l_\text{B}=7.5$, $w_\text{B}=2$.
\textbf{d},~Configuration~4: rod~A, $l_\text{A}=20$, $w_\text{A}=14$; rod~B, $l_\text{B}=w_\text{B}=7.5$ (rod~B is straight when load is applied).
(Scale bars in~\textbf{a}, \textbf{b}, \textbf{c} and~\textbf{d}:~10mm.)
\textbf{e},~Opening angles~$\theta_\text{A}$ and \textbf{f}~$\theta_\text{B}$ plotted versus the normalized applied vertical force $f_z$. The experimental opening angles (data points, measured through $\mu$CT tomography) and the FEM simulations (solid lines) are in excellent agreement. In all cases, the opening angles systematically decrease as the load~$f_z$ increases.
A simplified capstan computation of opening angles derived from a single elastic rod wrapped around a rigid straight cylinder in the absence of friction (dashed lines) delivers rather poor predictions.
}
\label{fig:Fig2}
\end{figure}

\begin{figure*}[b!]
\centering
\includegraphics[width=13cm]{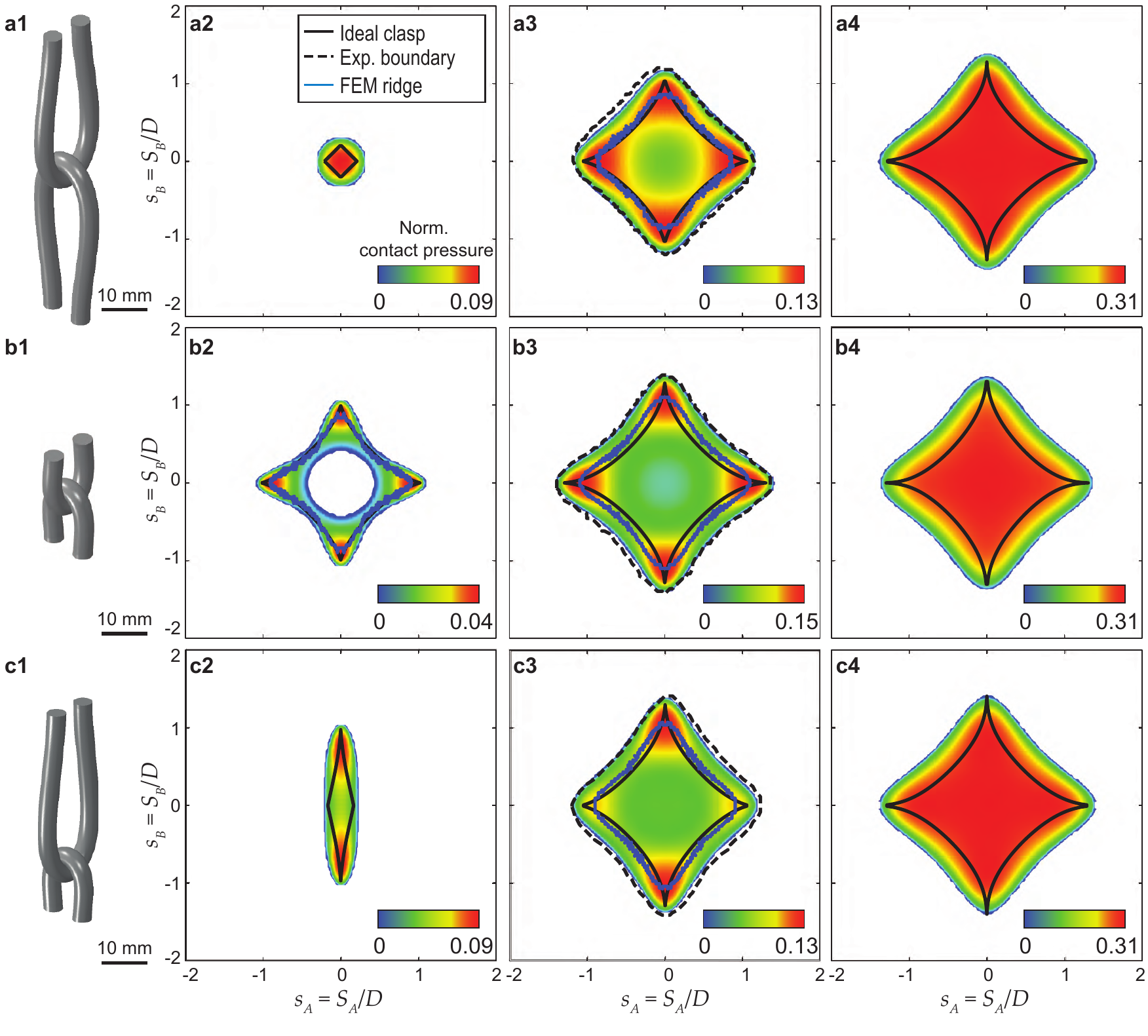}
\caption{\textbf{Contact pressure maps for nine different elastic orthogonal clasp equilibria.} FEM-computed contact pressure fields of elastic orthogonal clasps subject to three different sets of geometrical boundary conditions at three different values of the applied load: 
(\textbf{a1}) $l_\text{A}=l_\text{B}=20$,
\textbf{b1}, $l_\text{A}=l_\text{B}=7.5$, and
\textbf{c1}, $l_\text{A}=20$, $l_\text{B}=7.5$ (with the three physical configurations illustrated corresponding to the applied load $f_z=0.005$). The pressure field data for different values of normalized load $f_z=\{0.005,\,0.05,\,0.15\}$ are presented in (\textbf{a2,\,b2,\,c2}), 
(\textbf{a3,\,b3,\,c3}), and (\textbf{a4,\,b4,\,c4}), respectively. (Blue solid lines): Ridges  extracted from the FEM-computed pressure field.
(Dashed lines): Experimentally observed boundaries of the contact surfaces. (Black solid lines): Prediction of contact curves from ideal orthogonal clasp theory.}
\label{fig:Fig3}
\end{figure*}

In Fig.~\ref{fig:Fig2}\textbf{a}-\textbf{d}, we present four different configurations of displacement boundary conditions that we employed to explore a range of opening angles (the details of each configuration are provided in the caption of Fig.~\ref{fig:Fig2}). The corresponding curves for $\theta_\text{A}$ and $\theta_\text{B}$, as functions of the normalized vertical load~$f_z$, for these four configurations, are provided in  Figs.~\ref{fig:Fig2}\textbf{e} and~\textbf{f} for experiment (datapoints) and simulation (solid lines). We note that the local opening angles computed using FEM are in excellent agreement with the experimental measurement. As expected, in all four cases, the opening angles of rods~A and~B decrease monotonically as $f_z$ increases; the clasp configurations become tighter.

Interestingly, the $\theta_\text{A}(f_z)$ response is largely independent of the geometric conditions imposed on rod~B and {\em vice versa}. Indeed, in configurations~1, 2 and~3, rod~A has the same imposed boundary conditions; even though it is in contact with rods B of different lengths subject to different boundary conditions, it exhibits a nearly identical $\theta_\text{A}(f_z)$ curve across these three configurations. Motivated by this observation, we sought to predict the opening angle response based on a simplified theory of a single Kirchhoff elastic rod wrapped around a rigid, right circular, cylinder of diameter~$D$. 
Full details of this simplified theory are provided in the \textit{Methods} and in the \textit{SI text}, Sec.~4.A and Sec.~4.C. The Kirchhoff-based predictions, dashed lines in Fig.~\ref{fig:Fig2}\textbf{e} and~\textbf{f}, show a significant mismatch with the measured opening angles (experiments and FEM). We conclude that this simple theoretical approximation is inadequate, which will be consequential when we model the sliding clasp below.

The standard Kirchhoff rod model is based on the assumptions (or approximations) that cross-sections are undeformable and unshearable~\cite{antman05}. In the \textit{SI text}, Sec.~2, we quantify the cross-section deformation, the centerline curvature, and the shear strain along the centerline of rod~A, taking configuration~3 as a representative case.  We observe that, even at the relatively low load~$f_z=0.05$, all of these quantities reach substantial values in the neighborhood of contact. First, the cross-section can deform significantly, flattening from its original circular shape by up to~10\,\%. Secondly, the radius of the centerline curvature approaches the rod diameter. Finally, the shear strain reaches values up to~10\,\%. Therefore, in the elastic clasp, the Kirchhoff rod approximations are not accurate. Nevertheless, as we demonstrate below, the alternative one-dimensional geometric theory of the idealized orthogonal clasp does provide a framework to successfully rationalize the skeleton of the contact pressure distribution observed for the 3D elastic orthogonal clasp.

\section*{Distributions of the contact pressure}

We proceed by analyzing the pressure distribution in the contact region of elastic orthogonal clasp. By way of example, we focus on the three representative cases of imposed boundary conditions shown in Fig.~\ref{fig:Fig3}, each at three different values of the normalized applied loads ($f_z=\{0.005,\,0.05,\,0.15\}$). The three configurations considered are: a pair of long rods ($l_\text{A}=l_\text{B}=20$ and $w_\text{A}=w_\text{B}=2$; Fig.~\ref{fig:Fig3}\textbf{a1}), a pair of short rods ($l_\text{A}=l_\text{B}=7.5$ and $w_\text{A}=w_\text{B}=2$; Fig.~\ref{fig:Fig3}\textbf{b1}), and a combination of the long and short rods ($l_\text{A}=20$, $l_\text{B}=7.5$ and $w_\text{A}=w_\text{B}=2$; Fig.~\ref{fig:Fig3}\textbf{c1}), where the physical configurations shown correspond to the low applied load $f_z \simeq 0.005$. For each of the nine cases, the contact pressure maps are computed using FEM simulation and then compared with both the extent of the corresponding contact surface patch measured from the $\mu$CT experiments, and the contact line predicted by the ideal orthogonal clasp analysis. 

The contact pressure is computed from the FEM data as the normal force per unit area on the contact surface patch embedded in 3D. However, for visualization purposes, this scalar field is plotted as a color map in the 2D parameter space $(s_\text{A},\, s_\text{B})$; the arc lengths along the two rod centrelines provide a coordinate system for the contact surface patch (see Fig.~\ref{fig:Fig3}\textbf{a2}-\textbf{a4}, \textbf{b2}-\textbf{b4}, and \textbf{c2}-\textbf{c4}). The point  $s_\text{A}=s_\text{B}=0$ corresponds to the apices of the rods. With this parameterization, it is possible to overlay the boundaries of the contact region extracted from the $\mu$CT experiments of the elastic clasp configuration on the analogous FEM computations with the same opening angles (dashed lines); excellent agreement is found between the two. In these plots, we also juxtapose the contact line predicted by the ideal orthogonal clasp theory (solid black lines in Fig.~\ref{fig:Fig3}\textbf{a2-c2, a3-c3, a4-c4}). All three equilibria for the first two sets of boundary conditions exhibit four-fold symmetric contact pressure maps $(\theta_\text{A}=\theta_\text{B})$. By contrast, there is only two-fold symmetry for the third configuration $(\theta_\text{A} \neq \theta_\text{B})$. 
The contact pressures are highly heterogeneous. For instance, Fig.~\ref{fig:Fig3}\textbf{c3} shows four pressure peaks near the entrance and exit regions of contact along the two rods, connected by ridges of high-pressure values (solid blue lines). 

Remarkably, for all three cases of moderate loading $f_z \simeq 0.05$ presented in Fig.~\ref{fig:Fig3}, the pressure ridges closely follow the curvilinear diamond-shaped contact line predicted by the ideal clasp geometric theory (solid black lines) for the same pair of opening angles; the four peaks in pressure are close to the four corners of the diamond. In these three cases, the purely geometric ideal clasp model provides an accurate skeleton of the pressure map in the physical system. Due to the deformability of the rod cross-sections, the singular, linear pressure field (force per unit length) predicted from geometry is smoothed out to yield the ridges in the surface pressure (force per unit area) for the elastic orthogonal clasp. In the case of higher loading ($f_z \simeq 0.15$; Fig.~\ref{fig:Fig3}\textbf{a4-c4}), where there is significant deformation in the rod cross-sections, the surface pressure becomes more uniform over the interior of the region of contact, and the ridges are no longer observed. 

The case of small loading ($f_z \simeq 0.005$; Fig.~\ref{fig:Fig3}\textbf{a2-c2}) reveals a variety of contact pressure distributions depending on the global geometry of the equilibrium. Fig.~\ref{fig:Fig3}\textbf{a2} shows a highly localized pressure map, characterized by a single peak close to the apices $s_\text{A}=s_\text{B}=0$. However, in Fig.~\ref{fig:Fig3}\textbf{b2}, both pressure ridges and a central region of vanishing pressure are observed, corresponding closely to the predictions of the ideal orthogonal clasp with its predicted gap between the tube surfaces close to the two centerline tips. 

The hybrid long/short rod configuration in Fig.~\ref{fig:Fig3}\textbf{c2} exhibits a two-fold symmetric pressure corresponding to the asymmetry between rod A and rod B. There are two pressure peaks along rod~B, separated by a saddle with low pressure between the apices. Even in this extreme case, the ideal clasp diamond contact line yields a reasonable prediction of the extent of the actual contact surface in the elastic orthogonal clasp.

\section*{The elastic clasp with sliding friction}

Leveraging the physical understanding gained above for the elastic orthogonal clasp with negligible friction, we next impose a relative motion between two elastically deformable rods in a clasp configuration, with finite friction, to extend the problem to obtain the \textit{frictional sliding clasp}. There is a direct analogy between this system and the  \textit{capstan problem}, in which one deformable filament is wrapped in a planar configuration around a rigid drum, or capstan. The classic Euler-Eytelwein version of this problem~\cite{Euler1769, Eytelwein1832,Maddocks1987a} assumes that the capstan is a rigid cylinder and that the filament is perfectly flexible, with a negligible radius of curvature compared to the radius of the capstan. Then, the maximal possible ratio of high tension $T_1$ at one end to low tension $T_0$ at the other end is predicted by the well-known capstan relation,
\begin{equation}
    \frac{T_1}{T_0} = e^{\mu \varphi},
    \label{eq:capstan}
\end{equation}
where $\varphi$ is the winding angle around the capstan, and $\mu$ is the friction coefficient (either static or dynamic depending on context) between the filament and the drum. 

\begin{figure*}[t!]
\centering
\includegraphics[width=\textwidth]{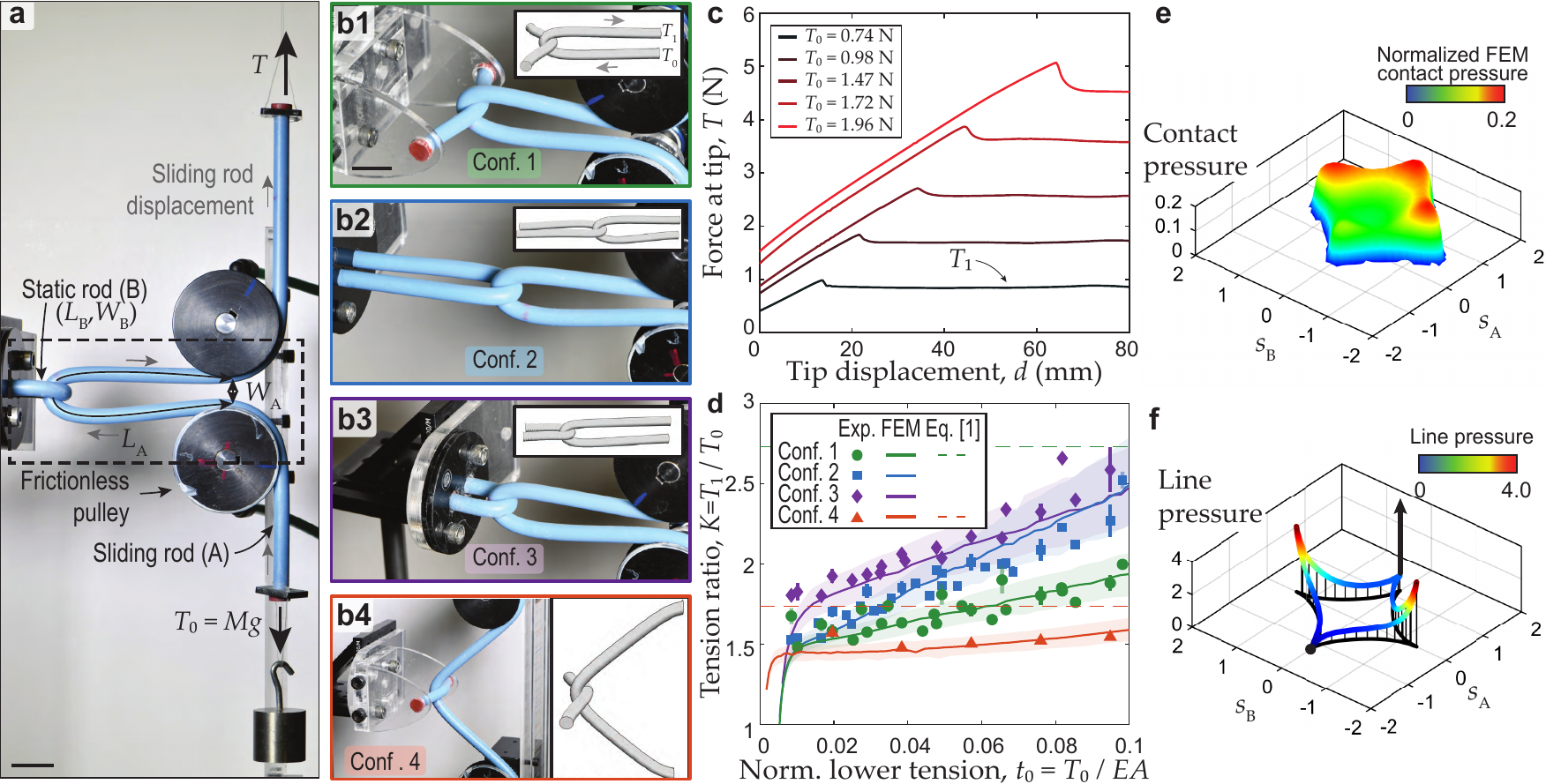}
    \caption{\textbf{Friction-induced tension gain in the sliding elastic clasp.}    \textbf{a},~Experimental apparatus used to measure the tension ratio in different geometric configurations. Scale bar: 20\,mm.
    \textbf{b1}-\textbf{b4}, Adaptation of configurations.~1, 2, 3 and~4 described in Fig.~\ref{fig:Fig2} to allow for the sliding of rod~A around the static rod~B.
    \textbf{c},~Experimentally measured upward force $T$ in the upper segment of the sliding rod A as a function of the upward displacement $d$ of the upper tip. 
    \textbf{d} Ratio $K=T_1/T_0$ between the steady-state value of $T$ (plateau in \textbf{c}) and the value of the dead load $T_0$ applied to the lower end of rod A, plotted as a function of the normalized dead load $t_0=T_0/EA$,  for the four different configurations. Experimental and FEM data are presented. The shaded regions correspond to the sensitivity of the FEM computation to the experimental uncertainty of the measured friction coefficient ($\mu=0.32\pm0.03$). The horizontal dashed lines is the prediction from Eq.~(\ref{eq:capstan}).
    \textbf{e},~Representative contact pressure map in the contact region obtained from the FEM simulation of configuration~3, at a normalized dead load of $t_0=0.05$ (the pressure is normalized by $EA/D^2$). 
    \textbf{f},~Linear pressure map of the ideal clasp, where it can be computed analytically that for a maximum tension growth equilibrium, the linear pressure vanishes at the touch-down point (black circle), but has a singular Delta function at the lift-off corner (represented by a black arrow).}
    \label{fig:Fig4}
\end{figure*}

Our frictional sliding clasp load-displacement experiment can be viewed as a generalized capstan problem in which the static, rigid drum is replaced by a second deformable rod, thereby adding elastic deformation in a nontrivial way. Motivated by this analogy, we seek to measure the change in the tension between the two ends of a rod sliding in steady-state while wrapped around a second static but deformable rod. In our experiments (see Fig.~\ref{fig:Fig4}\textbf{a}), rod~A is guided by two frictionless pulleys to thread around the clamped rod~B. The rods are surface-treated to ensure robust Amontons-Coulomb frictional behavior with $\mu=0.32\pm0.03$ (see details in the \textit{SI text}, Sec.~1.F). We measured the vertical force~$T$ necessary to displace the upper extremity of rod~A at a constant velocity subject to a dead load $T_0=Mg$, applied at the other extremity (where $M$ is the mass of dead weight and $g$ the gravitational acceleration). Upon pulling with a prescribed velocity, the measured tension $T$ first increases and then reaches a plateau value~$T_1$ (see Fig.~\ref{fig:Fig4}\textbf{c} and the \textit{Methods}). We then quantify the steady tension ratio $K=T_1/T_0$ between the high and low-tension strands, as a function of the non-dimensional lower tension $t_0=T_0/EA$, for different geometric configurations. Specifically, we adopt the same four sets of imposed global geometric parameters as for the static elastic orthogonal clasp experiments, namely configurations~1 to~4 described in Fig.~\ref{fig:Fig2}\textbf{a-d}), but with the presence of the pulleys instead of clamps for rod A (Fig.~\ref{fig:Fig4}\textbf{b1-b4}). In the \textit{SI text}, Sec.~4.B, we show that our long sliding rod (rod~A) mainly carries an internal tensile force (\textit{i.e.}, parallel to the tangent of the rod centerline) far from the contact region with rod~B.

In Fig.~\ref{fig:Fig4}\textbf{d}, we plot the experimental tension ratio $K$ (data points) as a function of the dimensionless lower tension $t_0$ imposed by the dead load. We observe that all $K(t_0)$ increase monotonically with $t_0$, albeit with significant differences for the various imposed geometrical configurations in either rod A or rod B.  In parallel to the experiments, we carried out FEM numerical simulations of the steady-state frictional sliding elastic clasp (solid lines; see \textit{Methods}, \textit{SI text}, Sec.~3 for details of the simulations), finding good agreement with the experimental data (mean difference of $\approx4.2\%$). The experimental and numerical data highlight the importance of the global opening angle imposed on the sliding rod; \textit{i.e.} the angle between the tangents to rod A in the nearly straight regions far from contact, set experimentally by the distance between the pulleys. Note that the local opening angles cannot be accessed from $\mu$CT for these experiments. Indeed, we observe that $K(t_0)$ is lower when the pulleys are far apart (configuration~4, orange triangles) than when they are close to each other (configuration~1, green circles). The overall ordering of this difference in tension ratio for varying global opening angle in rod A is correctly predicted in the capstan equation~(\ref{eq:capstan}) when applied with the wrapping angle $\varphi$ interpreted as the global opening angle of rod A (dashed lines). 
However, the simple capstan equation interpreted in this way is inadequate to explain our data because it predicts that the tension ratio should be independent of both $t_0$ and the imposed geometry on the static rod B; neither of these predictions is close to being accurate. The mismatch between the capstan equation and our results stems from the non-trivial dependence of the local opening angles of the sliding and static rods on the applied tension~$t_0$. Specifically, our results reveal the essential role of the static rod B; the tension ratio response is significantly larger when the static rod wraps further around the sliding rod. For example, the tension ratio of configuration~3 (where the static rod is initially short with ends close to each other) is larger than the tension ratio for configuration 1, by 21\% at $t_0=0.05$. 

The limitation of the simple capstan relation to predicting the data is consistent with the observation made above that a single elastic rod wrapped around a rigid cylindrical capstan is inadequate to accurately describe {\em local} opening angles of an elastic clasp configuration, and it is the local opening angles that are strongly related to the detailed contact mechanics that sets the maximal tension ratio $K(t_0)$.

The detail of the contact region interactions, including sliding friction, can be further probed via FEM. In Fig.~\ref{fig:Fig4}\textbf{e}, we plot the pressure field between the two rods for the frictional sliding clasp configuration of Fig.~\ref{fig:Fig4}\textbf{b3} (configuration~3). Just as in the static, frictionless case, the support of the contact pressure map in the presence of friction displays the familiar diamond shape, and again there are strong heterogeneities. Now, however, we observe only three pressure peaks in the $(s_\text{A},s_\text{B})$-plane; two at the extreme values of $s_\textbf{B}$, and one at the high-tension, or lift-off, end along $s_\text{A}$ where the pressure is largest. The symmetry along rod A is now broken, and the pressure at the touch-down end of the contact set is extremely low. As a first step towards understanding these qualitative features in the contact pressure distribution for the sliding elastic clasp, in the \textit{SI text}, Sec.~4.F, we present a hybrid analysis of the ideal orthogonal clasp. In this framework, the shape of tube B is frozen to form a rigid tubular capstan with its own intricate geometry and local opening angle, around which the flexible tube A can reptate, with the contact set remaining identical with the static case (\textit{i.e.}, a closed, curvilinear diamond-shape line). In this idealized context, and as a consequence of moment balance on the finite radius flexible tube A, it can be derived (\textit{SI text}, Sec.~4.E) that, in the presence of friction, the linear pressure distribution giving rise to the maximal possible tension ratio $K(t_0)$ vanishes at the touch-down point, and has a Dirac delta function singularity at the lift-off point. Despite the strong simplifying assumptions in this model, these predictions for the linear pressure density, as illustrated in  Fig.~\ref{fig:Fig4}\textbf{f}, have a striking qualitative similarity to the surface pressure distributions as computed in the FEM simulations Fig.~\ref{fig:Fig4}\textbf{e}.

\section*{Conclusions}
Despite the apparent simplicity of having only two elastic filaments in contact, our experiments and FEM computations on the orthogonal clasp system have  revealed highly nontrivial behavior. Specifically, we find that the contact pressure field between the two filaments is highly heterogeneous, with a curvilinear diamond ridge skeleton that can be explained qualitatively using a version of a primarily geometrical, one-dimensional theory for the {\em ideal orthogonal clasp}. The accuracy of this approximation for a wide range of loadings is all the more remarkable given the scale of the observed deformations of cross-sections in the two rods forming the {\em elastic orthogonal clasp}. Furthermore, a version of a V-belt capstan problem constructed from the ideal orthogonal clasp problem rationalizes the asymmetric roles of touch-down and lift-off points in the pressure distribution computed in the physical case when the two filaments are made to slide with respect to each other, in the presence of friction.

We hope that our findings will instigate future theoretical model developments that consider the intricate coupling of elasticity and contact geometry of filaments with a small but non-vanishing diameter. Future one-dimensional models should extend the geometric ideal tube description with the incorporation of elasticity, both through a bending stiffness for the centerline and in allowing cross-sectional deformation. Such efforts would significantly impact the homogenization schemes necessary to predictively describe more intricate networks of filaments such as knits, knots, and weaves, starting from the elastic clasp as a building block. 

\section*{Methods}

\subsection*{Fabrication of the rods for $\mu$CT imaging} For the \textit{static elastic orthogonal clasp} experiment, elastomeric rods were fabricated through casting using silicone-based polymers, with a coaxial geometry comprising (i) a bulk core, (ii) a thin physical centerline fiber, and (iii) an outer coating layer. The bulk core was fabricated out of Vinyl PolySiloxane (VPS-16, Elite Double 16 Fast, Zhermack), whereas the physical centerline fiber and the outer coating were made out of the Solaris polymer (Smooth-On). The overall diameter of the resulting rod was~$D=8.5$~mm. Additional fabrication details are provided in the \textit{SI text}, Sec.~1.A. 

\subsection*{X-ray micro-computed tomography ($\mu$CT)} Tomographic imaging of the elastic clasp configurations was performed using a  $\mu$CT~100 (Scanco) machine for the static elastic orthogonal clasps in configurations~1 and~3, and an Ultra-Tom (RX-Solutions) machine for configurations~2 and~4. An in-house post-processing algorithm (written in \texttt{MATLAB 2019b, MathWorks}) was used to extract the coordinates of the centerline of the rods, as well as those of the rod-rod contact surface, from the volumetric $\mu$CT images (see \textit{SI text}, Sec.~1.B).

\subsection*{Fabrication of the rods for mechanical testing} For mechanical testing, the rods were fabricated out of Vinyl PolySiloxane of two different grades: VPS-16 (Elite double 16 Fast, Zhermack) and VPS-32 (Elite double 32, Zhermack). The rods were cast in straight acrylic tubes of inner diameter $D=8$\,mm. Further details are provided in the \textit{SI text}, Section 1.E. 

\subsection*{Mechanical testing} For the \textit{sliding orthogonal clasp} experiment, we quantified the high-to-low tension ratio in the two ends of a rod sliding around another by performing force-displacement experiments on a mechanical testing machine (Instron~5943 with a 50\,N load-cell). These tests used frictionless rotating pulleys to guide the sliding rod. The upper end of the sliding rod was pulled upward at constant velocity $U=3$\,mm/s, while measuring the pulling force $T$. See further details in the \textit{SI text}, Sec.~1.F. 
 
\subsection*{Finite Element Modeling} Finite element method (FEM) simulations were performed using the software package ABAQUS/STANDARD to numerically compute the equilibria of the \textit{elastic static} and \textit{sliding orthogonal clasps}. The elastic rods were meshed with 3D brick elements with reduced integration and hybrid formulation (C3D8RH), with a Neo-Hookean hyperelastic material model. The geometry, meshing and loading sequences are detailed in the \textit{SI text}, Sec.~3.

\section*{Acknowledgement}
The authors thank B.~Audoly and S.~Neukirch for fruitful discussions as well as
G. Perrenoult and P. Turberg for advice on $\mu$CT tomography.
P.J. was supported by the Fonds National de la Recherche, Luxembourg 12439430. H.S. A.F. and J.H.M. were partially supported by Swiss National Science Foundation Grant 200020-18218 to J.H.M. and T.G.S. acknowledges financial support in the form of Grants-in-Aid for JSPS Overseas Research Fellowship (2019-60059).

\section*{Bibliography}
\bibliography{references_main}

\begin{thebibliography}{30}%
\makeatletter
\providecommand \@ifxundefined [1]{%
 \@ifx{#1\undefined}
}%
\providecommand \@ifnum [1]{%
 \ifnum #1\expandafter \@firstoftwo
 \else \expandafter \@secondoftwo
 \fi
}%
\providecommand \@ifx [1]{%
 \ifx #1\expandafter \@firstoftwo
 \else \expandafter \@secondoftwo
 \fi
}%
\providecommand \natexlab [1]{#1}%
\providecommand \enquote  [1]{``#1''}%
\providecommand \bibnamefont  [1]{#1}%
\providecommand \bibfnamefont [1]{#1}%
\providecommand \citenamefont [1]{#1}%
\providecommand \href@noop [0]{\@secondoftwo}%
\providecommand \href [0]{\begingroup \@sanitize@url \@href}%
\providecommand \@href[1]{\@@startlink{#1}\@@href}%
\providecommand \@@href[1]{\endgroup#1\@@endlink}%
\providecommand \@sanitize@url [0]{\catcode `\\12\catcode `\$12\catcode
  `\&12\catcode `\#12\catcode `\^12\catcode `\_12\catcode `\%12\relax}%
\providecommand \@@startlink[1]{}%
\providecommand \@@endlink[0]{}%
\providecommand \url  [0]{\begingroup\@sanitize@url \@url }%
\providecommand \@url [1]{\endgroup\@href {#1}{\urlprefix }}%
\providecommand \urlprefix  [0]{URL }%
\providecommand \Eprint [0]{\href }%
\providecommand \doibase [0]{https://doi.org/}%
\providecommand \selectlanguage [0]{\@gobble}%
\providecommand \bibinfo  [0]{\@secondoftwo}%
\providecommand \bibfield  [0]{\@secondoftwo}%
\providecommand \translation [1]{[#1]}%
\providecommand \BibitemOpen [0]{}%
\providecommand \bibitemStop [0]{}%
\providecommand \bibitemNoStop [0]{.\EOS\space}%
\providecommand \EOS [0]{\spacefactor3000\relax}%
\providecommand \BibitemShut  [1]{\csname bibitem#1\endcsname}%
\let\auto@bib@innerbib\@empty
\bibitem [{\citenamefont {Hardy}\ \emph {et~al.}(2020)\citenamefont {Hardy},
  \citenamefont {Moncel}, \citenamefont {Kerfant}, \citenamefont {Lebon},
  \citenamefont {Bellot-Gurlet},\ and\ \citenamefont
  {M{\'{e}}lard}}]{Hardy2020}%
  \BibitemOpen
  \bibfield  {author} {\bibinfo {author} {\bibfnamefont {B.~L.}\ \bibnamefont
  {Hardy}}, \bibinfo {author} {\bibfnamefont {M.-H.}\ \bibnamefont {Moncel}},
  \bibinfo {author} {\bibfnamefont {C.}~\bibnamefont {Kerfant}}, \bibinfo
  {author} {\bibfnamefont {M.}~\bibnamefont {Lebon}}, \bibinfo {author}
  {\bibfnamefont {L.}~\bibnamefont {Bellot-Gurlet}},\ and\ \bibinfo {author}
  {\bibfnamefont {N.}~\bibnamefont {M{\'{e}}lard}},\ }\bibfield  {title}
  {\enquote {\bibinfo {title} {Direct evidence of neanderthal fibre technology
  and its cognitive and behavioral implications},}\ }\href
  {https://doi.org/10.1038/s41598-020-61839-w} {\bibfield  {journal} {\bibinfo
  {journal} {Sci. Rep.}\ }\textbf {\bibinfo {volume} {10}} (\bibinfo {year}
  {2020}),\ 10.1038/s41598-020-61839-w}\BibitemShut {NoStop}%
\bibitem [{\citenamefont {Buckner}\ \emph {et~al.}(2020)\citenamefont
  {Buckner}, \citenamefont {Bilodeau}, \citenamefont {Kim},\ and\ \citenamefont
  {Kramer-Bottiglio}}]{Buckner2020}%
  \BibitemOpen
  \bibfield  {author} {\bibinfo {author} {\bibfnamefont {T.~L.}\ \bibnamefont
  {Buckner}}, \bibinfo {author} {\bibfnamefont {R.~A.}\ \bibnamefont
  {Bilodeau}}, \bibinfo {author} {\bibfnamefont {S.~Y.}\ \bibnamefont {Kim}},\
  and\ \bibinfo {author} {\bibfnamefont {R.}~\bibnamefont {Kramer-Bottiglio}},\
  }\bibfield  {title} {\enquote {\bibinfo {title} {Roboticizing fabric by
  integrating functional fibers},}\ }\href
  {https://doi.org/10.1073/pnas.2006211117} {\bibfield  {journal} {\bibinfo
  {journal} {Proc. Natl. Acad. Sci. U.S.A.}\ }\textbf {\bibinfo {volume} {-}},\
  \bibinfo {pages} {202006211} (\bibinfo {year} {2020})}\BibitemShut {NoStop}%
\bibitem [{\citenamefont {Euler}(1952)}]{euler1952methodus}%
  \BibitemOpen
  \bibfield  {author} {\bibinfo {author} {\bibfnamefont {L.}~\bibnamefont
  {Euler}},\ }\href@noop {} {\emph {\bibinfo {title} {Methodus inveniendi
  lineas curvas maximi minimive proprietate gaudentes sive solutio problematis
  isoperimetrici latissimo sensu accepti}}},\ Vol.~\bibinfo {volume} {1}\
  (\bibinfo  {publisher} {Springer Science \& Business Media},\ \bibinfo {year}
  {1952})\BibitemShut {NoStop}%
\bibitem [{\citenamefont {Landau}\ and\ \citenamefont
  {Lifshitz}(1986)}]{landau1986lifshitz}%
  \BibitemOpen
  \bibfield  {author} {\bibinfo {author} {\bibfnamefont {L.~M.}\ \bibnamefont
  {Landau}}\ and\ \bibinfo {author} {\bibfnamefont {E.~M.}\ \bibnamefont
  {Lifshitz}},\ }\bibfield  {title} {\enquote {\bibinfo {title} {Theory of
  elasticity},}\ }\href@noop {} {\bibfield  {journal} {\bibinfo  {journal}
  {Course of theoretical physics}\ }\textbf {\bibinfo {volume} {7}} (\bibinfo
  {year} {1986})}\BibitemShut {NoStop}%
\bibitem [{\citenamefont {Kirchhoff}(1859)}]{Kirchhoff1859}%
  \BibitemOpen
  \bibfield  {author} {\bibinfo {author} {\bibfnamefont {G.}~\bibnamefont
  {Kirchhoff}},\ }\bibfield  {title} {\enquote {\bibinfo {title} {Ueber das
  gleichgewicht und die bewegung eines unendlich dünnen elastischen stabes.}}\
  }\href {http://eudml.org/doc/147766} {\bibfield  {journal} {\bibinfo
  {journal} {Journal f{\"u}r die reine und angewandte Mathematik}\ }\textbf
  {\bibinfo {volume} {56}},\ \bibinfo {pages} {285--313} (\bibinfo {year}
  {1859})}\BibitemShut {NoStop}%
\bibitem [{\citenamefont {Kirchhoff}(1876)}]{Kirchhoff1876}%
  \BibitemOpen
  \bibfield  {author} {\bibinfo {author} {\bibfnamefont {G.}~\bibnamefont
  {Kirchhoff}},\ }\href@noop {} {\emph {\bibinfo {title} {Vorlesungen {\"u}ber
  mathematische physik: mechanik}}},\ Vol.~\bibinfo {volume} {1}\ (\bibinfo
  {publisher} {BG Teubner},\ \bibinfo {year} {1876})\BibitemShut {NoStop}%
\bibitem [{\citenamefont {Sano}, \citenamefont {Yamaguchi},\ and\ \citenamefont
  {Wada}(2017)}]{Sano2017Elastica}%
  \BibitemOpen
  \bibfield  {author} {\bibinfo {author} {\bibfnamefont {T.~G.}\ \bibnamefont
  {Sano}}, \bibinfo {author} {\bibfnamefont {T.}~\bibnamefont {Yamaguchi}},\
  and\ \bibinfo {author} {\bibfnamefont {H.}~\bibnamefont {Wada}},\ }\bibfield
  {title} {\enquote {\bibinfo {title} {Slip morphology of elastic strips on
  frictional rigid substrates},}\ }\href
  {https://doi.org/10.1103/PhysRevLett.118.178001} {\bibfield  {journal}
  {\bibinfo  {journal} {Phys. Rev. Lett.}\ }\textbf {\bibinfo {volume} {118}},\
  \bibinfo {pages} {178001} (\bibinfo {year} {2017})}\BibitemShut {NoStop}%
\bibitem [{\citenamefont {Goriely}\ and\ \citenamefont
  {Neukirch}(2006)}]{GorielyNeukirch2006Plant}%
  \BibitemOpen
  \bibfield  {author} {\bibinfo {author} {\bibfnamefont {A.}~\bibnamefont
  {Goriely}}\ and\ \bibinfo {author} {\bibfnamefont {S.}~\bibnamefont
  {Neukirch}},\ }\bibfield  {title} {\enquote {\bibinfo {title} {Mechanics of
  climbing and attachment in twining plants},}\ }\href
  {https://doi.org/10.1103/PhysRevLett.97.184302} {\bibfield  {journal}
  {\bibinfo  {journal} {Phys. Rev. Lett.}\ }\textbf {\bibinfo {volume} {97}},\
  \bibinfo {pages} {184302} (\bibinfo {year} {2006})}\BibitemShut {NoStop}%
\bibitem [{\citenamefont {Poincloux}, \citenamefont {Adda-Bedia},\ and\
  \citenamefont {Lechenault}(2018)}]{Poincloux2018}%
  \BibitemOpen
  \bibfield  {author} {\bibinfo {author} {\bibfnamefont {S.}~\bibnamefont
  {Poincloux}}, \bibinfo {author} {\bibfnamefont {M.}~\bibnamefont
  {Adda-Bedia}},\ and\ \bibinfo {author} {\bibfnamefont {F.}~\bibnamefont
  {Lechenault}},\ }\bibfield  {title} {\enquote {\bibinfo {title} {Geometry and
  elasticity of a knitted fabric},}\ }\href
  {https://doi.org/10.1103/physrevx.8.021075} {\bibfield  {journal} {\bibinfo
  {journal} {Phys. Rev. X.}\ }\textbf {\bibinfo {volume} {8}} (\bibinfo {year}
  {2018}),\ 10.1103/physrevx.8.021075}\BibitemShut {NoStop}%
\bibitem [{\citenamefont {Ayres}, \citenamefont {Martin},\ and\ \citenamefont
  {Zwierzycki}(2018)}]{ayres2018beyond}%
  \BibitemOpen
  \bibfield  {author} {\bibinfo {author} {\bibfnamefont {P.}~\bibnamefont
  {Ayres}}, \bibinfo {author} {\bibfnamefont {A.~G.}\ \bibnamefont {Martin}},\
  and\ \bibinfo {author} {\bibfnamefont {M.}~\bibnamefont {Zwierzycki}},\
  }\bibfield  {title} {\enquote {\bibinfo {title} {Beyond the basket case: a
  principled approach to the modelling of kagome weave patterns for the
  fabrication of interlaced lattice structures using straight strips},}\ }in\
  \href@noop {} {\emph {\bibinfo {booktitle} {Advances in Architectural
  Geometry 2018}}}\ (\bibinfo {organization} {Chalmers University of
  Technology},\ \bibinfo {year} {2018})\ pp.\ \bibinfo {pages}
  {72--93}\BibitemShut {NoStop}%
\bibitem [{\citenamefont {Vekhter}\ \emph {et~al.}(2019)\citenamefont
  {Vekhter}, \citenamefont {Zhuo}, \citenamefont {Fandino}, \citenamefont
  {Huang},\ and\ \citenamefont {Vouga}}]{vekhter2019weaving}%
  \BibitemOpen
  \bibfield  {author} {\bibinfo {author} {\bibfnamefont {J.}~\bibnamefont
  {Vekhter}}, \bibinfo {author} {\bibfnamefont {J.}~\bibnamefont {Zhuo}},
  \bibinfo {author} {\bibfnamefont {L.~F.~G.}\ \bibnamefont {Fandino}},
  \bibinfo {author} {\bibfnamefont {Q.}~\bibnamefont {Huang}},\ and\ \bibinfo
  {author} {\bibfnamefont {E.}~\bibnamefont {Vouga}},\ }\bibfield  {title}
  {\enquote {\bibinfo {title} {Weaving geodesic foliations},}\ }\href@noop {}
  {\bibfield  {journal} {\bibinfo  {journal} {ACM Trans. Graph.}\ }\textbf
  {\bibinfo {volume} {38}},\ \bibinfo {pages} {1--22} (\bibinfo {year}
  {2019})}\BibitemShut {NoStop}%
\bibitem [{\citenamefont {Baek}\ \emph {et~al.}(2018)\citenamefont {Baek},
  \citenamefont {Sageman-Furnas}, \citenamefont {Jawed},\ and\ \citenamefont
  {Reis}}]{baek2018form}%
  \BibitemOpen
  \bibfield  {author} {\bibinfo {author} {\bibfnamefont {C.}~\bibnamefont
  {Baek}}, \bibinfo {author} {\bibfnamefont {A.~O.}\ \bibnamefont
  {Sageman-Furnas}}, \bibinfo {author} {\bibfnamefont {M.~K.}\ \bibnamefont
  {Jawed}},\ and\ \bibinfo {author} {\bibfnamefont {P.~M.}\ \bibnamefont
  {Reis}},\ }\bibfield  {title} {\enquote {\bibinfo {title} {Form finding in
  elastic gridshells},}\ }\href@noop {} {\bibfield  {journal} {\bibinfo
  {journal} {Proc. Natl. Acad. Sci. U.S.A.}\ }\textbf {\bibinfo {volume}
  {115}},\ \bibinfo {pages} {75--80} (\bibinfo {year} {2018})}\BibitemShut
  {NoStop}%
\bibitem [{\citenamefont {Baek}\ and\ \citenamefont
  {Reis}(2019)}]{baek2019rigidity}%
  \BibitemOpen
  \bibfield  {author} {\bibinfo {author} {\bibfnamefont {C.}~\bibnamefont
  {Baek}}\ and\ \bibinfo {author} {\bibfnamefont {P.~M.}\ \bibnamefont
  {Reis}},\ }\bibfield  {title} {\enquote {\bibinfo {title} {Rigidity of
  hemispherical elastic gridshells under point load indentation},}\ }\href@noop
  {} {\bibfield  {journal} {\bibinfo  {journal} {J. Mech. Phys. Solids}\
  }\textbf {\bibinfo {volume} {124}},\ \bibinfo {pages} {411--426} (\bibinfo
  {year} {2019})}\BibitemShut {NoStop}%
\bibitem [{\citenamefont {Yamaguchi}, \citenamefont {Onoue},\ and\
  \citenamefont {Sawae}(2020)}]{Yamaguchi2020Network}%
  \BibitemOpen
  \bibfield  {author} {\bibinfo {author} {\bibfnamefont {T.}~\bibnamefont
  {Yamaguchi}}, \bibinfo {author} {\bibfnamefont {Y.}~\bibnamefont {Onoue}},\
  and\ \bibinfo {author} {\bibfnamefont {Y.}~\bibnamefont {Sawae}},\ }\bibfield
   {title} {\enquote {\bibinfo {title} {Topology and toughening of sparse
  elastic networks},}\ }\href {https://doi.org/10.1103/PhysRevLett.124.068002}
  {\bibfield  {journal} {\bibinfo  {journal} {Phys. Rev. Lett.}\ }\textbf
  {\bibinfo {volume} {124}},\ \bibinfo {pages} {068002} (\bibinfo {year}
  {2020})}\BibitemShut {NoStop}%
\bibitem [{\citenamefont {Stoop}, \citenamefont {Wittel},\ and\ \citenamefont
  {Herrmann}(2008)}]{Stoop2008Wire}%
  \BibitemOpen
  \bibfield  {author} {\bibinfo {author} {\bibfnamefont {N.}~\bibnamefont
  {Stoop}}, \bibinfo {author} {\bibfnamefont {F.~K.}\ \bibnamefont {Wittel}},\
  and\ \bibinfo {author} {\bibfnamefont {H.~J.}\ \bibnamefont {Herrmann}},\
  }\bibfield  {title} {\enquote {\bibinfo {title} {Morphological phases of
  crumpled wire},}\ }\href {https://doi.org/10.1103/PhysRevLett.101.094101}
  {\bibfield  {journal} {\bibinfo  {journal} {Phys. Rev. Lett.}\ }\textbf
  {\bibinfo {volume} {101}},\ \bibinfo {pages} {094101} (\bibinfo {year}
  {2008})}\BibitemShut {NoStop}%
\bibitem [{\citenamefont {Grason}(2009)}]{grason2009braided}%
  \BibitemOpen
  \bibfield  {author} {\bibinfo {author} {\bibfnamefont {G.~M.}\ \bibnamefont
  {Grason}},\ }\bibfield  {title} {\enquote {\bibinfo {title} {Braided bundles
  and compact coils: The structure and thermodynamics of hexagonally packed
  chiral filament assemblies},}\ }\href@noop {} {\bibfield  {journal} {\bibinfo
   {journal} {Phys. Rev. E.}\ }\textbf {\bibinfo {volume} {79}},\ \bibinfo
  {pages} {041919} (\bibinfo {year} {2009})}\BibitemShut {NoStop}%
\bibitem [{\citenamefont {Ward}\ \emph {et~al.}(2015)\citenamefont {Ward},
  \citenamefont {Hilitski}, \citenamefont {Schwenger}, \citenamefont {Welch},
  \citenamefont {Lau}, \citenamefont {Vitelli}, \citenamefont {Mahadevan},\
  and\ \citenamefont {Dogic}}]{Ward:2015fc}%
  \BibitemOpen
  \bibfield  {author} {\bibinfo {author} {\bibfnamefont {A.}~\bibnamefont
  {Ward}}, \bibinfo {author} {\bibfnamefont {F.}~\bibnamefont {Hilitski}},
  \bibinfo {author} {\bibfnamefont {W.}~\bibnamefont {Schwenger}}, \bibinfo
  {author} {\bibfnamefont {D.}~\bibnamefont {Welch}}, \bibinfo {author}
  {\bibfnamefont {A.~W.~C.}\ \bibnamefont {Lau}}, \bibinfo {author}
  {\bibfnamefont {V.}~\bibnamefont {Vitelli}}, \bibinfo {author} {\bibfnamefont
  {L.}~\bibnamefont {Mahadevan}},\ and\ \bibinfo {author} {\bibfnamefont
  {Z.}~\bibnamefont {Dogic}},\ }\bibfield  {title} {\enquote {\bibinfo {title}
  {{Solid friction between soft filaments}},}\ }\href@noop {} {\bibfield
  {journal} {\bibinfo  {journal} {Nat. Mater.}\ }\textbf {\bibinfo {volume}
  {14}},\ \bibinfo {pages} {583--588} (\bibinfo {year} {2015})}\BibitemShut
  {NoStop}%
\bibitem [{\citenamefont {Panaitescu}, \citenamefont {Grason},\ and\
  \citenamefont {Kudrolli}(2017)}]{panaitescu2017measuring}%
  \BibitemOpen
  \bibfield  {author} {\bibinfo {author} {\bibfnamefont {A.}~\bibnamefont
  {Panaitescu}}, \bibinfo {author} {\bibfnamefont {G.~M.}\ \bibnamefont
  {Grason}},\ and\ \bibinfo {author} {\bibfnamefont {A.}~\bibnamefont
  {Kudrolli}},\ }\bibfield  {title} {\enquote {\bibinfo {title} {Measuring
  geometric frustration in twisted inextensible filament bundles},}\
  }\href@noop {} {\bibfield  {journal} {\bibinfo  {journal} {Phys. Rev. E}\
  }\textbf {\bibinfo {volume} {95}},\ \bibinfo {pages} {052503} (\bibinfo
  {year} {2017})}\BibitemShut {NoStop}%
\bibitem [{\citenamefont {Panaitescu}, \citenamefont {Grason},\ and\
  \citenamefont {Kudrolli}(2018)}]{panaitescu2018persistence}%
  \BibitemOpen
  \bibfield  {author} {\bibinfo {author} {\bibfnamefont {A.}~\bibnamefont
  {Panaitescu}}, \bibinfo {author} {\bibfnamefont {G.~M.}\ \bibnamefont
  {Grason}},\ and\ \bibinfo {author} {\bibfnamefont {A.}~\bibnamefont
  {Kudrolli}},\ }\bibfield  {title} {\enquote {\bibinfo {title} {Persistence of
  perfect packing in twisted bundles of elastic filaments},}\ }\href@noop {}
  {\bibfield  {journal} {\bibinfo  {journal} {Phys. Rev. Lett.}\ }\textbf
  {\bibinfo {volume} {120}},\ \bibinfo {pages} {248002} (\bibinfo {year}
  {2018})}\BibitemShut {NoStop}%
\bibitem [{\citenamefont {Warren}, \citenamefont {Ball},\ and\ \citenamefont
  {Goldstein}(2018)}]{Warren2018}%
  \BibitemOpen
  \bibfield  {author} {\bibinfo {author} {\bibfnamefont {P.~B.}\ \bibnamefont
  {Warren}}, \bibinfo {author} {\bibfnamefont {R.~C.}\ \bibnamefont {Ball}},\
  and\ \bibinfo {author} {\bibfnamefont {R.~E.}\ \bibnamefont {Goldstein}},\
  }\bibfield  {title} {\enquote {\bibinfo {title} {Why clothes don't fall
  apart: Tension transmission in staple yarns},}\ }\href
  {https://doi.org/10.1103/physrevlett.120.158001} {\bibfield  {journal}
  {\bibinfo  {journal} {Phys. Rev. Lett.}\ }\textbf {\bibinfo {volume} {120}}
  (\bibinfo {year} {2018}),\ 10.1103/physrevlett.120.158001}\BibitemShut
  {NoStop}%
\bibitem [{\citenamefont {Maddocks}\ and\ \citenamefont
  {Keller}(1987)}]{Maddocks1987a}%
  \BibitemOpen
  \bibfield  {author} {\bibinfo {author} {\bibfnamefont {J.~H.}\ \bibnamefont
  {Maddocks}}\ and\ \bibinfo {author} {\bibfnamefont {J.~B.}\ \bibnamefont
  {Keller}},\ }\bibfield  {title} {\enquote {\bibinfo {title} {Ropes in
  equilibrium},}\ }\href {https://doi.org/10.1137/0147080} {\bibfield
  {journal} {\bibinfo  {journal} {SIAM J. Appl. Math.}\ }\textbf {\bibinfo
  {volume} {47}},\ \bibinfo {pages} {1185--1200} (\bibinfo {year}
  {1987})}\BibitemShut {NoStop}%
\bibitem [{\citenamefont {Audoly}, \citenamefont {Clauvelin},\ and\
  \citenamefont {Neukirch}(2007)}]{Audoly2007}%
  \BibitemOpen
  \bibfield  {author} {\bibinfo {author} {\bibfnamefont {B.}~\bibnamefont
  {Audoly}}, \bibinfo {author} {\bibfnamefont {N.}~\bibnamefont {Clauvelin}},\
  and\ \bibinfo {author} {\bibfnamefont {S.}~\bibnamefont {Neukirch}},\
  }\bibfield  {title} {\enquote {\bibinfo {title} {Elastic knots},}\ }\href
  {https://doi.org/10.1103/physrevlett.99.164301} {\bibfield  {journal}
  {\bibinfo  {journal} {Phys. Rev. Lett.}\ }\textbf {\bibinfo {volume} {99}}
  (\bibinfo {year} {2007}),\ 10.1103/physrevlett.99.164301}\BibitemShut
  {NoStop}%
\bibitem [{\citenamefont {Jawed}\ \emph {et~al.}(2015)\citenamefont {Jawed},
  \citenamefont {Dieleman}, \citenamefont {Audoly},\ and\ \citenamefont
  {Reis}}]{Jawed2015}%
  \BibitemOpen
  \bibfield  {author} {\bibinfo {author} {\bibfnamefont {M.~K.}\ \bibnamefont
  {Jawed}}, \bibinfo {author} {\bibfnamefont {P.}~\bibnamefont {Dieleman}},
  \bibinfo {author} {\bibfnamefont {B.}~\bibnamefont {Audoly}},\ and\ \bibinfo
  {author} {\bibfnamefont {P.~M.}\ \bibnamefont {Reis}},\ }\bibfield  {title}
  {\enquote {\bibinfo {title} {Untangling the mechanics and topology in the
  frictional response of long overhand elastic knots},}\ }\href
  {https://doi.org/10.1103/physrevlett.115.118302} {\bibfield  {journal}
  {\bibinfo  {journal} {Phys. Rev. Lett.}\ }\textbf {\bibinfo {volume} {115}}
  (\bibinfo {year} {2015}),\ 10.1103/physrevlett.115.118302}\BibitemShut
  {NoStop}%
\bibitem [{\citenamefont {Patil}\ \emph {et~al.}(2020)\citenamefont {Patil},
  \citenamefont {Sandt}, \citenamefont {Kolle},\ and\ \citenamefont
  {Dunkel}}]{Patil2020}%
  \BibitemOpen
  \bibfield  {author} {\bibinfo {author} {\bibfnamefont {V.~P.}\ \bibnamefont
  {Patil}}, \bibinfo {author} {\bibfnamefont {J.~D.}\ \bibnamefont {Sandt}},
  \bibinfo {author} {\bibfnamefont {M.}~\bibnamefont {Kolle}},\ and\ \bibinfo
  {author} {\bibfnamefont {J.}~\bibnamefont {Dunkel}},\ }\bibfield  {title}
  {\enquote {\bibinfo {title} {Topological mechanics of knots and tangles},}\
  }\href {https://doi.org/10.1126/science.aaz0135} {\bibfield  {journal}
  {\bibinfo  {journal} {Science}\ }\textbf {\bibinfo {volume} {367}},\ \bibinfo
  {pages} {71--75} (\bibinfo {year} {2020})}\BibitemShut {NoStop}%
\bibitem [{\citenamefont {Gonzalez}\ and\ \citenamefont
  {Maddocks}(1999)}]{Gonzalez1999}%
  \BibitemOpen
  \bibfield  {author} {\bibinfo {author} {\bibfnamefont {O.}~\bibnamefont
  {Gonzalez}}\ and\ \bibinfo {author} {\bibfnamefont {J.}~\bibnamefont
  {Maddocks}},\ }\bibfield  {title} {\enquote {\bibinfo {title} {Global
  curvature, thickness, and the ideal shapes of knots},}\ }\href@noop {}
  {\bibfield  {journal} {\bibinfo  {journal} {Proc. Natl. Acad. Sci. U.S.A.}\
  }\textbf {\bibinfo {volume} {96}} (\bibinfo {year} {1999})}\BibitemShut
  {NoStop}%
\bibitem [{\citenamefont {Carlen}\ \emph {et~al.}(2005)\citenamefont {Carlen},
  \citenamefont {Laurie}, \citenamefont {Maddocks},\ and\ \citenamefont
  {Smutny}}]{Carlen2005}%
  \BibitemOpen
  \bibfield  {author} {\bibinfo {author} {\bibfnamefont {M.}~\bibnamefont
  {Carlen}}, \bibinfo {author} {\bibfnamefont {B.}~\bibnamefont {Laurie}},
  \bibinfo {author} {\bibfnamefont {J.~H.}\ \bibnamefont {Maddocks}},\ and\
  \bibinfo {author} {\bibfnamefont {J.}~\bibnamefont {Smutny}},\ }\bibfield
  {title} {\enquote {\bibinfo {title} {Biarcs, global radius of curvature, and
  the computation of ideal knot shapes},}\ }in\ \href
  {https://doi.org/10.1142/9789812703460_0005} {\emph {\bibinfo {booktitle}
  {Physical and Numerical Models in Knot Theory and Their Application to the
  Life Sciences}}}\ (\bibinfo  {publisher} {World Scientific Publishing Co.
  Pte. Ltd.},\ \bibinfo {year} {2005})\BibitemShut {NoStop}%
\bibitem [{\citenamefont {Starostin}(2003)}]{Starostin2003}%
  \BibitemOpen
  \bibfield  {author} {\bibinfo {author} {\bibfnamefont {E.~L.}\ \bibnamefont
  {Starostin}},\ }\bibfield  {title} {\enquote {\bibinfo {title} {A
  constructive approach to modelling the tight shapes of some linked
  structures},}\ }\href@noop {} {\bibfield  {journal} {\bibinfo  {journal}
  {Forma}\ }\textbf {\bibinfo {volume} {18}},\ \bibinfo {pages} {263--293}
  (\bibinfo {year} {2003})}\BibitemShut {NoStop}%
\bibitem [{\citenamefont {Antman}(2005)}]{antman05}%
  \BibitemOpen
  \bibfield  {author} {\bibinfo {author} {\bibfnamefont {S.~S.}\ \bibnamefont
  {Antman}},\ }\href@noop {} {\emph {\bibinfo {title} {Nonlinear Problems of
  Elasticity}}}\ (\bibinfo  {publisher} {Springer},\ \bibinfo {address} {New
  York},\ \bibinfo {year} {2005})\BibitemShut {NoStop}%
\bibitem [{\citenamefont {Euler}(1769)}]{Euler1769}%
  \BibitemOpen
  \bibfield  {author} {\bibinfo {author} {\bibfnamefont {L.}~\bibnamefont
  {Euler}},\ }\bibfield  {title} {\enquote {\bibinfo {title} {Remarques sur
  l'effet du frottement dans l'equilibre},}\ }\href@noop {} {\bibfield
  {journal} {\bibinfo  {journal} {Memoires de l'academie des sciences de
  Berlin}\ }\textbf {\bibinfo {volume} {-}},\ \bibinfo {pages} {265--278}
  (\bibinfo {year} {1769})}\BibitemShut {NoStop}%
\bibitem [{\citenamefont {Eytelwein}(1832)}]{Eytelwein1832}%
  \BibitemOpen
  \bibfield  {author} {\bibinfo {author} {\bibfnamefont {J.~A.}\ \bibnamefont
  {Eytelwein}},\ }\href@noop {} {\emph {\bibinfo {title} {Handbuch der Statik
  fester K{\"o}rper: mit vorz{\"u}glicher R{\"u}cksicht auf ihre Anwendung in
  der Architektur}}},\ Vol.~\bibinfo {volume} {1}\ (\bibinfo  {publisher}
  {Reimer},\ \bibinfo {year} {1832})\BibitemShut {NoStop}%
\end{thebibliography}%


\begin{thebibliography}{23}%
\makeatletter
\providecommand \@ifxundefined [1]{%
 \@ifx{#1\undefined}
}%
\providecommand \@ifnum [1]{%
 \ifnum #1\expandafter \@firstoftwo
 \else \expandafter \@secondoftwo
 \fi
}%
\providecommand \@ifx [1]{%
 \ifx #1\expandafter \@firstoftwo
 \else \expandafter \@secondoftwo
 \fi
}%
\providecommand \natexlab [1]{#1}%
\providecommand \enquote  [1]{``#1''}%
\providecommand \bibnamefont  [1]{#1}%
\providecommand \bibfnamefont [1]{#1}%
\providecommand \citenamefont [1]{#1}%
\providecommand \href@noop [0]{\@secondoftwo}%
\providecommand \href [0]{\begingroup \@sanitize@url \@href}%
\providecommand \@href[1]{\@@startlink{#1}\@@href}%
\providecommand \@@href[1]{\endgroup#1\@@endlink}%
\providecommand \@sanitize@url [0]{\catcode `\\12\catcode `\$12\catcode
  `\&12\catcode `\#12\catcode `\^12\catcode `\_12\catcode `\%12\relax}%
\providecommand \@@startlink[1]{}%
\providecommand \@@endlink[0]{}%
\providecommand \url  [0]{\begingroup\@sanitize@url \@url }%
\providecommand \@url [1]{\endgroup\@href {#1}{\urlprefix }}%
\providecommand \urlprefix  [0]{URL }%
\providecommand \Eprint [0]{\href }%
\providecommand \doibase [0]{https://doi.org/}%
\providecommand \selectlanguage [0]{\@gobble}%
\providecommand \bibinfo  [0]{\@secondoftwo}%
\providecommand \bibfield  [0]{\@secondoftwo}%
\providecommand \translation [1]{[#1]}%
\providecommand \BibitemOpen [0]{}%
\providecommand \bibitemStop [0]{}%
\providecommand \bibitemNoStop [0]{.\EOS\space}%
\providecommand \EOS [0]{\spacefactor3000\relax}%
\providecommand \BibitemShut  [1]{\csname bibitem#1\endcsname}%
\let\auto@bib@innerbib\@empty
\bibitem [{\citenamefont {Landau}\ and\ \citenamefont
  {Levich}(1988)}]{Landau1988}%
  \BibitemOpen
  \bibfield  {author} {\bibinfo {author} {\bibfnamefont {L.}~\bibnamefont
  {Landau}}\ and\ \bibinfo {author} {\bibfnamefont {B.}~\bibnamefont
  {Levich}},\ }\href {https://doi.org/10.1016/b978-0-08-092523-3.50016-2}
  {\emph {\bibinfo {title} {Dynamics of Curved Fronts}}}\ (\bibinfo
  {publisher} {Elsevier},\ \bibinfo {year} {1988})\ pp.\ \bibinfo {pages}
  {141--153}\BibitemShut {NoStop}%
\bibitem [{\citenamefont {Mayer}\ and\ \citenamefont
  {Krechetnikov}(2012)}]{Mayer2012}%
  \BibitemOpen
  \bibfield  {author} {\bibinfo {author} {\bibfnamefont {H.~C.}\ \bibnamefont
  {Mayer}}\ and\ \bibinfo {author} {\bibfnamefont {R.}~\bibnamefont
  {Krechetnikov}},\ }\bibfield  {title} {\enquote {\bibinfo {title}
  {Landau-levich flow visualization: Revealing the flow topology responsible
  for the film thickening phenomena},}\ }\href
  {https://doi.org/10.1063/1.4703924} {\bibfield  {journal} {\bibinfo
  {journal} {Phys. Fluids}\ }\textbf {\bibinfo {volume} {24}},\ \bibinfo
  {pages} {052103} (\bibinfo {year} {2012})}\BibitemShut {NoStop}%
\bibitem [{\citenamefont {Grundke}\ \emph {et~al.}(2008)\citenamefont
  {Grundke}, \citenamefont {Michel}, \citenamefont {Knispel},\ and\
  \citenamefont {Grundler}}]{Grundke2008}%
  \BibitemOpen
  \bibfield  {author} {\bibinfo {author} {\bibfnamefont {K.}~\bibnamefont
  {Grundke}}, \bibinfo {author} {\bibfnamefont {S.}~\bibnamefont {Michel}},
  \bibinfo {author} {\bibfnamefont {G.}~\bibnamefont {Knispel}},\ and\ \bibinfo
  {author} {\bibfnamefont {A.}~\bibnamefont {Grundler}},\ }\bibfield  {title}
  {\enquote {\bibinfo {title} {Wettability of silicone and polyether impression
  materials: Characterization by surface tension and contact angle
  measurements},}\ }\href {https://doi.org/10.1016/j.colsurfa.2007.11.046}
  {\bibfield  {journal} {\bibinfo  {journal} {Colloids Surf. A Physicochem.
  Eng. Asp.}\ }\textbf {\bibinfo {volume} {317}},\ \bibinfo {pages} {598--609}
  (\bibinfo {year} {2008})}\BibitemShut {NoStop}%
\bibitem [{\citenamefont {Rio}\ and\ \citenamefont {Boulogne}(2017)}]{Rio2017}%
  \BibitemOpen
  \bibfield  {author} {\bibinfo {author} {\bibfnamefont {E.}~\bibnamefont
  {Rio}}\ and\ \bibinfo {author} {\bibfnamefont {F.}~\bibnamefont {Boulogne}},\
  }\bibfield  {title} {\enquote {\bibinfo {title} {Withdrawing a solid from a
  bath: How much liquid is coated?}}\ }\href
  {https://doi.org/10.1016/j.cis.2017.01.006} {\bibfield  {journal} {\bibinfo
  {journal} {Adv. Colloid Interface Sci.}\ }\textbf {\bibinfo {volume} {247}},\
  \bibinfo {pages} {100--114} (\bibinfo {year} {2017})}\BibitemShut {NoStop}%
\bibitem [{\citenamefont {Goucher}\ and\ \citenamefont
  {Ward}(1922)}]{Goucher1922}%
  \BibitemOpen
  \bibfield  {author} {\bibinfo {author} {\bibfnamefont {F.}~\bibnamefont
  {Goucher}}\ and\ \bibinfo {author} {\bibfnamefont {H.}~\bibnamefont {Ward}},\
  }\bibfield  {title} {\enquote {\bibinfo {title} {A problem in viscosity},}\
  }\href@noop {} {\bibfield  {journal} {\bibinfo  {journal} {Phil. Mag}\
  }\textbf {\bibinfo {volume} {44}},\ \bibinfo {pages} {1002} (\bibinfo {year}
  {1922})}\BibitemShut {NoStop}%
\bibitem [{\citenamefont {Herman}(2009)}]{Herman2009}%
  \BibitemOpen
  \bibfield  {author} {\bibinfo {author} {\bibfnamefont {G.~T.}\ \bibnamefont
  {Herman}},\ }\href {https://doi.org/10.1007/978-1-84628-723-7} {\emph
  {\bibinfo {title} {Fundamentals of Computerized Tomography}}}\ (\bibinfo
  {publisher} {Springer London},\ \bibinfo {year} {2009})\BibitemShut {NoStop}%
\bibitem [{\citenamefont {Schindelin}\ \emph {et~al.}(2012)\citenamefont
  {Schindelin}, \citenamefont {Arganda-Carreras}, \citenamefont {Frise},
  \citenamefont {Kaynig}, \citenamefont {Longair}, \citenamefont {Pietzsch},
  \citenamefont {Preibisch}, \citenamefont {Rueden}, \citenamefont {Saalfeld},
  \citenamefont {Schmid}, \citenamefont {Tinevez}, \citenamefont {White},
  \citenamefont {Hartenstein}, \citenamefont {Eliceiri}, \citenamefont
  {Tomancak},\ and\ \citenamefont {Cardona}}]{Schindelin2012}%
  \BibitemOpen
  \bibfield  {author} {\bibinfo {author} {\bibfnamefont {J.}~\bibnamefont
  {Schindelin}}, \bibinfo {author} {\bibfnamefont {I.}~\bibnamefont
  {Arganda-Carreras}}, \bibinfo {author} {\bibfnamefont {E.}~\bibnamefont
  {Frise}}, \bibinfo {author} {\bibfnamefont {V.}~\bibnamefont {Kaynig}},
  \bibinfo {author} {\bibfnamefont {M.}~\bibnamefont {Longair}}, \bibinfo
  {author} {\bibfnamefont {T.}~\bibnamefont {Pietzsch}}, \bibinfo {author}
  {\bibfnamefont {S.}~\bibnamefont {Preibisch}}, \bibinfo {author}
  {\bibfnamefont {C.}~\bibnamefont {Rueden}}, \bibinfo {author} {\bibfnamefont
  {S.}~\bibnamefont {Saalfeld}}, \bibinfo {author} {\bibfnamefont
  {B.}~\bibnamefont {Schmid}}, \bibinfo {author} {\bibfnamefont {J.-Y.}\
  \bibnamefont {Tinevez}}, \bibinfo {author} {\bibfnamefont {D.~J.}\
  \bibnamefont {White}}, \bibinfo {author} {\bibfnamefont {V.}~\bibnamefont
  {Hartenstein}}, \bibinfo {author} {\bibfnamefont {K.}~\bibnamefont
  {Eliceiri}}, \bibinfo {author} {\bibfnamefont {P.}~\bibnamefont {Tomancak}},\
  and\ \bibinfo {author} {\bibfnamefont {A.}~\bibnamefont {Cardona}},\
  }\bibfield  {title} {\enquote {\bibinfo {title} {Fiji: an open-source
  platform for biological-image analysis},}\ }\href
  {https://doi.org/10.1038/nmeth.2019} {\bibfield  {journal} {\bibinfo
  {journal} {Nat. Methods}\ }\textbf {\bibinfo {volume} {9}},\ \bibinfo {pages}
  {676--682} (\bibinfo {year} {2012})}\BibitemShut {NoStop}%
\bibitem [{\citenamefont {Persson}(2000)}]{persson_book}%
  \BibitemOpen
  \bibfield  {author} {\bibinfo {author} {\bibfnamefont {B.~N.~J.}\
  \bibnamefont {Persson}},\ }\href@noop {} {\emph {\bibinfo {title} {Sliding
  Friction}}}\ (\bibinfo  {publisher} {Springer-Verlag},\ \bibinfo {year}
  {2000})\BibitemShut {NoStop}%
\bibitem [{\citenamefont {Gent}(2001)}]{Gent2001ed}%
  \BibitemOpen
  \bibfield  {author} {\bibinfo {author} {\bibfnamefont {A.~N.}\ \bibnamefont
  {Gent}},\ }\href@noop {} {\emph {\bibinfo {title} {Engineering with
  rubber}}}\ (\bibinfo  {publisher} {Carl Hanser Verlag, Munich},\ \bibinfo
  {year} {2001})\BibitemShut {NoStop}%
\bibitem [{AST(2014)}]{ASTM-friction}%
  \BibitemOpen
  \href {https://doi.org/10.1520/d1894-14} {\enquote {\bibinfo {title} {Test
  method for static and kinetic coefficients of friction of plastic film and
  sheeting -- {ASTM(D1894-14)}},}\ } (\bibinfo {year} {2014})\BibitemShut
  {NoStop}%
\bibitem [{\citenamefont {Audoly}\ and\ \citenamefont
  {Pomeau}(2010)}]{AudolyBook}%
  \BibitemOpen
  \bibfield  {author} {\bibinfo {author} {\bibfnamefont {B.}~\bibnamefont
  {Audoly}}\ and\ \bibinfo {author} {\bibfnamefont {Y.}~\bibnamefont
  {Pomeau}},\ }\href {https://doi.org/10.1007/978-1-84628-723-7} {\emph
  {\bibinfo {title} {Elasticity and Geometry}}}\ (\bibinfo  {publisher} {Oxford
  University Press},\ \bibinfo {year} {2010})\BibitemShut {NoStop}%
\bibitem [{\citenamefont {Crandall}\ \emph {et~al.}(1978)\citenamefont
  {Crandall}, \citenamefont {Lardner}, \citenamefont {Archer}, \citenamefont
  {Cook},\ and\ \citenamefont {Dahl}}]{crandall1978introduction}%
  \BibitemOpen
  \bibfield  {author} {\bibinfo {author} {\bibfnamefont {S.~H.}\ \bibnamefont
  {Crandall}}, \bibinfo {author} {\bibfnamefont {T.~J.}\ \bibnamefont
  {Lardner}}, \bibinfo {author} {\bibfnamefont {R.~R.}\ \bibnamefont {Archer}},
  \bibinfo {author} {\bibfnamefont {N.~H.}\ \bibnamefont {Cook}},\ and\
  \bibinfo {author} {\bibfnamefont {N.~C.}\ \bibnamefont {Dahl}},\ }\href@noop
  {} {\emph {\bibinfo {title} {An introduction to the mechanics of solids}}}\
  (\bibinfo  {publisher} {McGraw-Hill},\ \bibinfo {year} {1978})\BibitemShut
  {NoStop}%
\bibitem [{\citenamefont {Antman}(2005)}]{antman05}%
  \BibitemOpen
  \bibfield  {author} {\bibinfo {author} {\bibfnamefont {S.~S.}\ \bibnamefont
  {Antman}},\ }\href@noop {} {\emph {\bibinfo {title} {Nonlinear Problems of
  Elasticity}}}\ (\bibinfo  {publisher} {Springer},\ \bibinfo {address} {New
  York},\ \bibinfo {year} {2005})\BibitemShut {NoStop}%
\bibitem [{\citenamefont {Dichmann}, \citenamefont {Li},\ and\ \citenamefont
  {Maddocks}(1996)}]{Dichmann1996}%
  \BibitemOpen
  \bibfield  {author} {\bibinfo {author} {\bibfnamefont {D.~J.}\ \bibnamefont
  {Dichmann}}, \bibinfo {author} {\bibfnamefont {Y.}~\bibnamefont {Li}},\ and\
  \bibinfo {author} {\bibfnamefont {J.~H.}\ \bibnamefont {Maddocks}},\
  }\bibfield  {title} {\enquote {\bibinfo {title} {Hamiltonian formulations and
  symmetries of rod mechanics},}\ }in\ \href@noop {} {\emph {\bibinfo
  {booktitle} {Mathematical Approaches to Biomolecular Structure and
  Dynamics}}},\ \bibinfo {editor} {edited by\ \bibinfo {editor} {\bibfnamefont
  {J.~P.}\ \bibnamefont {Mesirov}}, \bibinfo {editor} {\bibfnamefont
  {K.}~\bibnamefont {Schulten}},\ and\ \bibinfo {editor} {\bibfnamefont
  {D.~W.}\ \bibnamefont {Sumners}}}\ (\bibinfo  {publisher} {Springer New
  York},\ \bibinfo {year} {1996})\ pp.\ \bibinfo {pages} {71--113}\BibitemShut
  {NoStop}%
\bibitem [{\citenamefont {Kehrbaum}\ and\ \citenamefont
  {Maddocks}(1997)}]{Kehrbaummaddocks1997}%
  \BibitemOpen
  \bibfield  {author} {\bibinfo {author} {\bibfnamefont {S.}~\bibnamefont
  {Kehrbaum}}\ and\ \bibinfo {author} {\bibfnamefont {J.~H.}\ \bibnamefont
  {Maddocks}},\ }\bibfield  {title} {\enquote {\bibinfo {title} {Elastic rods,
  rigid bodies, quaternions and the last quadrature},}\ }\href@noop {}
  {\bibfield  {journal} {\bibinfo  {journal} {Philos. Trans. R. Soc. A}\
  }\textbf {\bibinfo {volume} {355}},\ \bibinfo {pages} {2117--2136} (\bibinfo
  {year} {1997})}\BibitemShut {NoStop}%
\bibitem [{\citenamefont {Hinch}(1991)}]{Hinch1991}%
  \BibitemOpen
  \bibfield  {author} {\bibinfo {author} {\bibfnamefont {E.~J.}\ \bibnamefont
  {Hinch}},\ }\href@noop {} {\emph {\bibinfo {title} {Perturbation Methods}}}\
  (\bibinfo  {publisher} {Cambridge University Press},\ \bibinfo {address}
  {Cambridge},\ \bibinfo {year} {1991})\BibitemShut {NoStop}%
\bibitem [{\citenamefont {Elasticity}\ and\ \citenamefont {geometry: from hair
  curls to the non-linear response~of shells}(2010)}]{Audoly2010}%
  \BibitemOpen
  \bibfield  {author} {\bibinfo {author} {\bibnamefont {Elasticity}}\ and\
  \bibinfo {author} {\bibnamefont {geometry: from hair curls to the non-linear
  response~of shells}},\ }\href@noop {} {\emph {\bibinfo {title} {B. Audoly and
  Y. Pomeau}}}\ (\bibinfo  {publisher} {Oxford University Press},\ \bibinfo
  {address} {Oxford},\ \bibinfo {year} {2010})\BibitemShut {NoStop}%
\bibitem [{\citenamefont {Miura}(2020)}]{Miura2020}%
  \BibitemOpen
  \bibfield  {author} {\bibinfo {author} {\bibfnamefont {T.}~\bibnamefont
  {Miura}},\ }\bibfield  {title} {\enquote {\bibinfo {title} {Elastic curves
  and phase transitions},}\ }\href@noop {} {\bibfield  {journal} {\bibinfo
  {journal} {Math. Ann.}\ }\textbf {\bibinfo {volume} {376}},\ \bibinfo {pages}
  {1629--1674} (\bibinfo {year} {2020})}\BibitemShut {NoStop}%
\bibitem [{\citenamefont {Maddocks}\ and\ \citenamefont
  {Keller}(1987)}]{maddockskeller1987}%
  \BibitemOpen
  \bibfield  {author} {\bibinfo {author} {\bibfnamefont {J.~H.}\ \bibnamefont
  {Maddocks}}\ and\ \bibinfo {author} {\bibfnamefont {J.~B.}\ \bibnamefont
  {Keller}},\ }\bibfield  {title} {\enquote {\bibinfo {title} {Ropes in
  equilibrium},}\ }\href@noop {} {\bibfield  {journal} {\bibinfo  {journal}
  {SIAM J. Appl. Math.}\ }\textbf {\bibinfo {volume} {47}},\ \bibinfo {pages}
  {1185--1200} (\bibinfo {year} {1987})}\BibitemShut {NoStop}%
\bibitem [{\citenamefont {O'Reilly}(2017)}]{oreilly2017}%
  \BibitemOpen
  \bibfield  {author} {\bibinfo {author} {\bibfnamefont {O.~M.}\ \bibnamefont
  {O'Reilly}},\ }\href@noop {} {\emph {\bibinfo {title} {Modeling Nonlinear
  Problems in the Mechanics of Strings and Rods}}}\ (\bibinfo  {publisher}
  {Springer},\ \bibinfo {address} {New York},\ \bibinfo {year}
  {2017})\BibitemShut {NoStop}%
\bibitem [{\citenamefont {Starostin}(2003)}]{Starostin2003}%
  \BibitemOpen
  \bibfield  {author} {\bibinfo {author} {\bibfnamefont {E.~L.}\ \bibnamefont
  {Starostin}},\ }\bibfield  {title} {\enquote {\bibinfo {title} {A
  constructive approach to modelling the tight shapes of some linked
  structures},}\ }\href@noop {} {\bibfield  {journal} {\bibinfo  {journal}
  {Forma}\ }\textbf {\bibinfo {volume} {18}},\ \bibinfo {pages} {263--293}
  (\bibinfo {year} {2003})}\BibitemShut {NoStop}%
\bibitem [{\citenamefont {Cantarella}\ \emph {et~al.}(2006)\citenamefont
  {Cantarella}, \citenamefont {Fu}, \citenamefont {R.~Kusner},\ and\
  \citenamefont {Wrinkle}}]{Cantarella2006}%
  \BibitemOpen
  \bibfield  {author} {\bibinfo {author} {\bibfnamefont {J.}~\bibnamefont
  {Cantarella}}, \bibinfo {author} {\bibfnamefont {J.~H.~G.}\ \bibnamefont
  {Fu}}, \bibinfo {author} {\bibfnamefont {J.~M.~S.}\ \bibnamefont
  {R.~Kusner}},\ and\ \bibinfo {author} {\bibfnamefont {N.~C.}\ \bibnamefont
  {Wrinkle}},\ }\bibfield  {title} {\enquote {\bibinfo {title} {Criticality for
  the {G}ehring link problem},}\ }\href@noop {} {\bibfield  {journal} {\bibinfo
   {journal} {Geom. Topol.}\ }\textbf {\bibinfo {volume} {10}},\ \bibinfo
  {pages} {2055--2115} (\bibinfo {year} {2006})}\BibitemShut {NoStop}%
\bibitem [{\citenamefont {Belofsky}(1976)}]{Belofsky1976}%
  \BibitemOpen
  \bibfield  {author} {\bibinfo {author} {\bibfnamefont {H.}~\bibnamefont
  {Belofsky}},\ }\bibfield  {title} {\enquote {\bibinfo {title} {On the theory
  of power transmission by {V}-belts},}\ }\href
  {https://doi.org/10.1016/0043-1648(76)90054-5} {\bibfield  {journal}
  {\bibinfo  {journal} {Wear}\ }\textbf {\bibinfo {volume} {39}},\ \bibinfo
  {pages} {263--275} (\bibinfo {year} {1976})}\BibitemShut {NoStop}%
\end{thebibliography}%

\end{document}


\preprint{AIP/123-QED}

\title[]{SUPPLEMENTARY INFORMATION -- Mechanics of two filaments in tight contact: The orthogonal clasp}%

\author{Paul Grandgeorge}
 \affiliation{
 Flexible Structures Laboratory, Institute of Mechanical Engineering, École Polytechnique Fédérale de Lausanne (EPFL), Lausanne, Switzerland
 }
 
\author{Changyeob Baek}%
\affiliation{ 
Department of Mechanical Engineering, Massachusetts Institute of Technology, Cambridge, USA
}%

\author{Harmeet Singh}
\affiliation{%
Laboratory for Computation and Visualization in Mathematics and Mechanics, Institute of Mathematics, École Polytechnique Fédérale de Lausanne (EPFL), Switzerland
}%

\author{Paul Johanns}
 \affiliation{
 Flexible Structures Laboratory, Institute of Mechanical Engineering, École Polytechnique Fédérale de Lausanne (EPFL), Lausanne, Switzerland
 }

 \author{Tomohiko G. Sano}
 \affiliation{
 Flexible Structures Laboratory, Institute of Mechanical Engineering, École Polytechnique Fédérale de Lausanne (EPFL), Lausanne, Switzerland
 }
 
 \author{Alastair Flynn}
\affiliation{%
Laboratory for Computation and Visualization in Mathematics and Mechanics, Institute of Mathematics, École Polytechnique Fédérale de Lausanne (EPFL), Switzerland
}%

\author{John H. Maddocks}
 \email{john.maddocks@epfl.ch}
\affiliation{%
Laboratory for Computation and Visualization in Mathematics and Mechanics, Institute of Mathematics, École Polytechnique Fédérale de Lausanne (EPFL), Switzerland
}%
\author{Pedro M. Reis}
 \email{pedro.reis@epfl.ch}
\affiliation{
 Flexible Structures Laboratory, Institute of Mechanical Engineering, École Polytechnique Fédérale de Lausanne (EPFL), Lausanne, Switzerland
 }%

\date{\today}

\maketitle

\section{EXPERIMENTAL PROTOCOLS -- X-ray tomographic imaging and mechanical testing of the orthogonal elastic clasp}
\label{sect:Exp}
In this section, we describe the experimental methods to perform X-ray micro-computed tomography ($\mu$CT) and the subsequent analysis of the \textit{static elastic orthogonal clasp} (\textit{i.e.}, no friction). We also present the procedure for the mechanical testing of the \textit{sliding elastic orthogonal clasp} (\textit{i.e.}, with friction). First, in \secref{sect:rod_fabrication_tomography}, we detail the fabrication protocol that we developed to produce composite (coaxial) elastomeric rods. These rods were made compatible with $\mu$CT imaging for the geometrical characterization of the \textit{elastic orthogonal clasp}. Then, in \secref{sect:micro_computed_tomography}, we describe the procedure to capture the $\mu$CT images of the \textit{elastic orthogonal clasp}. In \secref{sect:image_analysis}, we present the subsequent image-processing algorithm that we developed to track the coordinates of the centerlines of the touching rods, as well as their contacting region. Making use of the image-processing algorithm, in \secref{sect:geom_quantification}, we quantify the complex three-dimensional geometry of the contact in the elastic clasp. In \secref{sect:rod_fabrication_mechanical_testing}, we focus on the protocol that we designed to fabricate the rods used for mechanical testing of the elastic clasp in either the presence or the absence of friction. Finally, in \secref{sect:mechanical_characterization}, we report on the material characterization of the polymers used to fabricate the rods and the characterization of the frictional behavior of our polymeric rod samples. 

\subsection{Fabrication of the rods for $\mu$CT imaging}
\label{sect:rod_fabrication_tomography}
\begin{figure}[h!]
    \centering
    \includegraphics[width=0.8\textwidth]{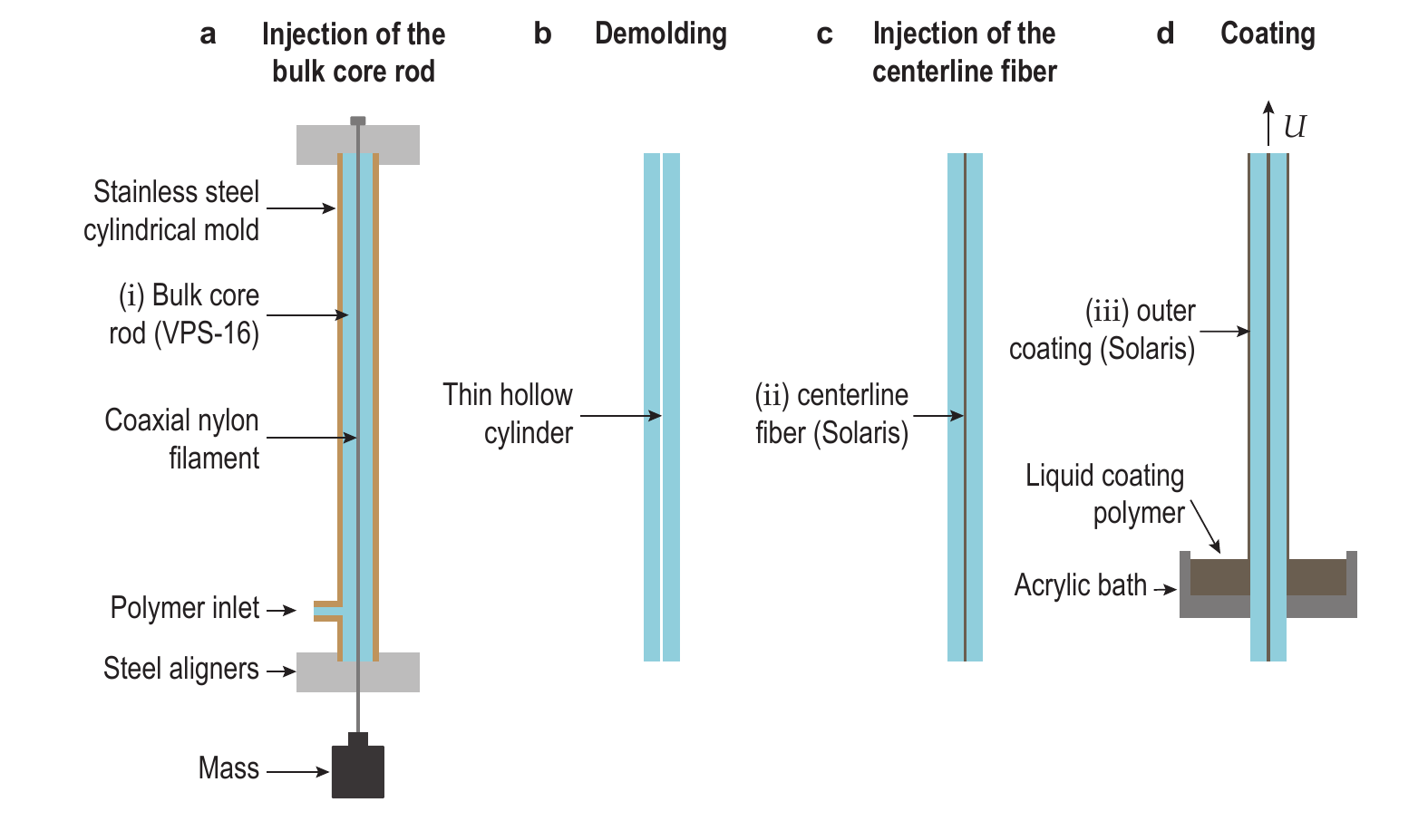}
    \caption{\textbf{Schematics of the fabrication protocol of the coaxial composite VPS rods used for $\mu$CT imaging.} \textbf{a},~The bulk core rod was cast inside a stainless steel cylindrical tube by injecting the liquid polymer (VPS-16) through the lower inlet of the mold. A weight of mass $M=200\,\text{g}$, together with machined aligners, ensured coaxiality of the nylon filament. \textbf{b},~Demolding of the bulk core rod from the mold and extraction of the nylon fiber yielded a monolithic polymeric rod with a void along its centerline. \textbf{c},~Injection of liquid polymer (Solaris) inside the void generated the material centerline. \textbf{d},~A dip-coating technique~\cite{Landau1988} was used to uniformly deposit an outer polymeric coating layer on the bulk core rod; the bulk core rod is pulled out vertically from a circular bath containing uncured Solaris, at a constant velocity $U$.} 
    \label{fig:rod_fabrication_uCT_summary}
\end{figure}

This subsection describes the protocol we developed to fabricate the composite elastomeric rods used for $\mu$CT volumetric imaging of the \textit{static elastic orthogonal clasp}. The main goals of this fabrication specifics were twofold. First, we seek to extract the physical centerline coordinates of the two rods of the \textit{elastic orthogonal clasp}. Second, we want to quantify the contact geometry between the two rods of the clasp. For this purpose, our composite rods comprised three different regions: (i) a bulk core rod embedded within (ii) a physical centerline fiber, and (iii) an outer coating layer. The physical centerline fiber and the outer coating layer were required to be made of a material with a sufficiently lower density than the density of the bulk core rod (i) (see \secref{sect:image_analysis}). In the following paragraphs, we describe the steps necessary to fabricate parts (i), (ii), (iii); each of these steps is summarized schematically in~\figref{fig:rod_fabrication_uCT_summary}).

Next, we describe, in more detail, the fabrication of each of the three components of the composite rod: (i) the bulk core rod, (ii) the physical centerline fiber, and (iii) the outer coating layer. 

\begin{figure}[h!] 
    \centering
    \includegraphics[width=0.7\textwidth]{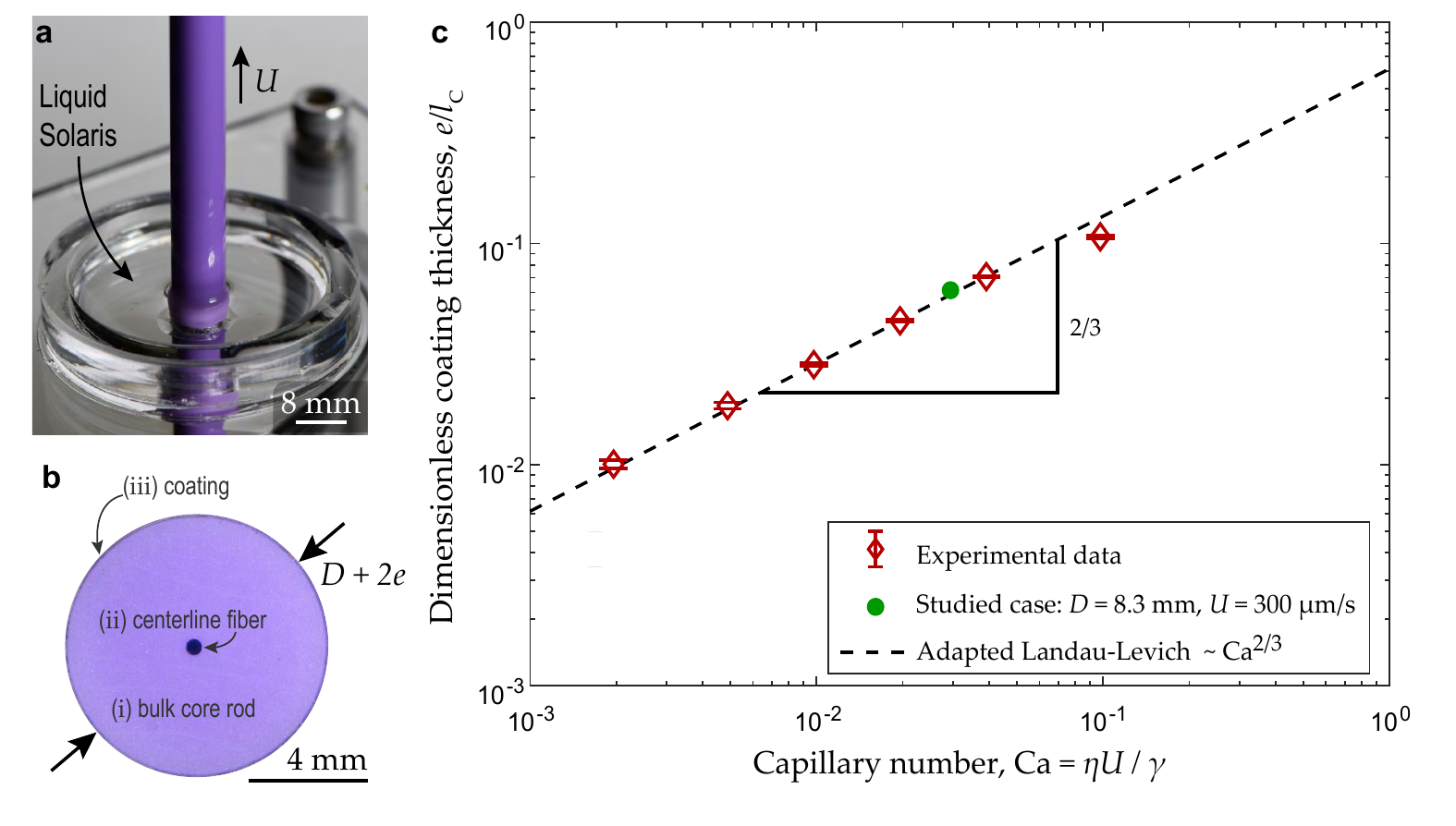}
    \caption{\textbf{Coating of the VPS-16 core rod with Solaris.} \textbf{a}, Experimental apparatus for dip-coating of the VPS-16 rods. \textbf{b}, Micrograph of a cross-sectional cut of the coated rod embedded with the physical centerline fiber (image obtained using a VHX 5000 digital microscope, Keyence). \textbf{c}, Dimensionless coating thickness, $e/l_c$ ($l_c$ the capillary length of the Solaris polymer), as a function of the Capillary number, $\text{Ca}$ (see text for the definitions of $l_c$ and $\text{Ca}$). The dashed line corresponds to the prediction from \eqref{eq:coating_thickness_vs_Ca}. The green datapoint refers to the coating conditions (velocity~$U=300\,\mu$m/s) selected to achieve a coating thickness~of $e=150\,\mu$m.}
\label{fig:CoatingCollage}
\end{figure}

\begin{enumerate}[label=(\roman*)]
\item The \textbf{bulk core rod} was cast with Vinylpolysiloxane (VPS-16, Zhermack) using a stainless-steel cylindrical mold (SS pipes, part number PSTS12A-400, Misumi). This casting tube had length $L_\text{m} = 400\,\text{mm}$ and inner diameter $D_\text{m}=8.3\,\text{mm}$, which set the outer diameter of the bulk core rod upon demolding (\figref{fig:rod_fabrication_uCT_summary}\textbf{a}). The density of cured VPS-16 was measured to be $\rho_\text{VPS16} = 1160\,$~kg/m$^3$ using a pycnometer (25~mL Gay-Lussac Pycnometer, Milian) and its Young's modulus was $E=520\,\text{kPa}$ (see \secref{sect:mechanical_characterization} for mechanical characterization).
\item A \textbf{physical fiber} was embedded into the bulk of the rod to act as a thin material centerline. The following procedure was used to fabricate this centerline fiber. First, custom-machined steel aligners were placed at both ends of the casting tube to align a straight nylon filament (diameter $D_\text{n} = 0.5\,\text{mm}$) concentrically. During casting, we kept the nylon filament under tension by hanging a weight of mass $M=200\,\text{g}$ at its lower extremity (see \figref{fig:rod_fabrication_uCT_summary}\textbf{a}). Liquid-phase VPS-16 was then injected into the assembly to cast the bulk core rod. After curing, the rod was demolded and the nylon thread was pulled out, leaving a thin hollow cylinder (of diameter $D_\text{n}=0.5\,\text{mm}$) along its central axis of the rod (\figref{fig:rod_fabrication_uCT_summary}\textbf{b}). Next, we filled up this void by injecting still-liquid platinum cure silicone rubber (Solaris, Smooth-On), (\figref{fig:rod_fabrication_uCT_summary}\textbf{c}). The Solaris polymer cured in approximately 24 hours after injection, with a final Young's modulus of $E_{\textrm{solaris}}=320\,\text{kPa}$ and density $\rho_{\textrm{solaris}} = 1001\,$kg/m$^3$. It is important to highlight that the density difference of $13\,\%$ between the cured Solaris and VPS-16 polymers was sufficiently large to enable the differentiation and segmentation of the material centerline fiber from the bulk core rod during $\mu$CT imaging (more details on this below). This alignment procedure ensured a satisfactory concentricity of the material centerline and the rod; the deviation of the centroid of the physical centerline from the centroid of the rod of diameter $D$ was less than $\delta=0.015D$.
\item An \textbf{outer coating} of Solaris polymer was deposited on the outer surface of the rod (\figref{fig:rod_fabrication_uCT_summary}\textbf{d}) to  enable the quantification of the contact regions during the $\mu$CT imaging (see \figref{fig:uCT_imaging}). This outer coating was fabricated using a dip-coating technique to reproduce a homogeneous thin film of uniform thickness along the rods~\cite{Landau1988,Mayer2012}. The photograph of our experimental coating setup is shown in \figref{fig:CoatingCollage}\textbf{a}. The VPS-16 core rod, including the centerline fiber described above in (ii), was pulled vertically out of a bath of liquid Solaris, at a constant velocity $U$, yielding a coating of thickness $e=150\,\mu\text{m}$ (see \figref{fig:CoatingCollage}\textbf{b}). To avoid using a large liquid bath, the bulk core rod was threaded through a concentric hole at the bottom of a shallow cylindrical acrylic bath (depth of the bath~$\approx 5\,\text{mm}$) and passed through the liquid Solaris before being extruded perpendicularly to the upper free surface. The hole at the bottom of the bath had the same diameter as the bulk core rod ($8.3\,\text{mm}$), thus avoiding leakage. The bath container was cylindrical (inner diameter $40\,\text{mm}$) to ensure an axisymmetric flow and, hence, a homogeneous deposition during coating. The withdrawal velocity was set to $U=300\,\mu \text{m/s}$, resulting in a coating thickness of $e=150\,\mu \text{m}$ (green point in \figref{fig:CoatingCollage}\textbf{c}). This value of $e$ was selected so that the coating layer was sufficiently thick to be captured in the $\mu$CT images (see \secref{sect:micro_computed_tomography}), while sufficiently small not to affect the overall mechanical behavior of the VPS rods. The small cross-sectional areas of both the Solaris centerline fiber and the outer coating, with respect to the VPS-16 bulk core, allowed us to assume a homogeneous mechanical behavior of the composite rod.
\end{enumerate}

In an effort to achieve reproducible and predictable coating layers, we systematically investigated the dependence of the withdrawal speed, $U$, on the coating thickness, $e$ (see \figref{fig:CoatingCollage}\textbf{c}). We coated multiple identical bulk core rods of diameter $D=8.3\,\text{mm}$ at different withdrawal velocities, $U = \{20,\,50,\,100,\,200,\,400,\,1000\}\,\mu\text{m/s}$. After curing, the resulting coating thickness was measured by analyzing digital microscope images of cross-sectional cuts (Keyence VHX 5000) at multiple locations (see \figref{fig:CoatingCollage}\textbf{b}). In \figref{fig:CoatingCollage}\textbf{c}, we present the experimental results (datapoints) for the coating thickness normalized by the capillary length
$l_c=\sqrt{\gamma/\rho_\text{solaris} g}$ 
as a function of the capillary number $\text{Ca}=\eta U / \gamma$. The parameters $\rho_\text{solaris}$, $\eta$ and $\gamma$ are respectively the density, the dynamic viscosity and the surface tension of the Solaris polymer, and $g=9.81\,\text{m/s}^2$ is the gravitational acceleration. Rheological measurements were carried out (rheometer Discovery HR-2, TA-Instruments) to determine the viscosity of Solaris, $\eta = 2.19\,\text{Pa}\cdot\text{s}$, the time dependence of which was found to be negligible during the entire coating process (the viscosity increases by 5.6\,\% during the coating process of 25~min). This slow time evolution of $\eta$ ensured a nearly constant thickness of the coating layer along the rod. The surface tension of Solaris,~$\gamma$ is $\gamma = 23\, \text{mN/m}$~\cite{Grundke2008}.

The classic Landau-Levich theory for the withdrawal of a plate from a viscous bath~\cite{Landau1988} provides a relationship between the final thickness, $e$ of the thin-film coating, and the relevant input parameters: the extraction velocity of the film $U$, and characteristic properties of the coating liquid (viscosity $\eta$, specific weight $\rho g$ and surface tension $\gamma$). Past an initial transient, the coating thickness reaches a plateau value that scales with the capillary length $l_c=\sqrt{\gamma/\rho g}$ and the capillary number $\textrm{Ca}=\eta U / \gamma$, such that $e/l_c \sim \text{Ca}^{2/3}$. These existing models accurately predict this constant thickness for coated fibers or plates~\cite{Rio2017}. However, our coated bulk core rod cannot be strictly considered as a fiber since its diameter is close to the capillary length of the coating liquid. Defining the Goucher number~\cite{Goucher1922} as $\textrm{Go} = D/(2 l_c)$, in our experiments, we have $\textrm{Go}\approx 2.7$, at the crossover between the plate limit ($\textrm{Go} \rightarrow \infty$) and the fiber regime ($\textrm{Go} \ll 1$)~\cite{Rio2017}. Moreover, the Landau-Levich theory neglects gravity and is only valid when $\textrm{Ca} < 10^{-3}$. In our case, gravity cannot be neglected since our velocity range $U \in [100\mu\textrm{m/s}; 1000\mu\textrm{m/s}]$ corresponds to capillary numbers $\textrm{Ca} \in [0.01; 0.95]$. Nevertheless, we observed that the same scaling law $e\sim \,\mathrm{Ca}^{2/3} l_c$ was still appropriate to describe our coating thicknesses, albeit with a prefactor that we had to determine through fitting of the experimental results,
\begin{equation}
e=0.651\,\text{Ca}^{2/3}~l_c.
\label{eq:coating_thickness_vs_Ca}
\end{equation}
In \figref{fig:CoatingCollage}\textbf{c} we juxtapose \eqref{eq:coating_thickness_vs_Ca} (dashed line) onto the experimental data, finding excellent agreement.
This semi-empirical scaling allowed us to achieve the desired coating thickness of $e=150\,\mu \text{m}$ by imposing a withdrawal velocity $U=300\,\mu \text{m/s}$.

\subsection{Tomographic imaging of the static elastic clasp}
\label{sect:micro_computed_tomography}
We used X-ray $\mu$CT to quantify the geometry of different configurations of the \textit{static elastic orthogonal clasp}. X-ray~$\mu$CT is a non-destructive imaging technique based on volumetric differential X-ray absorption; an X-ray generator irradiates the sample of interest, which partially absorbs the incoming X-ray beam~\cite{Herman2009}. The X-ray fraction passing through the sample is projected onto a detector, producing a 2D snapshot. Then, a collection of 2D images is obtained from multiple viewpoints (we performed scans with 1,000 projections), from which the 3D volumetric (tomographic) image of the sample is reconstructed.

The orthogonal elastic clasp samples were scanned ($\mu$CT~100, Scanco, and Ultra-Tom, RX-Solutions) and reconstructed using the built-in commercial software package. Subsequently, the raw tomographic images were further post-processed following the scheme presented in the following Subsection. 

\begin{figure}[h!]
    \centering
    \includegraphics[width=0.6\textwidth]{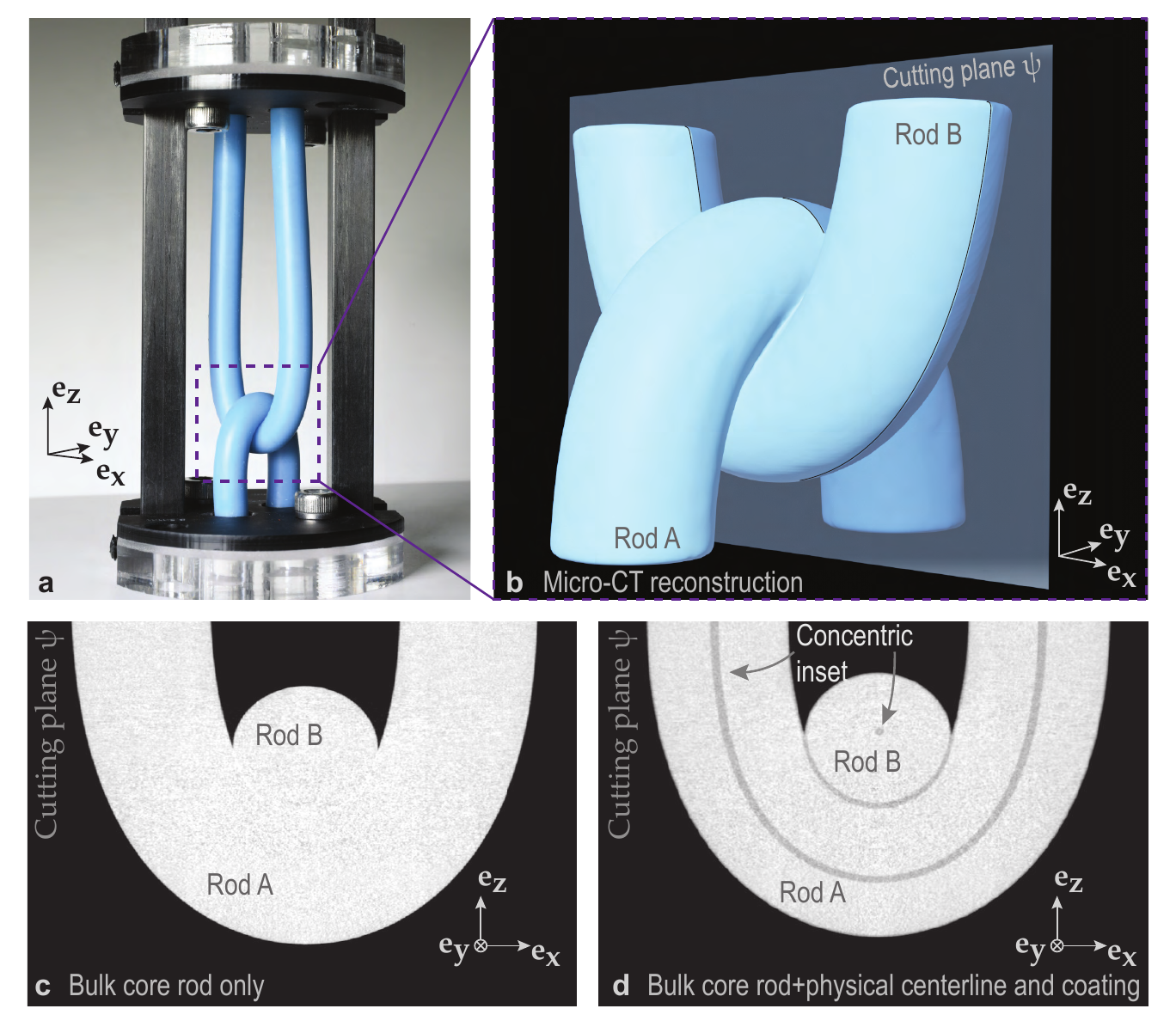}
    \caption{\textbf{Volumetric $\mu$CT imaging of the static elastic clasp.} \textbf{a},~Optical photograph of an elastic clasp. \textbf{b},~X-ray tomographic reconstruction of the configuration in \textbf{a}. \textbf{c-d},~Images of a cut (along the plane $\psi$ defined in~\textbf{b}) of a 3D volumetric image for a~\textbf{c} monolithic elastomeric VPS-16 rod (the contact regions are not visible) and~\textbf{d}~for a VPS-16 bulk core rod coated with Solaris polymer and including the material centerline (also made out of Solaris). In ~(\textbf{d}), the contact interface at the crossing of the two rods (dark gray region) is clearly visible, whereas not so in ~\textbf{c}. The rest diameter of the rods composing the elastic clasps in~\textbf{a}, \textbf{b} and \textbf{c} is $D=8.3$~mm and in \textbf{d}, it is~$D=8.5$~mm (the extra thickness is due to the presence of the outer coating).}
    \label{fig:uCT_imaging}
\end{figure}

In \figref{fig:uCT_imaging}\textbf{a}, we present an optical photograph of an elastic clasp (the clasp configuration~3 defined in the main article, at a vertical load~$f_z=0.05$). The corresponding 3D reconstruction obtained from $\mu$CT scanning is provided in \figref{fig:uCT_imaging}\textbf{b}.
Distinguishing the two rods composing the elastic clasp and their contact regions will be central to the subsequent image analysis presented in \secref{sect:image_analysis}. We observed that two VPS monolithic rods, without the outer coating layer, could not be told apart in the $\mu$CT images. The reason is that there would be no differential X-ray absorption across the contact interface of a homogeneous material (\figref{fig:uCT_imaging}\textbf{c}). Empirically, we found that a density difference of $10-15\%$ was required to differentiate two touching volumes. As such, to visualize the contact regions and the material centerlines, we had to use the composite rods introduced in \secref{sect:rod_fabrication_tomography}. Also, a minimum size of 3-5\,pixels of one material volume is required to be accurately captured in the subsequent image analysis. Combined with the finite resolution of the volumetric images (typically $\sim 30\,\mu \text{m}$/voxel), the above requirements set the limit thickness of both the outer coating and the physical centerline fiber. These constraints rationalize our choices of $e=150\,\mu \text{m}$ for the coating thickness and $D_\text{n} = 500\,\mu$m) for the diameter of the centerline fiber. With these parameters, as demonstrated in the cross-sectional cut of the clasp shown in \figref{fig:uCT_imaging}\textbf{d}, we were able to achieve a clear visualization of both the contact region and the material centerline.

Determining the optimal $\mu$CT scanning settings was crucial to obtain high-quality volumetric images of our elastic clasps. For all the configurations of the \textit{elastic orthogonal clasp}, we consistently performed the $\mu$CT scans at an X-ray source energy of 70\,kV and intensity 200\,$\mu$A. Each projection was produced with an acquisition time of 300\,ms. Furthermore, we made use of the built-in 0.5\,mm-thick aluminum filter to correct beam hardening effects~\cite{Herman2009}. The total scanning time of each elastic clasp was approximately 1 hour. The spatial resolution of \textit{elastic orthogonal clasp} configurations typically ranged from 25 to $35\,\mu$m (size of the voxels).

\subsection{Extraction of the centerline coordinates and the contact regions from the $\mu$CT data}
\label{sect:image_analysis} 

Once the tomographic data of the \textit{elastic orthogonal clasps} was acquired using the $\mu$CT and reconstructed into 3D volumetric images (\secref{sect:micro_computed_tomography}), it was necessary to process them further to digitize the material centerline coordinates and locate the contact region. This section describes the image post-processing algorithm that we developed in-house to obtain both of these quantities. Specifically, this algorithm aims to relate the location of the contact regions to the corresponding arc length coordinates of the centerlines of the respective rods. The boundaries of the experimental two-dimensional contact maps presented in Fig.~3 of the main article are examples of the output of this procedure.

The reconstructed 3D images obtained from the $\mu$CT data comprise a stack of grayscale \textit{DICOM} images. First, these images were treated using the open-source image analysis platform (ImageJ~1.52b, National Institutes of Health~\cite{Schindelin2012}) for contrast enhancement; voxels corresponding to the surrounding air down were shifted down to black (0) and the brightest regions to white (255). The 16-bit data was then downscaled to 8-bit to reduce the file sizes and further alleviate the computational load of the subsequent image post-processing routine (detailed below). Secondly, Matlab's Imaging Processing Toolbox (\texttt{Matlab 2019b, MathWorks}) was used to convert the 8-bit stack of images into a dense 3D matrix, whose voxels had values ranging from 0 to 1, representing the local material density. The voxel values were $\approx$0.9 and $\approx$0.8, for the VPS ($\rho_\text{VPS16} = 1160\,$~kg/m$^3$) and Solaris regions, respectively. As a reminder, VPS-16 was used for the bulk core rod and Solaris for both the centerline fiber and the outer coating layer.

The subsequent post-processing algorithm to digitize both the centerlines and the contact surface between the two rods involved the following two stages:
\textbf{Stage~(I)} Extraction of a coarsely distributed set of centerline coordinates, and \textbf{Stage~(II)} Refinement of the centerline and digitization of the contact surfaces by iterating between the coarse set of centerline locations, both of which are detailed next. 

\begin{figure}[h!]
    \centering
    \includegraphics[width=0.7\textwidth]{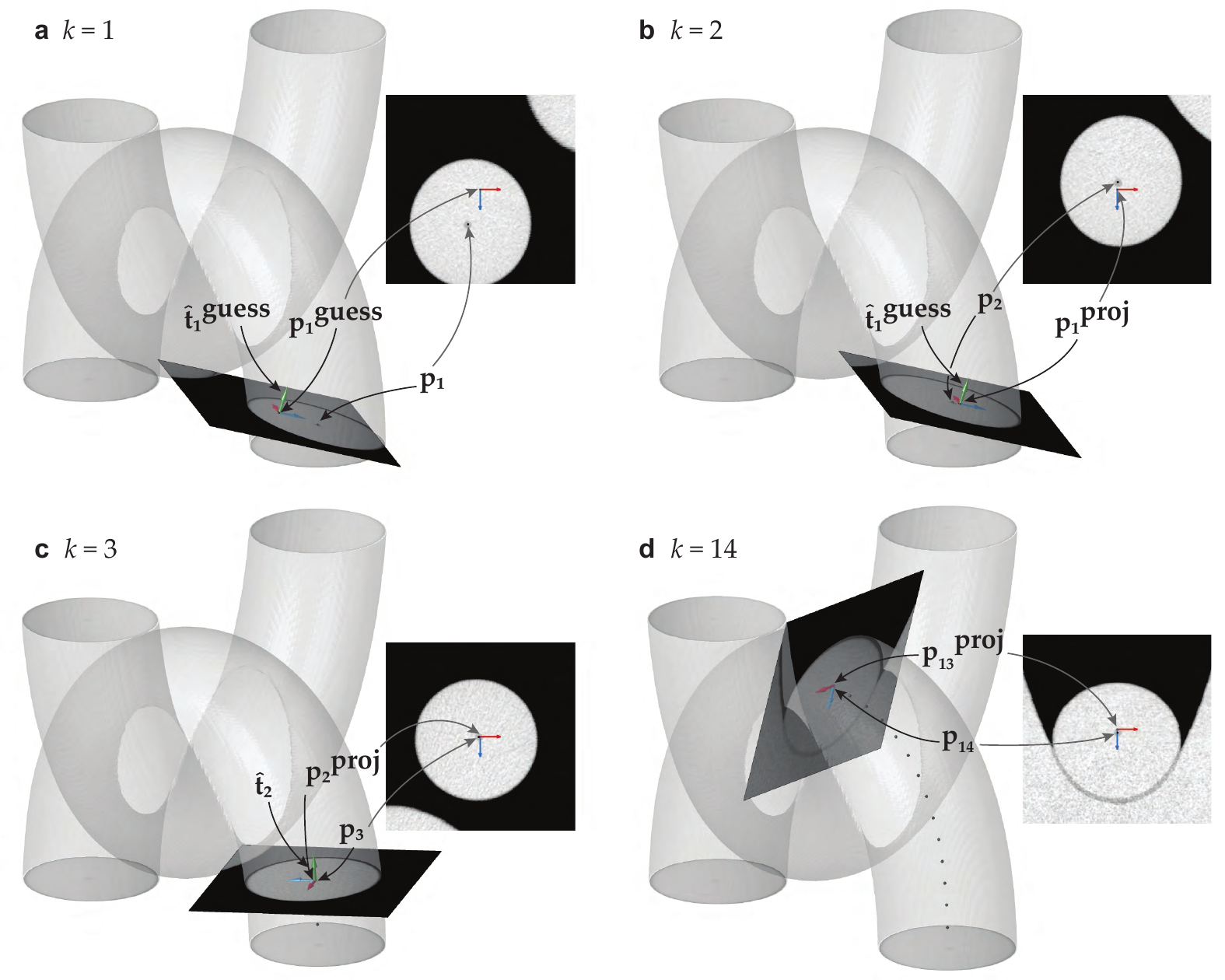}
    \caption{\textbf{Procedures for digitizing the coarse centerline.} \textbf{a.}~Initialization. The user inputs ansatzes for a  point $\mathbf{p}_{1}^\text{guess}$ in 3D (represented by starting point of the green arrow) and the tangent $\hat{\mathbf{t}}_1^\text{guess}$  (represented by the green arrow). Using $\mathbf{p}_{1}^\text{guess}$ and $\hat{\mathbf{t}}_1^\text{guess}$, the material frame is constructed (cross-section presented in the inset). The user then picks the dark region (physical centerline) to indicate its location in the 2D material frame (black dot in the inset). This location is then translated back to the 3D laboratory frame ($\mathbf{p_1}$ in the 3D representation). \textbf{b.}~Identification of the second point on the centerline ($k=2$). The point $\mathbf{p_1}$ is projected along $\hat{\mathbf{t}}_1^\text{guess}$ to construct the second cross-sectional view (normal to $\hat{\mathbf{t}}_1^\text{guess}$). The user again picks the region of the physical centerline to obtain $\mathbf{p_2}$. \textbf{c.}~Beyond the third point on the centerline ($k\ge3$), the centerline tracking is fully aumated; the centerline fiber is tracked by isolating the darker (less dense) regions in the material frame using the \texttt{regionprops} command in the Matlab 2019b environment. At each step, the tangent vector is constructed using the two previous centerline points. \textbf{d.}~Example of the tracking algorithm at $k=14$. The 14 first coarse centerline coordinates are represented by the black dots in the image. In all four panels, the unit binormal~($\hat{\mathbf{b}}$) and normal~($\hat{\mathbf{n}}$) vectors are respectively represented as red and blue arrows. 
    }
    \label{fig:Image_analysis_coarse}
\end{figure}

\vspace{.5cm}
\noindent
\textbf{Stage (I) -- Coarse discretization of the rod centerline}: The goal of this first step is to build a coarse (approximate) discrete centerline, $\mathbf{p} = \{\mathbf{p}_k\}$, where $k=1,\,2,\,\cdots$ represents the index of a point along the discrete centerline. Each point $\mathbf{p}_k$ is obtained from cross-sectional cuts of the 3D volumetric image of the elastic clasp (see \figref{fig:Image_analysis_coarse}). We iterate through the centerline by following these steps:

\begin{enumerate}
    \setlength{\itemsep}{0cm}
    \item[(I).1] Initial ansatzes for both the starting point of the centerline ($\mathbf{p}_{1}^\text{guess}$) and the tangent of the centerline at that point ($\hat{\mathbf{t}}_1^\text{guess}$) are input by the user. A square cross-sectional view of size $1.5D\times1.5D$ ($D$ is the rest diameter of the rod), normal to $\hat{\mathbf{t}}_1^\text{guess}$  at $\mathbf{p}_{1}^\text{guess}$ is investigated. A Gaussian filter was applied to this frame using Matlab's \texttt{imgaussfilt} function, with the standard deviation \texttt{SIGMA=3}. This smoothing allowed us to decrease the noise of the grayscale material frame, without loss of fine details.
    
    \item[(I).2] From the cross-sectional view at $\mathbf{p}_{1}^\text{guess}$, and along $\hat{\mathbf{t}}_1^\text{guess}$, the user is prompted to input the refined position of $\mathbf{p}_{1}$ by locating the darker area (see inset of \figref{fig:Image_analysis_coarse}\textbf{a}) corresponding to the physical material centerline.
    
    \item[(I).3] The second point of the material centerline ($\mathbf{p}_{2}$) is now to be determined. The cross-sectional view at the projected second point, $\mathbf{p}_{2}^\text{proj} = \mathbf{p}_1 + \xi \, \hat{\mathbf{t}}_1^\text{guess}$, with a normal $\hat{\mathbf{t}}_1^\text{guess}$, is investigated (see \figref{fig:Image_analysis_coarse}\textbf{b}). We typically took the incremental distance between two successive coarsely distributed centerline locations $\xi = D/ 10$.
    
    \item[(I).4] The user is prompted to input the position of $\mathbf{p_2}$ by locating the dark area of the cross-sectional view. From $\mathbf{p}_{1}$ and $\mathbf{p}_{2}$, the first tangent vector is updated by $\hat{\mathbf{t}}_1=(\mathbf{p}_2 - \mathbf{p}_1) / \| \mathbf{p}_2 - \mathbf{p}_1 \|$. 
    
    \item[(I).5] From this step onward, the iterative loop runs automatically without any further user inputs. For $k\ge 3$, the $k^\mathrm{th}$ point is projected from the $(k-1)^\mathrm{th}$ point such that $\mathbf{p}_{k}^\text{proj} = \mathbf{p}_{k-1} + \xi \, \hat{\mathbf{t}}_{k-1}$ and $\hat{\mathbf{t}}_{k-1} = (\mathbf{p}_k - \mathbf{p}_{k-1}) / \| \mathbf{p}_k - \mathbf{p}_{k-1} \|$. Since the 2D cross-sectional cuts are in close proximity ($\xi=0.1D$), the image generated at~$\mathbf{p}_{k}^\text{proj}$, at each step, is sufficiently similar to the previous one, such that the dark region of the physical centerline can be tracked automatically by locating the small darker region of the physical centerline in the close vicinity of the center of the generated 2D image (see \figref{fig:Image_analysis_coarse}\textbf{c} and~\textbf{d}). To track this darker region, we first binarize the image of the cross-sectional cut and then use the Matlab 2019 function~\texttt{regionprops} to locate the centroid of the dark centerline region.
    
    \item[(I).6] Iteration on $k$, step (I).5., is repeated until the other rod extremity is reached.
\end{enumerate}

The final output of Stage I of the algorithm is a discrete set of 3D points $\mathbf{p} = \{\mathbf{p}_k\}$ and a corresponding set of unit tangent vectors $\hat{\mathbf{t}} = \{\hat{\mathbf{t}}_k\}$, representing the material centerline of the rod. 
We define the unit normal~$\hat{\mathbf{n}}$ and binormal vectors~$\hat{\mathbf b}$ such that
    \begin{equation}
       \hat{ \mathbf{b}} = \frac{(\mathbf{e_z} \times \hat{\mathbf{t}})} {\| \mathbf{e_z} \times \hat{\mathbf{t}} \|},
        \label{eq:binormal_vector}
    \end{equation}
and
    \begin{equation}
       \hat{ \mathbf{n} }= \hat{\mathbf{b}} \times\hat{ \mathbf{t}},
        \label{eq:normal_vector}
    \end{equation}
where $\mathbf{e_z}=[0,0,1]$ is the unit vector along the $z$-axis. We used the plane spanned by~$\hat{\mathbf{n}}$ and~$\hat{\mathbf{b}}$ to construct a material frame using only a central point $\mathbf{p}$ and a tangent vector $\hat{\mathbf{t}}$ (allowing us to construct the unit normal and binormal vectors) at each step. 
Next, in Stage~(II) of the algorithm, the centerline is refined by interpolating the coarse centerline that had been previously obtained. 

\vspace{.5cm}
\noindent
\textbf{Stage (II) -- Refinement of the centerline coordinates and tracking of the rod surface}: During this second stage of the algorithm, we refine the density of the discrete arc lengths at which we locate the centerline fiber coordinates. This second stage also allows us to compute a discrete set of 3D coordinates of the rod's surface boundaries at each discrete arc length. This Stage~(II) is fully automated (\textit{i.e.}, there is no user input); the discrete centerline $\mathbf{p}_k$ and tangents~$\hat{\mathbf{t}}_k$ determined previously, in Stage~(I), are now used to construct intermediary material frames in between the coarsely distributed material frames.

\begin{figure}[ht!]
    \centering
    \includegraphics[width=0.7\textwidth]{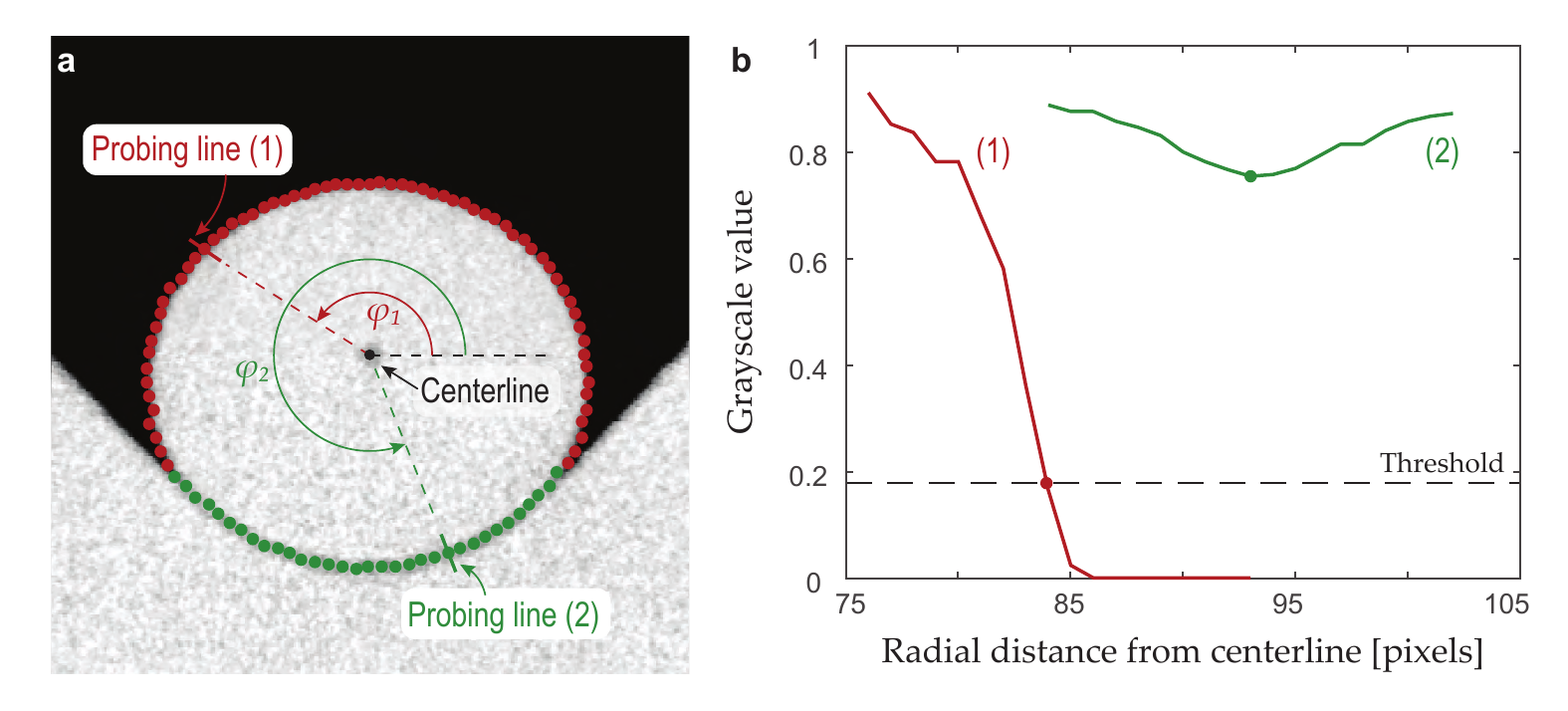}
    \caption{\textbf{Procedure for digitizing the rod surface from a cross-sectional cut.} 
    \textbf{a,}~The cross-sectional cut along the centerline is constructed, and the probing lines are generated. We represent two representative probing lines at angles with the horizontal axis $\varphi_1$ and $\varphi_2$ (red and green solid lines respectively). We also present the result of the rod-surface tracking procedure. The red points correspond to free rod surfaces (contact with surrounding air), and the green points correspond to contact location with another rod. 
    \textbf{b,}~The grayscale profiles are generated along the probing lines (1) and (2) and represented as a function of the radial distance to the centerline location (black point in \textbf{a}). When the rod is in contact with air (probing line~(1)), we determine the location of the rod surface as the location where the grayscale profile measured along this probing line drops below a threshold value (determined dynamically using the average of the gray value of the bulk rod at each frame). On the other hand, when the rod is in contact with another rod (probing line~(2)), we define the rod surface at the minimum of the grayscale profile.   
    }
    \label{fig:edge_detection}
\end{figure}

\begin{enumerate}
    \setlength{\itemsep}{0cm}
    \item[(II).1]  The~$m^\mathrm{th}$ intermediary frame is constructed between the coarse centerline locations~$\mathbf{p}_k$ and~$\mathbf{p}_{k+1}$ determined in Stage (I), such that $1\le m \le M$ where $M$ is the number of intermediate centerline locations in between each coarse centerline location. To construct this~$m^\text{th}$ material frame, we linearly interpolate $\mathbf{p}_k$ to provide a guess for the location of the current centerline such that $\mathbf{p}^\mathrm{guess}_{k,m}=(M-m+1)/M \cdot \mathbf{p}_k + (m-1)/M \cdot \mathbf{p}_{k+1}$. We make use of the same linear interpolation scheme to generate the~$m^\mathrm{th}$ intermediary guessed tangent vector~$\hat{\mathbf{t}}^\mathrm{guess}_{k,m}$. 
    
    \item[(II).2] By automatically tracking the darker region of the centerline fiber in these intermediary frames (using the Matlab 2019 function \texttt{regionprops}), the refinement of the centerline coordinates is straightforward to achieve. We obtain all the refined centerline coordinates~$\mathbf{p}_{n}$, where~$n$ is the overall index of the finely distributed centerline coordinates such that $n=M\, (k-1)+m$, with $1\le n \le N$ (where $N$ is the total number of finely distributed locations of the centerline coordinates, typically we have $N\simeq250$). 
    
    \item[(II).3] At the end of this refinement step, we smoothed to the centerline coordinates by making use of the Matlab 2019 command \texttt{smoothdata} with a window size of $\simeq \texttt{round}(N/7)$. These smoothed centerline locations allow us to generate the discrete $N-1$ final tangent vectors~$\hat{\mathbf{t}}_{n}$ of the discrete centerline such that~$\hat{\mathbf{t}}_{n}= (\mathbf{p}_{n+1} - \mathbf{p}_{n}) / \|\mathbf{p}_{n+1} - \mathbf{p}_{n} \|$. Without this smoothing step, the resulting tangent vectors would be too noisy to allow for the subsequent processing steps. 

    \item[(II).4] In addition to the centerline refinement, we also developed an edge-detection algorithm to track the surface boundaries of the deformed rod, yielding the full  3D envelope (discrete) of the rod of interest. Note that locating the rod boundaries is straightforward in the absence of rod-rod contact. In this case, the material frame is binarized and, then the boundaries of the rod cross-section are located using the Matlab function \texttt{regionprops}. By contrast, when there is rod-rod contact along some locations of its arc length (see example \figref{fig:Image_analysis_coarse}\textbf{d}), the boundary of the rod must be determined with an alternative way. This edge-detection strategy is detailed in \figref{fig:edge_detection}. After constructing the intermediary 2D material frame of interest (centered at the centerline location~$\mathbf{p}_{n}$), we analyze the gray values of this material frame along a \textit{probing line} that emanates radially from the material centerline, with a specified angle~$\varphi$. This angle was varied discretely within the range $0 \leq\varphi\leq 360^\circ$, in 100 steps. If the probing line ends in the surrounding air, the gray values along this line will jump from a finite value (inside the rod) to zero (surrounding air). If the probing line ends in another rod strand (in case of contact), the gray values along the probing line will not drop to zero, but the gray values along the probing line will be affected by the darker surface coating of the rod. When contact is present, the gray values will therefore exhibit a local minimum along the probing line, the location of which is attributed to the rod surface. 
\end{enumerate}

\subsection{Geometric quantification of the elastic rods in the orthogonal clasp configuration}
\label{sect:geom_quantification} 

Once the locations of the material centerline and the contact surface have been extracted, we can quantify the deformation of the rods of the elastic orthogonal clasp. Specifically, we computed the (i) cross-section flattening of the cross section of the rod (see \figref{fig:LargeDef}\textbf{a}-\textbf{b}), (ii) the centerline curvature of the rod (see \figref{fig:LargeDef}\textbf{c}), and (iii) the boundary of the contact region parametrized by the arc length coordinate of the rod, $(s_\text{A}$,~$s_\text{B})$ (see \figref{fig:LargeDef}\textbf{d}), each of which is detailed next. 

\begin{enumerate}
\setlength{\itemsep}{0cm}
\item[(i)] \textbf{Cross-sectional flattening:} The protocol described above allowed us to extract the shape of the rod cross-section at each constructed material frame, from which we quantified the cross-section flattening of the rod along its physical centerline. We define this flattening using the geometric quantity $1-b/a$, where $a$ and $b$ are the major and minor axes of the rod cross-section, respectively. Our 2D material frames were constructed in a way that $a$ and $b$ naturally corresponded to the horizontal and vertical span of the rod cross-section (see \figref{fig:LargeDef}\textbf{a}). 
Indeed, the binormal unit vector~$\hat{\mathbf{b}}$ (as defined in \eqref{eq:binormal_vector}) points along the horizontal direction of the constructed material frame, orthogonally to the $z$-axis (by construction). Moreover, since the scanned elastic clasps were consistently oriented in the $z$-direction (see \figref{fig:uCT_imaging}\textbf{b}), the cross-sections always appeared flattened along the vertical direction of the material frame. 

\item[(ii)] \textbf{Discrete curvature of the extracted centerline:} The centerline curvature along its arc length was computed from the tangent vectors $\hat{\mathbf{t}}_n$. Since each rod composing the elastic orthogonal clasp is planar, we could use 2D methods to compute the curvature of the centerlines. By construction, the normal vector~$\hat{\mathbf{n}}$ along a rod centerline (as defined in \eqref{eq:normal_vector}) was contained in the same plane as centerline of the rod. We therefore computed the $N-2$ discrete values of curvature $\kappa_n$ such that:
\begin{equation}
\kappa_n = \frac{1}{\text{d}s_n} (\hat{\mathbf{t}}_{n+1} - \hat{\mathbf{t}}_{n})  \cdot \hat{\mathbf{n}}_n
\end{equation} 
with the discrete arc length differential $\text{d}s_n = \| \mathbf{p}_{n+1} -  \mathbf{p}_{n} \|$. 

\item[(iii)] \textbf{Shape of the contact region between the two rods:} In addition to using mechanical testing experiments to validate our FEM procedure (see \secref{subsection:FEM_3} and \figref{fig:FEM_validation_macro}), we also attested that the shape of the FEM-computed contact region was in close agreement with the experimental $\mu$CT measurements of the elastic clasp. For this second comparison between FEM and experiments, we projected the experimental contact surface onto the ($s_\text{A}$,$s_\text{B}$)-plane by assigning each discrete contact point detected in step~(II).4 to the closest discrete centerline locations along the arc lengths~$s_\text{A}$ and~$s_\text{B}$ of rods~A and~B, respectively. In Figs.~3\textbf{a3}-\textbf{b3}-\textbf{c3} of the main article, instead of presenting the complete set of discrete contact points in the ($s_\text{A}$,$s_\text{B}$)-plane, we only report the boundary of the experimental contact map (black dashed lines in these figure pannels), so as to make a visual comparison clearer. The boundary of this set of 2D contact locations was determined using the Matlab~2019 command \texttt{boundary}.
\end{enumerate}

\subsection{Fabrication of the rods for mechanical testing (static and sliding elastic orthogonal clasp)}
\label{sect:rod_fabrication_mechanical_testing}

In parallel to the precise tomographic imaging of the orthogonal elastic clasp, we performed two sets of experiments to quantify the mechanical behavior of the orthogonal elastic clasp. The first set of experiments -- the mechanical characterization of the \textit{static orthogonal elastic clasp} -- allowed us to validate our FEM computations (the results are presented in \figref{fig:FEM_validation_macro}). In the second set of experiments -- the \textit{sliding elastic orthogonal clasp} -- the two rods of a clasp were set to slide against each other, as described in Fig.~4 of the main article. 

For these mechanical characterization experiments, we used elastomeric rods that were fabricated differently than those used for the X-ray $\mu$CT imaging. The primary difference was that, for mechanical testing, the rods required neither the centerline fibers nor the outer surface coating. The casting of the rods used in these experiments was done using smooth acrylic tubes (Plexiglas XT tubes incolore 0A070, Röhm, Switzerland) of inner diameter $D=8$\,mm. To vary the Young's Modulus of the rods, we used two different grades of VPS: VPS-16 (Elite Double 16 fast, shore hardness 16A) and VPS-32 (Elite Double 32, shore hardness 32A). The casting protocol allowed us to fabricate $40\,\text{cm}$ long homogeneous elastic rods; these rods were then cut at the desired lengths using a cutter blade. The mechanical properties of VPS-16 and VPS-32 are quantified in \secref{sect:mechanical_characterization}. Next, we  describe the specific steps that we followed before performing the \textit{elastic orthogonal clasp} and the \textit{sliding clasp} experiments.

\paragraph{Experimental details of the \textit{elastic orthogonal clasp} tests (no friction):} The macroscopic response of the \textit{elastic orthogonal clasp} was characterized to validate the FEM computations (see \figref{fig:FEM_validation_macro}). In these experiments, even though no sliding was imposed between the two rods, we deactivated rod-rod friction by applying a few drops of liquid soap (Palmolive Original from Migros, Switzerland) in the contacting region. The liquid soap generated a thin lubrication layer between the two rods, which was sufficient to reduce local tangential forces significantly. We performed the mechanical testing of the static orthogonal clasp with rods made out of VPS-16.

\paragraph{Experimental details for the \textit{sliding clasp}:} When performing the \textit{sliding clasp} experiment (Fig.~4 of the main article), we observed that the tension ratio between the high and low-tension strands ($T_1/T_0$) was strongly influenced by the frictional behavior of the two VPS rods in contact. We further observed that the VPS-to-VPS contact could not be described by the Amontons-Coulomb friction; the tangential force~$F_\text{t}$ applied between the samples is not linearly proportional to the normal force~$F_\text{n}$ between them. However, in many real-life applications of filaments in frictional contact, the Amontons-Coulomb friction law is a reasonable description~\cite{persson_book}. To ensure the Amontons-Coulomb friction throughout, we surface-treated our VPS rods by applying a talcum-powder (Milette baby powder, Migros, Switzerland) and waited for 24~hours. During this waiting time, part of the talcum particles was absorbed at the VPS surface, increasing its effective roughness. Before each experimental run, the excess talcum powder was gently wiped off from the surface with a fine cloth. This surface treatment protocol allowed us to achieve a robust dynamic friction coefficient of $\mu=0.32\pm0.03$ (see more details in \secref{sect:mechanical_characterization}). 

\subsection{Characterization of material properties of VPS and of the frictional behavior of the powder-treated VPS surfaces}
\label{sect:mechanical_characterization}
\begin{figure}[h!]
    \centering
    \includegraphics[width=0.8\textwidth]{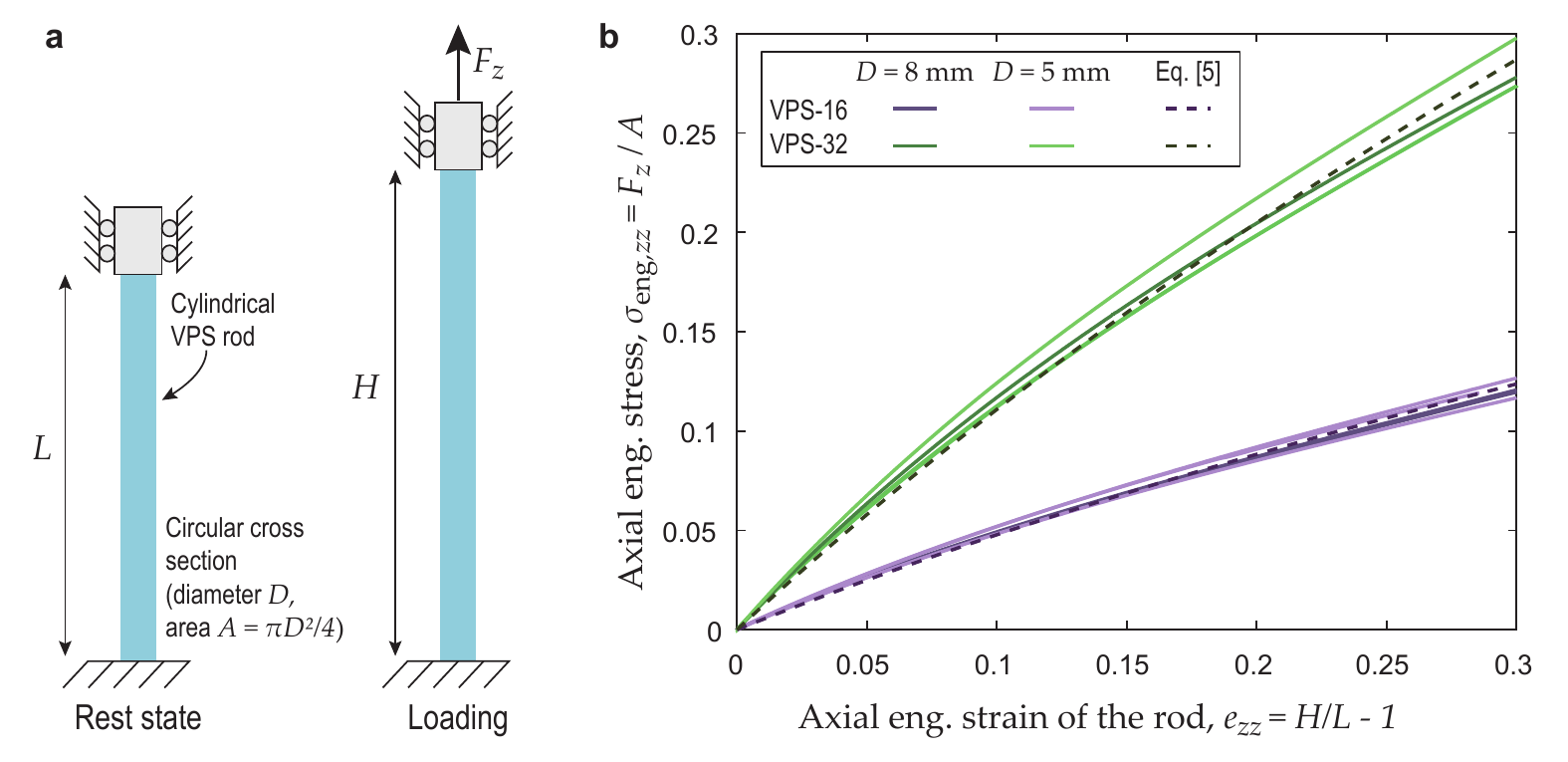}
    \caption{\textbf{Mechanical testing of the VPS-16 and VPS-32 rods.} \textbf{a},~Schematic of the tensile tests. \textbf{b},~Engineering stress-strain curves of the cylindrical. To validate the measurements at different length scales, we performed the tests on cylindrical rods of diameters $D=5\,$and 8\,mm, and length $L=20D$; no significant difference was observed between these two rod diameters. The dashed line corresponds to the one-dimensional engineering stress-strain predicted for incompressible Neo-Hookean material; \eqref{eq:neo_hookean_constitutive_relation}.}
    \label{fig:mechanical_characterization}
\end{figure}

\paragraph{Mechanical testing of rods:} We characterized the mechanical properties of VPS-16 and VPS-32 using a Universal Mechanical Testing Machine (Instron 5943) with rods of diameters $D=5$ and~8\,mm. The length of the rods was chosen to preserve a constant aspect ratio $L/D=20$ (with $L$ the rod length); \textit{i.e.} $L=100$ and~160\,mm, respectively. Before performing any of the experiments described in the main article and the mechanical material tests, we left the VPS samples to rest for one week after curing to ensure their mechanical properties had reached a steady state. We measured the force $F_z$ necessary to impose a distance $H>L$ between the rod extremities (see \figref{fig:mechanical_characterization}\textbf{a}) and related the axial engineering stress $\sigma_{\text{eng,}zz}=F_z/A$ (where $A=\pi D^2/4$ is the cross-section area of the rod at rest) and the engineering strain $e_{zz} = H/L - 1$. The corresponding experimental curves are reported  in \figref{fig:mechanical_characterization}\textbf{b} (solid lines). We then fitted each experimental curve with the one-dimensional engineering stress-strain prediction from the Neo-Hookean model for incompressible hyperelastic materials~\cite{Gent2001ed}:
\begin{equation}
\sigma_{\text{eng,}zz} = \frac{F_z}{A}= \frac{E}{3}\left(  (1+e_{zz})  - \frac{1}{(1+e_{zz})^2}  \right)
\label{eq:neo_hookean_constitutive_relation}
\end{equation}
Fitting the above expression to the experimental data yielded the ground-state Young's modulus: $E = 520\pm0.02$~kPa for VPS-16 and $E=1.22\pm0.05$~MPa for VPS-32.
The dashed lines in \figref{fig:mechanical_characterization}\textbf{b}) correspond to  \eqref{eq:neo_hookean_constitutive_relation} using the mean value of $E$ across five experiments for VPS-16 and three experiments for VPS-32. The uncertainty corresponds to the standard deviation of the fitted parameter $E$ across the different samples.

\paragraph{Characterization of the friction behavior:}
\label{sect:friction_coefficient}

The frictional behavior of the powder-treated VPS-16 and VPS-32 elastomers  (see \secref{sect:rod_fabrication_mechanical_testing} for details of the surface treatment procedure) was characterized using an adapted version of the standard ASTM D1894-14 protocol~\cite{ASTM-friction}. This  ASTM  protocol is designed to measure the dynamic friction coefficient, $\mu$, from the normal force $F_\text{n}$ and tangential force $F_\text{t}$ by the relation $F_\text{t}=\mu F_\text{n}$. 

\begin{figure}[h!]
\centering
\includegraphics[width=0.7\textwidth]{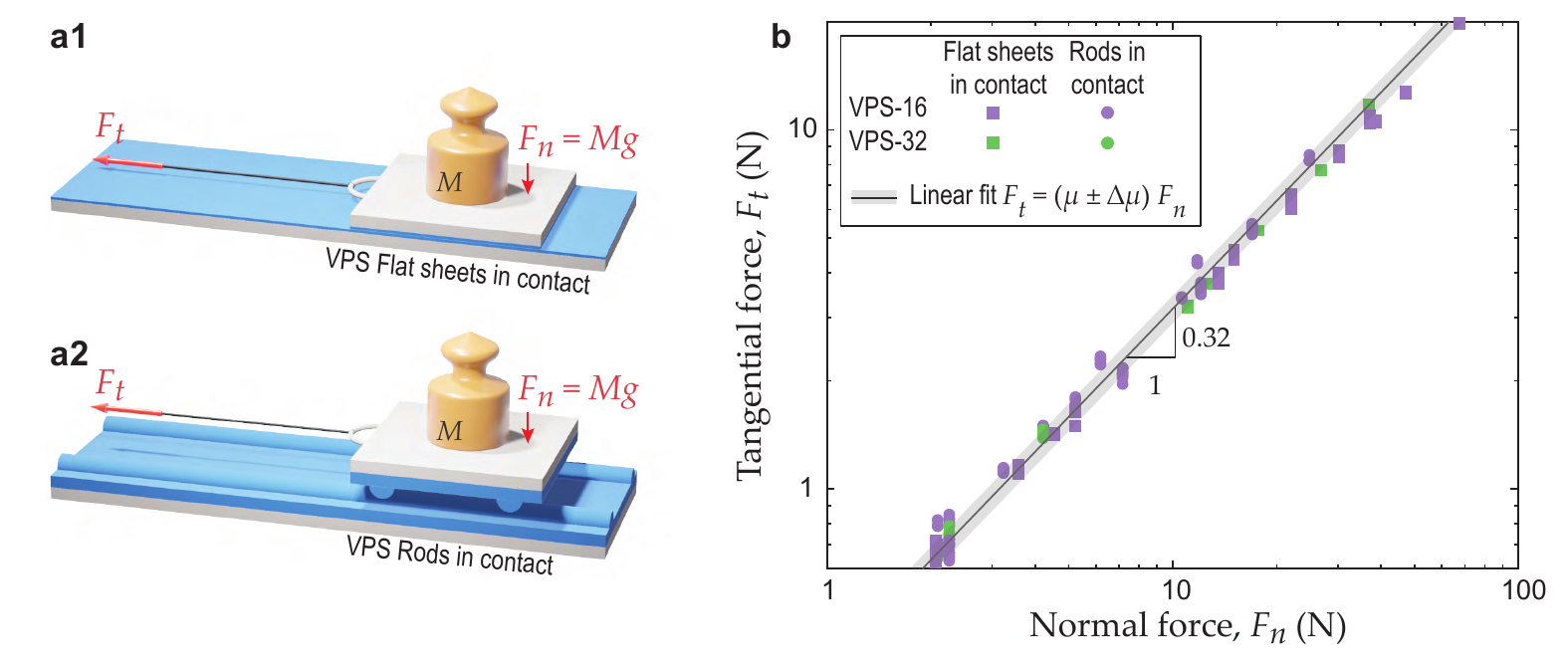}
    \caption{\textbf{Frictional behavior of the powder-treated VPS surfaces .} \textbf{a1-a2},~Schematics of the two sets of friction experiments that we considered: two VPS flat slabs in frictional contact and two VPS rods in frictional contact with two other VPS rods set orthogonally. In the second set of experiments, the rods are embedded in thick VPS pads. \textbf{b},~Tangential force~$F_\text{t}$ during the displacement of the upper sleigh made of VPS-16 (light purple symbols) and VPS-32 (light green symbols) between the two flat sheets in contact (square symbols) and between the two rods sliding orthogonally on two rods in the same configuration (circles). The solid line and shaded regions correspond to a linear fit of our experimental results; the uncertainty (shaded regions) is based on the standard deviation of the data.}
\label{fig:friction_coefficient}
\end{figure}

Two sets of frictional experiments were performed to verify that the frictional behavior was not affected by the geometry of the samples in contact.  First, a 60$\times$60\,mm$^2$ steel sleigh covered by a 1\,mm thick VPS sheet was dragged at a velocity of 3\,mm/s on a second VPS layer of the same thickness (\figref{fig:friction_coefficient}\textbf{a1}). A calibrated weight imposed the normal force $F_\text{n}$ between the two flat sheets of VPS on the sliding sleigh.
The corresponding tangential force $F_\text{t}$ was measured using a Universal Mechanical Testing Machine (Instron 5943). In the second set of experiments, we quantified the frictional relation between $F_\text{n}$ and $F_\text{t}$ using the same setup but with a different contact geometry: a pair of VPS rods attached to the sleigh was dragged perpendicular to the other pair of VPS rods attached to the stationary substrate (\figref{fig:friction_coefficient}\textbf{a2}). 

The results of our frictional tests are presented in \figref{fig:friction_coefficient}\textbf{b}. From these data, we quantified the friction coefficient as $\mu = \bar \mu \pm \Delta \mu = 0.32 \pm 0.03$ ($\bar \mu = 0.32$ and $\Delta \mu=0.03$ correspond to the average and standard deviation of the measurement $F_\text{t}/F_\text{n}$ from 143 independent experiments, respectively). We found that the difference in geometry between the two friction experiments (flat slabs and two rods in contact) yielded nearly identical results. We highlight that the approximation of Amontons-Coulomb friction law is valid over a large range of normal force, $2 \le F_\text{n}\,[\text{N}] \le 70$.

\section{QUANTIFICATION OF THE DEFORMATION OF THE RODS IN THE ELASTIC ORTHOGONAL CLASP}
\label{sec:largedeformation}

In this section, we quantify the deformations that elastic rods undergo when brought into tight contact, in the orthogonal clasp configuration. Specifically, by making use of our $\mu$CT imaging and FEM measurements (as detailed in the \secref{sect:FEM}), we measured the cross-section deformation, the centerline curvature, as well as the shear strain along the centerlines of the rods. Even at relatively low values of the loading forces~$f_z=F_z/EA$, our results indicate that there is significant cross-sectional deformation, in violation of the base assumptions of the conventional Kirchhoff-rod model~\cite{AudolyBook}.

\begin{figure}[h!]
\centering
\includegraphics[width=\textwidth]{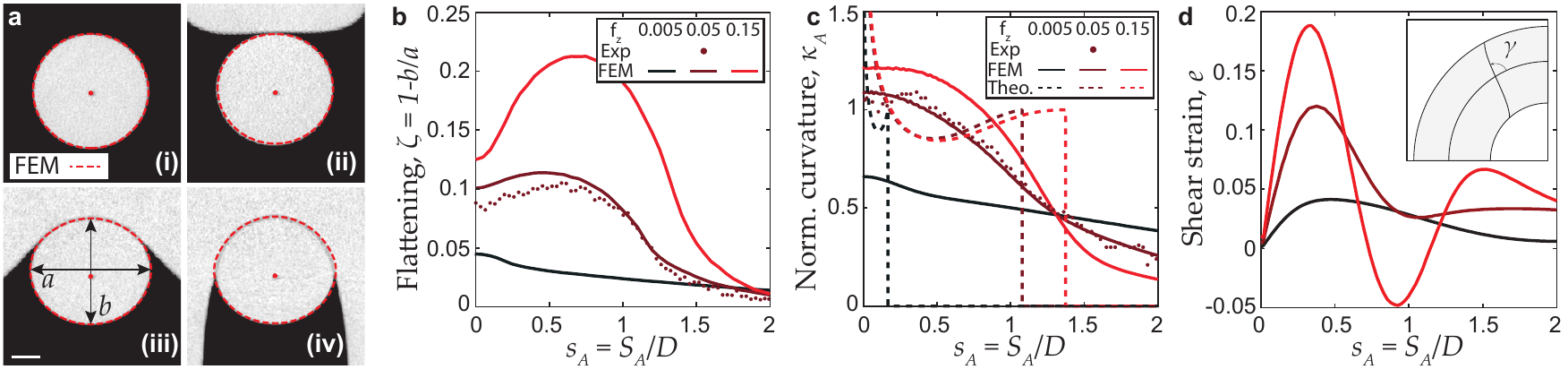}
\caption{\textbf{Cross-section deformation, centerline curvature, and shear deformation of the elastic rod~A in clasp configuration~3.} \textbf{a},~Experimental $\mu$CT snapshots of the normal cross-sections to the physical centerline, at four different locations, for $f_z=0.05$; $s_\mathrm{A}=2.0,1.2,0.4$~and~0 for snapshots (\textbf{i}) to (\textbf{iv}), respectively. In the $\mu$CT data, the deformed centerline and the rod surface are visible by the darker gray pixels (due to the different material densities between VPS and Solaris). The FEM-computed material centerline and cross-section boundaries are superposed, respectively, as a red point and red dashed lines. The lengths $a$ and $b$ are the width and height of the flattened cross-section. Scale bar: 2\,mm. \textbf{b},~Flattening~$\zeta$ of the cross-section as a function of centerline arc length obtained from FEM (three solid lines corresponding to different applied loads), and from the $\mu$CT images (dots). \textbf{c},~Normalized curvature~$\kappa_\mathrm{A}=K_\mathrm{A}\,D$ as a function of centerline arc length. For the FEM data (three solid lines) correspond to the different values of the applied load. The data points represent the $\mu$CT data. \textbf{d},~Shear strain,~$e:=1/2 (\pi/2 - \gamma)$ as a function of centerline arc length, taken from FEM data.}
\label{fig:LargeDef}
\end{figure}

We proceed to examine the local deformations of the elastic rods within the contact region of the orthogonal clasp. For this purpose, we take advantage of the 3D data provided by both the $\mu$CT tomographic images and the FEM computations. We take the configuration 3 introduced above as a representative case throughout this section. Even at the moderate load $f_z=0.05$, the rods undergo significant cross-section deformations in the contact regions. The anisotropy of the cross-section is enhanced as the rod B embraces rod A (See the snapshots provided in \figref{fig:LargeDef}\textbf{a} of cross-sections normal to the physical centerline). 

We further quantified the cross-sectional deformations, picking three different loading conditions, $f_z=\{0.005,\,0.05,\,0.15\}$. In \figref{fig:LargeDef}\textbf{b}, we plot the cross-section flattening,~$\zeta:= 1-b/a$, of the rod along its arc length $s_\text{A}$ ($a$ and $b$ are, respectively, the width and height of the flattened cross-section boundaries; see \figref{fig:LargeDef}\textbf{a}). An almost circular cross section has $\zeta=0$ and $\zeta$ increases as the cross section deviates from a circular shape through flattening. We observe that $\zeta$ is close to zero outside the contact regions, but substantial cross-section deformations arise in the contact regions (\textit{e.g.}, more than~10\% at the loading $f_z=0.05$). 

Next, in \figref{fig:LargeDef}\textbf{c}, we present the normalized curvature $\kappa = K D$ of the deformed centerline of rod~A of configuration 3. Again, the results obtained from FEM (solid lines) and experiments (data points) are in good agreement. At $f_z=0.005$, the normalized curvature profile remains relatively low, but for the higher values of~$f_z$ ($0.05$ and~$0.15$), the normalized curvature reaches and even exceeds 1. In these high-load regimes, the radius of curvature is, therefore, of the order of the rod diameter. As a comparison, we plotted the theoretical curves based on the geometrical solutions for the corresponding opening angles (\secref{sect:ideal_clasp}), which does not agree with the experimental nor the FEM observations.

Finally, in \figref{fig:LargeDef}\textbf{d}, we plot the shear strain $e:=1/2\,(\pi/2-\gamma)$~\cite{crandall1978introduction} along the rod centerline, which can only be accessed in FEM simulation data via an angle~$\gamma$ between the tangent of the deformed material centerline and the cross-section surface that was initially planar and normal to the centerline before loading ($e = 0$). We observe significant values for the shear strain ($e \simeq 0.08$) in the vicinity of the contact region, even at moderate loading~$f_z=0.05$.

From the data presented above, one may be tempted to conclude that the complexity of the problem studied in the main article derives from the 3D nature of the rod deformation, which in turn could call into question the applicability of a classic one-dimensional centerline based theory such as a Kirchhoff-like rod description. However, as we demonstrate in the main article (Sections ``\textit{The geometrical theory of idealized orthogonal clasps}" and ``\textit{Distributions of the contact pressure}"), much of the nonlinear geometry and mechanics of the contact region can be captured by an appropriate one-dimensional theory provided that the small, but positive, rod thickness is appropriately incorporated.  The details of this geometric theory for two tubes in contact is provided below in \secref{sect:Theory}.

\section{SIMULATIONS USING THE FINITE ELEMENT METHOD}\label{sect:FEM}

Throughout our study, we performed extensive simulations using the finite element method (FEM) to compute the equilibrium shapes of the static and sliding elastic clasps. These FEM computations were fully validated against experiments and enabled us to access quantities not available experimentally. Particular focus was given to the contact pressure distribution of the contact region between the two rods composing the orthogonal elastic clasp. 

In this section, we detail the procedure followed for these FEM simulations. First, in \secref{subsection:FEM_1}, we provide an overview of the computational protocol employed to simulate both the static and the sliding (\textit{i.e.} frictional) elastic orthogonal clasps . In \secref{subsection:FEM_2}, we then describe the loading sequences specialized to each of the two elastic clasp cases. Finally, in \secref{subsection:FEM_3}, we validate the FEM results by comparing the numerical and experimental force-displacement curves of the four elastic orthogonal clasp configurations presented in Fig.~2 of the main article.

\subsection{Overview of the FEM simulations}
\label{subsection:FEM_1}

We used the commercially available finite element software (ABAQUS/STANDARD~6.14-1, Dassault Syst\`emes~2014) for all FEM simulations. For the static elastic clasp case, we performed a nonlinear \textit{static} analysis, whereas a nonlinear \textit{dynamic-implicit} analysis was performed for the frictional clasp case. The latter was beneficial to facilitate the convergence of the simulations, even if our main interest was on the quasi-static response of the frictional clasp (no inertial nor time-dependent effects).
The loads were applied slowly enough to exclude any undesired inertial forces (see Subsection~\secref{subsection:FEM_2}) to ensure that these simulations were quasi-static. 

The rods were meshed with reduced hybrid 3D brick elements (C3D8RH) with $\sim 500\,\text{mm}^{-1}$ elements per unit length; this number was selected from a separate mesh-sensitivity analysis. We exploited the symmetry of the problem and meshed only a quarter of the system for the (quasi-)static clasp case and one-half of the system for the frictional clasp case. This symmetry-enabled reduction of the system size significantly reduced the computational cost of the full 3D simulations. 
An incompressible neo-Hookean constitutive description was used to model the hyperelastic response of the VPS elastomer. The Young's modulus $E=0.52\,\text{MPa}$ of VPS-16 was quantified separately through mechanical tests (see~\secref{sect:mechanical_characterization}). The contact between rods was modeled with penalty normal force, with specified Amontons-Coulomb dynamic friction coefficient $\mu$. For the frictionless clasp, $\mu$ was set to zero. For the frictional clasp, we set to $\mu = 0.32$, whose value was measured experimentally (see~\secref{sect:friction_coefficient}). 

The computational cost for a static clasp simulation, in which the wall-to-wall distance $H$ increased with an increment of $\Delta H = 1\,\text{mm}$ until the nondimensional vertical force $f_z=0.2$ is reached, is approximately 40 hours on a desktop workstation with an octa-core processor (Intel Xeon processor 6136 3.20\,GHz) and 32\,GB of RAM. For a frictional clasp simulation, in which $H$ increased with $\Delta H = 1\,\text{mm}$ until $f_z=0.1$ is reached, the simulation took approximately 60 hours with the same computational power.

\subsection{Loading sequence}
\label{subsection:FEM_2}
\begin{figure}[h!]
    \centering
    \includegraphics[width=0.9\textwidth]{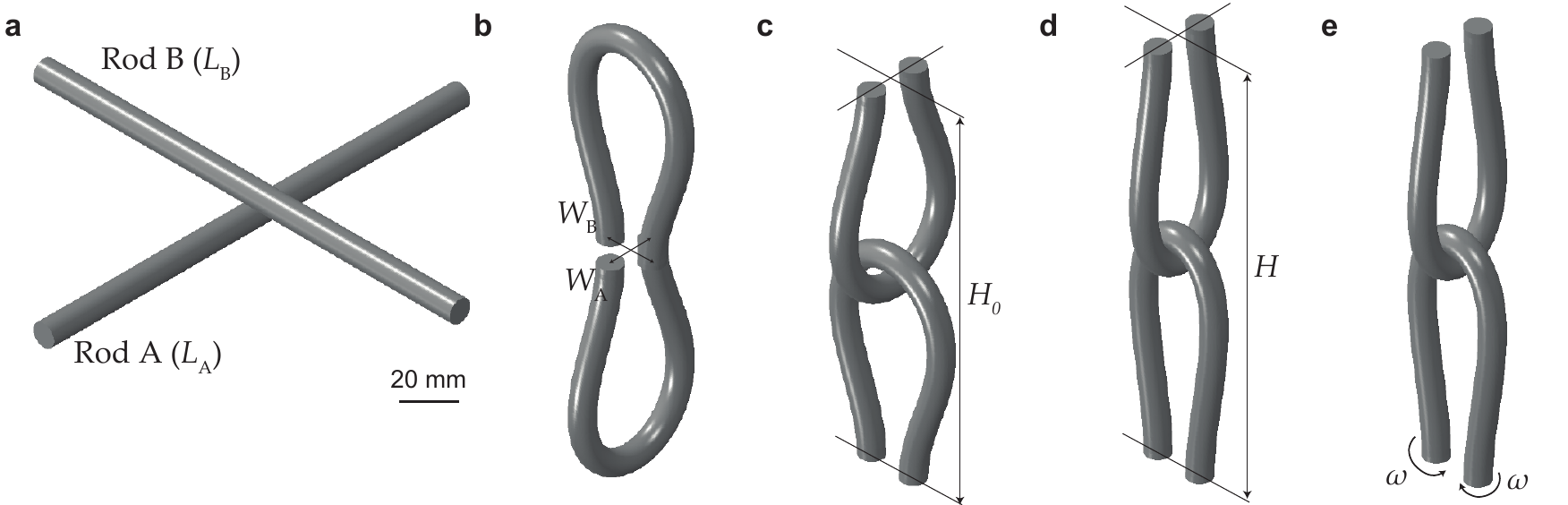}
    \caption{\textbf{Loading sequence of the FEM elastic clasp} (\textbf{a}),~Rod~A and rod~B are laid orthogonal to each other. (\textbf{b}),~The extremities of each rod are brought together, with clamp-to-clamp distances of $W_\text{A}$ and $W_\text{B}$. (\textbf{c}),~The onset of contact of the two rods occurs when the wall-to-wall distance is $H_o$.  (\textbf{d}),~The wall-to-wall distance, $H$, is increased gradually during loading of the elastic clasp. (\textbf{e}),~Rotation of the extremities of rod~A. A constant rotational speed $\omega$) is imposed to simulate a sliding of the rods in the elastic clasp.}
    \label{fig:FEM_loadingseq}
\end{figure}

We applied the following sequence of prescribed translational and rotational displacements at each extremity of the two rods to achieve the desired boundary conditions. First, two initially straight rods of specified rest length~$(L_\text{A},\,L_\text{B})$ and rest diameter $D$ (\figref{fig:FEM_loadingseq}\textbf{a}) were brought together, with specified clamp-to-clamp distance $(W_\text{A},\,W_\text{B})$ (\figref{fig:FEM_loadingseq}\textbf{b}). During this first step, the rods were not yet in contact. 

Contact was only established when the wall-to-wall distance reached $H_o$ (see \figref{fig:FEM_loadingseq}\textbf{c}). The onset of the contact region was a single contact point when the imposed boundary conditions induced low values centerline curvature at the apices. By contrast, for the cases where the centerline curvature was significant before contact, double-contact could be observed at onset (see Fig.~3 of the main article). We then gradually increased the wall-to-wall distance $H$ from $H_o$ ($H>H_o$), in increments of $\Delta H=1\,\text{mm}$ (see \figref{fig:FEM_loadingseq}\textbf{d}). At each value of $H$, the quantities of interest were recorded (\textit{e.g.}, local opening angles, reaction forces, and  shape of the contact region). 

For the frictional clasp simulations, an additional step was required to impose the sliding of the rod~A. Recall that, in the experiments, a long rod~A was fed into the system through pulleys until the engagement of sliding friction (Fig.~4 of the main manuscript). By contrast, in the FEM simulations, it was computationally prohibitive to consider a configuration identical to the experiments and mesh the region of rod~A outside the pulleys. Alternatively, to simulate the relative displacement between rod~A and rod~B, we fixed the extremities of rod~A and applied a rotational displacement at the extremities of rod~B (see \figref{fig:FEM_loadingseq}\textbf{e}), instead of directly displacing rod~A. The prescribed rotational velocity at the ends of rod~B was set to $\omega=\pi/30\,\text{rad/s}$, which was still sufficiently slow to avoid inertial effects. This strategy enabled us to impose sliding friction between the two rods without requiring us to simulate the full system (including the far-end region of rod~A and the pulleys). Once the reaction forces exerted on rod~A reached a steady-state, all the relevant quantities of the frictional clasps were recorded.

\subsection{Validation of the FEM simulations against the experimental macroscopic mechanical response of the elastic clasp}
\label{subsection:FEM_3}

Throughout the main article, we presented resounding evidence for the quantitative trustworthiness of the FEM simulations to predictively reproduce a variety of features of the elastic clasp problem. The following quantities were compared directly between FEM and experiments: the opening angles (Fig.~2\textbf{e-f}), the cross-sectional flattening (Fig.~3\textbf{b}), the centerline curvature (Fig.~3\textbf{c}), the boundary of the contact map (Fig.~3), and the tension ratio (Fig.~4\textbf{c}); all figures mentioned correspond to the main article. Here, we now provide an additional layer of validation of the FEM by performing a comparison with experimental macroscopic mechanical tests of the elastic clasp.

\begin{figure}[h!]
    \centering
        \includegraphics[width=0.9\textwidth]{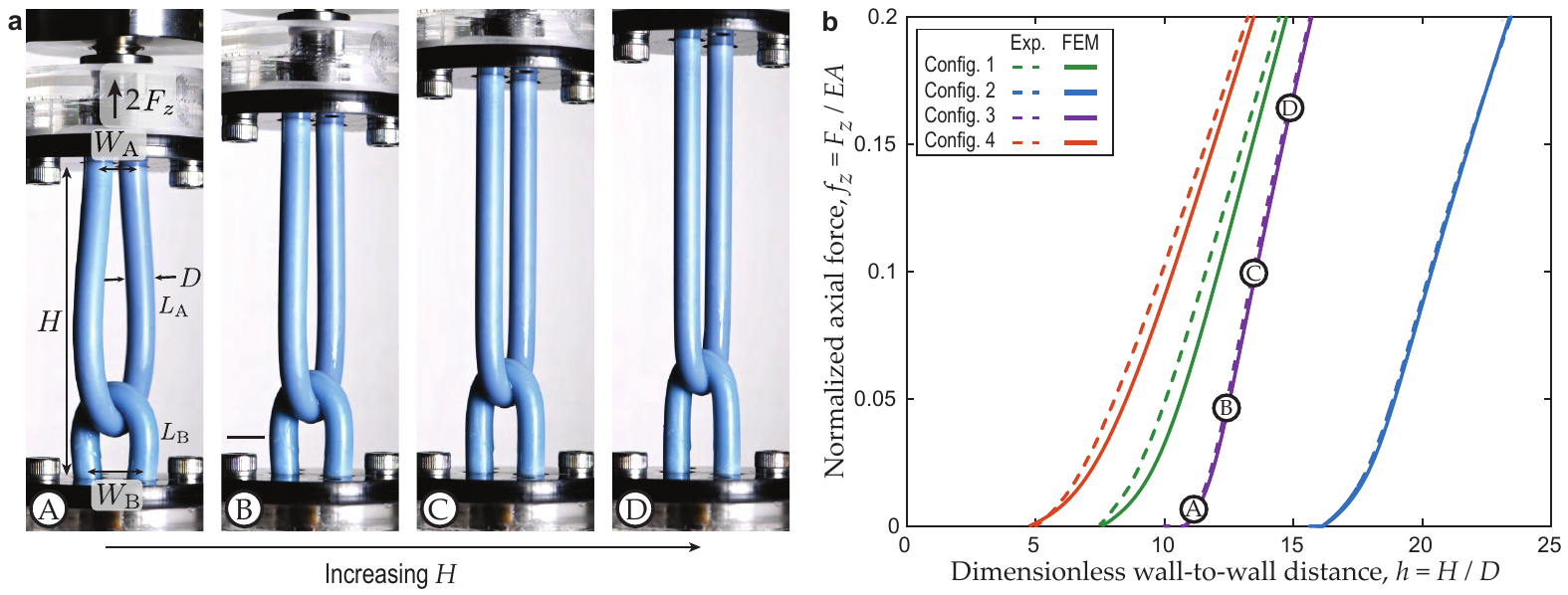}
    \caption{\textbf{Macroscopic mechanical testing of the elastic clasps.} \textbf{a},~Schematic of the mechanical testing apparatus. The elastic clasp configuration is first established, and, then, the value of $H$ is gradually increased by vertically moving the upper clamped boundaries of rod~A (while the lower clamp of rod~B remains fixed). During this displacement-imposed test, the reaction force, $2F_z$, on the upper boundary is recorded. Scale bar is $10\,\text{mm}$. \textbf{b},~Force-displacement curves for four different configurations presented in Fig.~2 of the main article. There is an excellent agreement between the experiments (dashed lines) and FEM simulations (solid lines).}
    \label{fig:FEM_validation_macro}
\end{figure}

In a setting identical to that of Fig.~2 of the main article, we clamp two rods at rigid supports (upper and lower clamps), which impose the global geometry $(L_\text{A}, W_\text{A}, L_\text{B}, W_\text{B})$. Then, we quasi-statically vary the wall-to-wall distance, $H$ (see  \figref{fig:FEM_validation_macro}\textbf{a}) and measure the corresponding vertical axial force, $2F_z$, at the upper clamp as a function of $H$. In \figref{fig:FEM_validation_macro}\textbf{b}, results are provided for the configurations \,1,\,2,\,3,\,and~4 defined in Fig.~2 of the main article. Again, there is excellent agreement between the numerical (thick solid lines) and experimental (thin solid lines) force-displacement curves. These results further highlight the validity and accuracy of the FEM simulations conducted throughout the study.

\section{THEORY -- Elastic rods and tubes}\label{sect:Theory}

In this section, we discuss 1D mathematical models for the orthogonal clasp in which the polymer rods are treated as tubular regions; the centerline is allowed to deform, but the cross-sections are approximated as undeformable. Given the large aspect ratio of the polymer rods, we can anticipate that these approximations are reasonable, at least in an appropriate loading regime. Our objective in introducing a set of reduced-order models is to gain insights into the observations from experiment and FEM simulation. 

This theory section is organized as follows. In \secref{sect:kirchhoff_rod_theory}, for completeness, we provide an overview of Kirchhoff rod theory specialized to the planar centerline configurations relevant in our study. In \secref{sect:boundary_layers}, we discuss the boundary layers that arise in tensile solutions. 
In \secref{sect:opening_angle_computation}, we describe the computations done using Kirchhoff rod theory to obtain the local opening angles, as a function of the normalized vertical force in the \textit{orthogonal elastic clasp} experiments in Fig.~2 of the main article. Next, in \secref{sect:ideal_clasp}, we present the details of the theory used to compute the equilibrium configurations of two flexible tubes of finite thickness in tight frictionless contact. The results from this theory are shown in the main article in Fig.~1\textbf{d} (geometric configuration of the ideal clasp) and Fig.~3 (contact sets superposed onto the FEM pressure maps). Finally, in \secref{sect:friction}, we incorporate friction into the ideal clasp model under appropriate assumptions and describe the computations done to obtain the dimensionless pressure map shown in Fig.~5\textbf{f} of the main article.

\subsection{Overview of Kirchhoff rod theory}\label{sect:kirchhoff_rod_theory}

In Kirchhoff rod theory~\cite{antman05}, any configuration of an elastic rod is identified with a centerline curve $\br(s)\in \mathbb{R}^3$ along with a right handed orthonormal frame of directors $\bd_i(s)$,  $i=1,2,3$, each parameterised by arc length $s$ (in a reference state). We will consider only the constrained case of inextensible, unshearable rods in which $\br'(s) = \bd_3(s)$ (so that $s$ is arc length in any configuration). By orthonormality, there is a Darboux vector $\bu(s)$  such that $\bd'_i = \bu\times\bd_i$, $i=1,2,3$, and the components $u_i:=\bu\cdot \bd_i$ are the (two) bending and twist strains.
The pointwise equilibrium conditions for the balance of forces and moments acting on each segment of the rod are:
\begin{align}
\bn' + \bp  &= {\mathbf{0}}\, ,\label{eq:force_balance_3D}\\
\bmo' + \br'\times\bn + \bl &= {\mathbf{0}}\, ,\label{eq:moment_balance_3D}
\end{align}
where $\bn(s)$ and $\bmo(s)$ are the total internal force and moment acting across the rod cross-section at $s$. In \eqref{eq:moment_balance_3D}, prime denotes derivative with respect to $s$. Both $\bn(s)$ and $\bmo(s)$ are computed as the appropriate averages of the stresses exerted by the material in $s^+$ on the material in $s^-$, and $\bp(s)$ and $\bl(s)$ are the external force and moment \emph{densities} (per unit $s$) acting on the rod. In the absence of external loads we immediately obtain the integrated form of the balance laws
\begin{align}
\bn(s) &= \bn(0)\, ,\label{eq:force_balance_int}\\
\bmo(s) + \br(s)\times\bn(s) &= \bmo(0) + \br(0)\times\bn(0)\, .\label{eq:moment_balance_int}
\end{align}

To obtain a closed system, the kinematics and balance laws must be supplemented by appropriate constitutive relations. Consistent with the FEM simulations, we will assume that the tubes in the experiment are formed from an isotropic, homogeneous, linearly elastic 3D material (in the limit of small strains), with its unstressed reference configuration having a straight centerline with normal cross-sections all being circular of the same diameter $D$. The Kirchhoff theory then arrives at linear constitutive relations between the director basis components of moment $m_i:=\bmo\cdot \bd_i$
\begin{align}
m_1 = B u_1\qquad m_2 = B u_2\qquad\ m_3 = C u_3 ,\label{eq:constitutive_relation_3D}
\end{align}
where $B=EI$ and $C=GJ$, with $I=\pi D^4/64$  
the geometric moment of inertia of the circular cross-section of diameter $D$, $J=2I$, and $E$ and $G$ are the Young and shear moduli of the material. Analogous constitutive relations between the director components of the force $n_i:=\bn\cdot \bd_i$ and shear and extension strains can be derived. Instead, we assume that these strains are negligible so that the components of the internal force $\bn$ remain as basic unknowns of the problem that must be determined from the equilibrium conditions. The assumptions of inextensibility of the centerline of the rod and unshearability of its cross-section are reasonable approximations to make when the shear force and the tension are small compared to the shear and axial rigidity of the rod; \textit{i.e.}, $n_3\ll EA$ and $\sqrt{n_1^2 + n_2^2}\ll GA$, where $A=\pi D^2/4$ is the area of the circular cross-section. Given that the two moduli are of the same order of magnitude, it is reasonable to assume inextensibility and unshearability when $\epsilon={|\bn|}/{EA} \ll 1$. For the specific experiments and FEM simulations discussed in the main article, we have $\epsilon\lessapprox 0.2$. We note that the assumption of unshearability relates to the assumption that the circular, orthogonal cross-section of the tubular material volume does not deform substantially. This assumption renders our analysis of the 3D contact geometry of two tubes tractable. Then the additional assumption of inextensibility ensures that the tube volume is invariant under any deformation.

For any hyperelastic constitutive relations, the system of kinematic and equilibrium conditions can be written in the form of a Hamiltonian system of ODE~\cite{Dichmann1996,Kehrbaummaddocks1997}.
In the case at hand, the specific form of the Hamiltonian is $\tfrac{m_1^2+m_2^2}{2B} +\tfrac{m_3^2}{2C} +\bn\cdot\bd_3$.
We will make use of the fact that as our rod is assumed uniform, the Hamiltonian is constant along all equilibria. Note that, since the constitutive relations in [\ref{eq:constitutive_relation_3D}] are transversely isotropic, we have the further integral that the twist moment $m_3$ is also constant along equilibria, but this fact does not enter our analysis.

We will only be concerned with equilibria assembled from tubes with planar centerlines. For planar centerlines the equilibrium conditions in \eqref{eq:force_balance_3D} and \eqref{eq:moment_balance_3D} can be simplified by assuming $\bn = n_1\bd_1 + n_3\bd_3$, $\bp=p_1\bd_1+p_3\bd_3$, $\bmo= m_2\bd_2$ and $\bl=l_2\bd_2$. The in-plane directors $\bd_1$ and $\bd_3$ can then be conveniently parametrised as $\bd_3 = \left(\cos\theta,\sin\theta\right)$ and $\bd_1 = \left(-\sin\theta,\cos\theta\right)$, where the function $\theta(s)$ is the angle between the tangent $\bd_3(s)$ and a Cartesian $y$ axis as shown in \figref{fig:elastica_schematics}. We choose the diameter $D$ of the rod as length scale, and a force scale $F$ as the characteristic highest applied end load in the experiments. We then arrive at  the dimensionless equations describing planar equilibria
\begin{align}
\theta' & = -\kappa\, ,\label{eq:theta}\\
y' &= \cos\theta\, ,\label{eq:Y}\\
z'&=\sin\theta\, ,\label{eq:Z}\\
n_1'&=n_3 \kappa - p_1\, ,\label{eq:n1}\\
n_3'&=-n_1 \kappa - p_3\, ,\label{eq:n3}\\
\BH \kappa'&=n_1+l_2\, .\label{eq:u2}
\end{align}
Here $\BH \equiv B/FD^2 = {1}/{16\epsilon}\lessapprox 0.3$ is a dimensionless bending modulus, and prime now denotes the derivative with respect to the dimensionless arclength $s\to s/D$. The above system of ODE is to be solved on an interval $0<s<L$, where $L$ is the dimensionless rod length. Typically, we have $L\gg 1$ corresponding to the assumption that the tubular volume has a high aspect ratio, subject to appropriate boundary conditions. 
Equations \eqref{eq:n1} and \eqref{eq:n3} are the components of the force balance \eqref{eq:force_balance_3D} in the director frame, \eqref{eq:u2} is the only nontrivial component of the moment balance, and \eqref{eq:theta} to \eqref{eq:Z} are kinematic relations that determine the components of the rod centerline in a fixed coordinate bases $(y, z)$. The only nonvanishing component $u_2$ of the Darboux vector has been identified as the negative of the curvature $\kappa$ of the centerline. 

This first-order system of ODE determines the equilibria of the version of the classic Euler elastica that arises when external distributed loads are applied. In the absence of external loads, \textit{i.e.}\ $p_1=p_3 = l_2=0$, $\kappa$ can be eliminated to obtain the classic second-order pendulum equation for the tangent angle $\theta$~\cite{antman05}
\begin{align}
\lambda^2\frac{d^2\theta}{\text{d}s^2} + \sin(\theta-\gamma) = 0\, ,\label{eq:elastica}
\end{align} 
where $n_1 = |\bn|\sin(\theta-\gamma)$, $n_3 = -|\bn|\cos(\theta-\gamma)$, and $\theta-\gamma$ is the angle between the tangent $\bd_3$ and the externally applied force at $s=0$ ($\gamma$ being the angle between the global $y$ axis and the applied force at $s=0$). The parameter $\lambda$ is defined as
\begin{align}
    \lambda=\sqrt{\frac{\BH}{|\bn|}}\, .\label{eq:boundary_layer_thickness}
\end{align}

Here we are concerned with tensile solutions where $n_3=-|\bn|\cos(\theta-\gamma)>0$ away from the boundaries, \textit{i.e.}\ $\cos(\theta-\gamma)\approx \pi$. We will show that such equilibria have boundary layers in the limit $\lambda/L\ll 1$, \textit{i.e.}\ the physical length of the rod is many multiples of its diameter.

We also note from \eqref{eq:force_balance_int} that in the absence of any external force densities, $|\bn|$ is constant. Then in the planar case at hand, equation \eqref{eq:elastica} has a first integral leading to the classic, cats eye, pendulum $(\theta,\theta')$ phase plane relation
\begin{align}
\frac{1}{2}\left(\lambda\frac{d\theta}{\text{d}s}\right)^2 - \cos(\theta-\gamma) = H\, ,\label{eq:hamiltonian}
\end{align} 
where the constant of integration is written as $H$, indicating that the left hand side of \eqref{eq:hamiltonian} is in fact the Hamiltonian expressed in the planar variables.

\subsection{Boundary value problems with boundary layers}
\label{sect:boundary_layers}

The bending rigidity of the VPS polymer rods and loadings used in the \textit{elastic orthogonal clasp} experiments, both with and without friction, leads to significant curvatures in the rods close to both the contact region and the clamped ends. Our objective in this subsection is to quantify the extent of such regions, hereafter referred to as \textit{boundary layers}. We will show that the typical length of these boundary layers is significantly smaller than the lengths of the rods used to conduct the experiments and FEM simulations. Hence, this separation of length scales ensures that the internal force in regions of the rod away from these boundary layers is primarily tensile. However, note that even though the boundary layers can be small compared to the overall rod length, they have an important impact on the local geometry of the contact regions. By contrast, in the absence of bending stiffness (as is the case in the ideal orthogonal clasp; see \secref{sect:ideal_clasp}), the rod strands would be straight anywhere outside the contact region, and the local contact geometry (\textit{e.g.} the opening angle) would be imposed directly by the geometry of the boundary conditions.

To evaluate the length of the boundary layers, we start from the dimensionless equation \eqref{eq:elastica} for the planar elastica and re-scale its independent variable by the total (dimensionless) length $L$  of the rod to obtain $\frac{\lambda^2}{L^2}\frac{d^2\theta}{d\bar{s}^2} + \sin(\theta-\gamma) = 0$, where  $\bar{s}\equiv s/L$, and $0<\bar{s}<1$. If we now assume the length $L$ to be large compared to $\lambda$, then the  elastica equation becomes a singular perturbation problem which is indicative of the existence of boundary layers~\cite{Hinch1991}. Given the tangent angles  at the two boundaries of a rod, as is the case in the contact free region of our \textit{elastic orthogonal clasp} experiments, boundary layers are expected to exist at both ends.

Outside the boundary layers, the `\textit{outer}' region, to leading order we obtain the algebraic relation $\sin(\theta_\text{out}-\gamma)=0$, which for a tensile solution $n_3>0$ implies that, in the outer approximation, we have $\theta_\text{out}=\gamma+\pi$.
The solution in the inner region of the boundary layer near $\bar{s}=0$ can be obtained by introducing a stretched arc-length~\cite{Hinch1991,Audoly2010} $\tilde{s} =\bar{s}L/\lambda$ to obtain $\frac{d^2\theta_\text{in}}{d\tilde{s}^2}+\sin(\theta_\text{in}-\gamma)=0$ (where $0\le\tilde{s}<\infty$).  The smooth matching of the inner and outer solution is ensured by enforcing the boundary conditions $\theta_\text{in}(0) = \theta_0$ and $\theta(\infty) = \theta_\text{out}$ on the ODE. A similar analysis at the other end of the rod would ensure the satisfaction of the boundary condition and a smooth matching of the outer solution with the boundary layer. We note that the only non-periodic solution of the elastica that connects two points on the $(\theta,\theta')$ plane with infinite arc length is the separatrix. Therefore, the shape of the rod inside the boundary layers at the two ends must be arbitrarily close to the separatrix solution. For a rigorous treatment of this assertion, see \cite{Miura2020}.

To compute the approximate numerical value of $\lambda$ in units relevant to our experiments (\textit{i.e.}, $EA$ for force and $D$ for lengths), we consider a case where the applied force $F$ is as low as $0.01\, EA$. For such low loads, \eqref{eq:boundary_layer_thickness} predicts a value of $\lambda\approx 2.5 D$, meaning that the bending rigidity influences the equilibrium shape of the rod for arc lengths smaller than~$\lambda$. On the other hand, past this length~$\lambda$, the rod becomes straight, and the internal force becomes aligned with the tangent of the rod.

Next, using experiments and numerical simulations, we verify that the length of the sliding rod ($L_\text{A} = 20D$) chosen in our \textit{elastic orthogonal clasp} experiments with friction is sufficient to ensure that the boundary layers occurring close to the contact region of the two rods do not have any appreciable effect on measured the tension ratios. To demonstrate the insensitivity of these experimentally measured tension ratios to the length $L_\text{A}$ of the sliding rod, we conducted an experiment using VPS-16 rods of diameter $D=8\,$mm with $w_\text{A} = W_\text{A}/D = 2$, $l_\text{B} = L_\text{B}/D=10$ and $w_\text{B} = W_\text{B}/D = 2$ (see Fig.~4\textbf{a} of the main article), and varied the length $L_\text{A}$ from $5D$ to $20D$. A dead load of $t_0=0.056$ (normalized by $EA$) was imposed at the lower end of the sliding rod (mass of the dead load $M=150$ g, see Fig.~4\textbf{a} of the main article).

\begin{figure}[h!]
	\centering
	\includegraphics[width=0.5\textwidth]{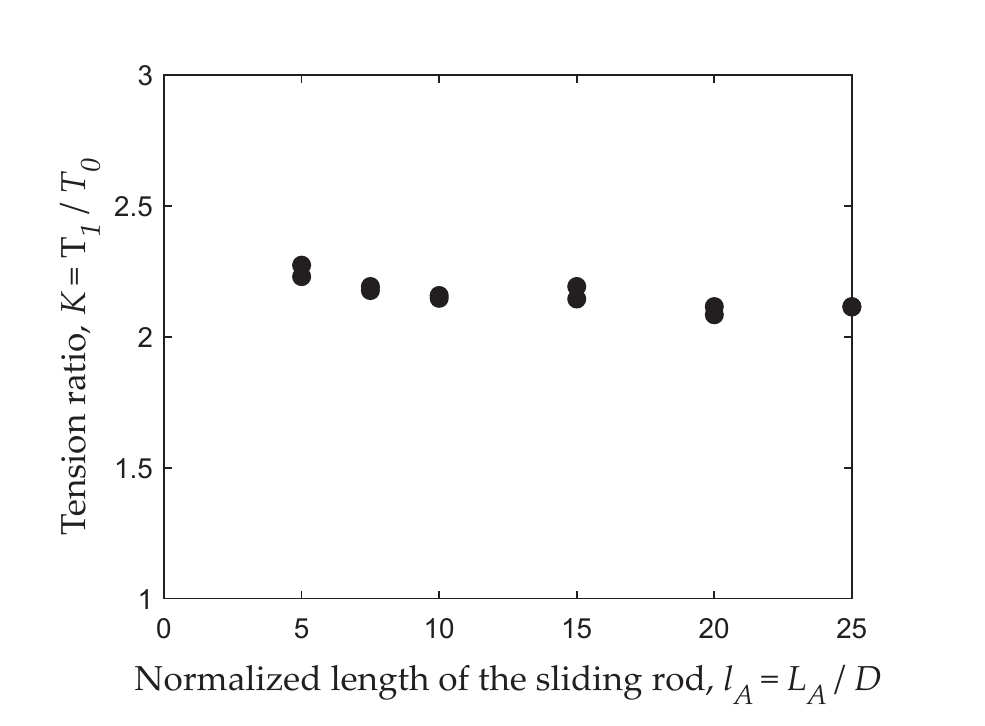}
	\caption{\textbf{Experimental high-to-low tension ratio of the sliding rod as function of the normalized sliding rod length, $l_\text{A}$, between the pulleys.} These experiments were performed in a geometrical configuration such that $w_\text{A}=W_\text{A}/D=2$, $l_\text{B}=L_\text{B}/D=10$ and  $w_\text{B}=W_\text{B}/D=2$ and with a dead load imposing a normalized lower force $t_0 = T_0/EA = 0.056$. We observe that for $l_\text{A}>10$, the measured force ratio $K$ shows little dependence on $l_\text{A}$. Hence, throughout this study, we  take $l_\text{A}$=20.} 
	\label{fig:K_vs_l}
\end{figure}

In \figref{fig:K_vs_l}, the measured tension ratio, $K=T_2/T_1$, is plotted as a function of the length of the sliding rod, in units of $D$. We observe that $K$ becomes nearly insensitive to the length $l_\text{A}$, as it increases. The relative difference between the tension ratio at $l_\text{A}=10$ and $l_\text{A} = 25$ is $1.6\%$.

\begin{figure}[h!]
	\centering
	\includegraphics[width=0.8\textwidth]{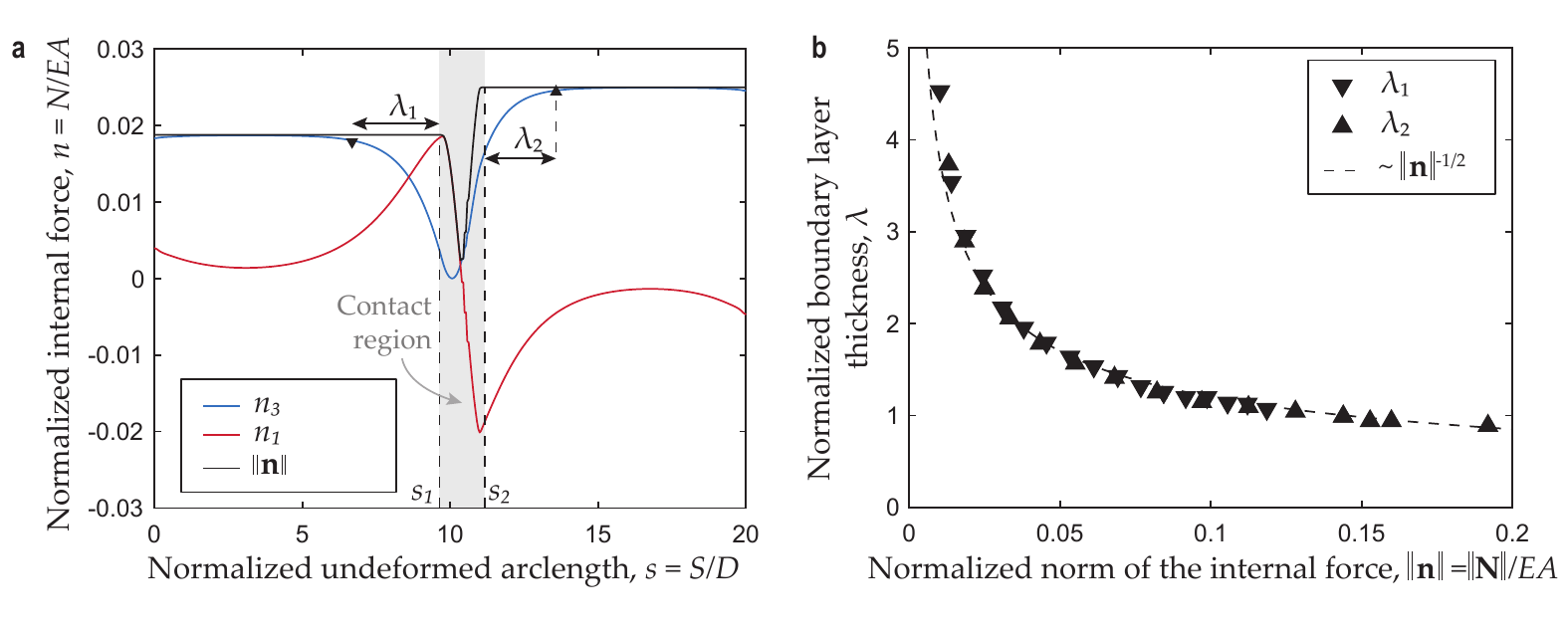}
	\caption{\textbf{Quantification of the boundary layer thickness using FEM to compute the tension and shear force profiles along the sliding rod~A.} \textbf{a}, Tension $n_3$ (solid blue line) and shear force components $n_1$ (solid red line) of the internal force of the sliding rod along its normalized undeformed arclength $s=S/D$, for the \textit{frictional elastic clasp} experiment.  The simulation parameters are $l_\text{A}=20$, $w_\text{A}=2$, $l_\text{B}=7.5$ and $w_\text{B}=7.5$.  \textbf{b}, Comparison of the scaling prediction of \eqref{eq:boundary_layer_thickness}, and the computed $\lambda_1$ and $\lambda_2$ from the FEM simulations, at different values of the dead load $f_1$. Specifically, employing the small arbitrary factor $\epsilon=0.05$ in \eqref{eq:boundary_layer_thickness_FEM}, we find  $\lambda =1/2.63 \sqrt{n}$ with $n=|\mathbf{n}|$. }
	\label{fig:Boundary_layer}
\end{figure}

We proceed by using FEM simulations of our \textit{orthogonal elastic clasp} experiments with friction to quantify the thickness of the boundary layer in our setup. We performed these simulation with $l_\text{A}=L_\text{A}/D=20$, $w_\text{A}=W_\text{A}/D=2$, $l_\text{B}=L_\text{B}/D=10$ and $w_\text{B}=W_\text{B}/D=2$, while imposing a normalized dead load $t_0=0.019 $ (normalized by $EA$) at the lower end of the sliding rod. In \figref{fig:Boundary_layer}, we show a typical profile of the dimensionless internal force $\bn$ (normalized by $EA$) near the contact region of the two tubes as a function of the non-dimensional arc-length $s=S/D$ ($0\le s \le l_\text{A} = 20$). The locations $s_1$ and $s_2$ correspond to the touch-down and lift-off point of rod~A with rod~B. We define the location of the boundary layer on the left (or incoming tail) in our simulations at a point $\lambda_1$, as shown in \figref{fig:Boundary_layer}\textbf{a}, where the difference between the tension and the norm of the total internal force becomes a certain small fraction $\varepsilon$ of their difference at the point of contact $s_1$,
\begin{align}
\varepsilon = \frac{|\bn| - n_3(s-\lambda_1)}{|\bn|-n_3(s_1)}\, .\label{eq:boundary_layer_thickness_FEM}
\end{align}
We chose the value $\varepsilon = 0.05$ as our criteria to define the onset of the boundary layer, and verified that varying $\varepsilon$ from $0.01$ to $0.1$ did not have a significant influence on the trend of $\lambda_1$. The same method was used to determine the boundary layer thickness $\lambda_2$ on the right (outgoing tail) side of the contact region. In \figref{fig:Boundary_layer}\textbf{b}, we plot  the values of $\lambda_1$ and $\lambda_2$ as functions of the norm of the internal force normalised by $EA$, observing a good agreement between FEM and \eqref{eq:boundary_layer_thickness} with $\lambda = 1/2.63\sqrt{n}$. We also found that away from the boundary layer, the internal force norm is close to the tension (\textit{i.e.}, $n_3$), as concluded from the analysis at the beginning of this subsection.

\subsection{Computation of the local opening angle}
\label{sect:opening_angle_computation}

The occurrence of boundary layers near the contact region of the two tubes in our \textit{elastic orthogonal clasp} experiments causes the local opening angles $\theta_\text{A}$ and $\theta_\text{B}$ (as defined in Fig.~1\textbf{b} of the main article) to deviate from the global tangent angles associated with the approximately straight rod segments far from any boundary. As we will discuss later in \secref{sect:ideal_clasp}, these local angles are key parameters in determining the extent and geometry of the contact region, while the global opening angles are the more easily observed quantities.

\begin{figure}[ht!]
	\begin{center}
		\includegraphics[]{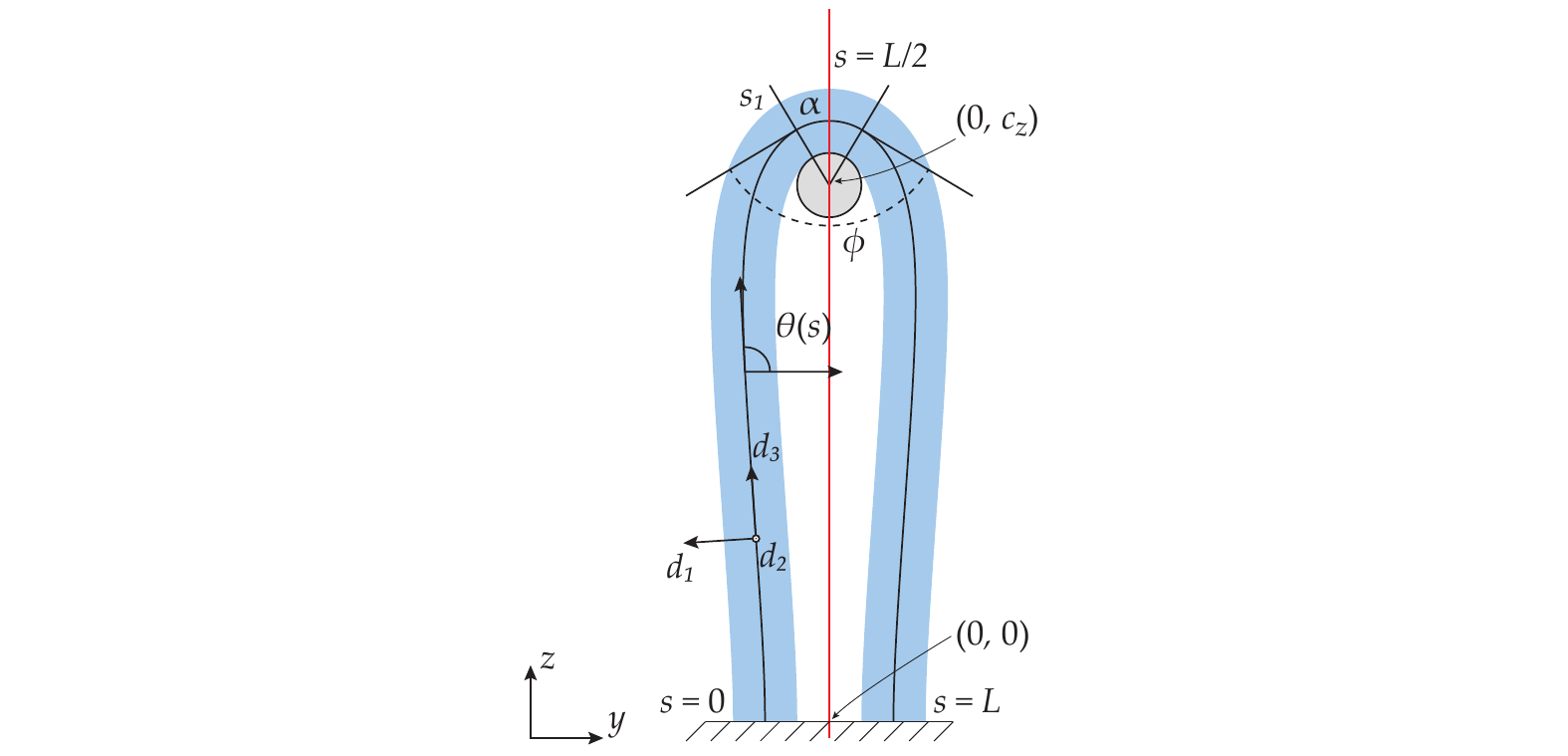}
		\caption{Schematic diagram of the contact of an elastic rod and a rigid cylinder, both of diameter $D$. The red line represents the plane of symmetry. The angle between the tangent, $\bd_3$ and the $y$-axis is given by $\theta(s)$, $\phi$ represents the local opening angle, $s_1$ is the unknown arc-length coordinate where the rod begins contact with the cylinder, and $\alpha$ is the angle subtended by half of the contact region at the center of the cylinder.}
		\label{fig:elastica_schematics}
	\end{center}
\end{figure}

In this section, we describe the computations done to obtain the local opening angles as a function of the normalized vertical force in the \textit{elastic orthogonal clasp} experiments, as shown in Fig.~2 of the main article. In doing so, we make the following strong simplifying assumption: each rod is assumed to turn in a circular arc around a single cross-section of the other rod (\figref{fig:elastica_schematics}). In other words, we consider the rod to be wrapped around a straight, rigid circular cylinder of the same diameter as the rod.  This assumption is often made to render such problems tractable~\cite{maddockskeller1987}. We will show that, under this assumption, the results obtained using the Kirchhoff rod theory do not compare well with the experimental and FEM results, as shown in Fig.~2\textbf{e} and \textbf{f} of the main article.

Consider a rod of diameter $D$ wrapped around a rigid cylinder of the same diameter as shown in \figref{fig:elastica_schematics}. The two far ends of the rod are assumed to be clamped at $s=0$ and $s=L$. Our objective is to formulate a boundary value problem that delivers the local opening angle $\phi$ of the rod, for a given distance $W_\text{A}$ between the clamped ends,  the total length $L$, and the vertical distance $c_z$ of the centre of the cylinder from the clamps. To formulate a well posed BVP, we would need to know the appropriate boundary conditions at the unknown point of contact at $s=s_1$, for which we must first discuss the regularity of various field quantities such as internal force and moment in the rod at that point.

Next, we recall the force and moment balance laws \eqref{eq:force_balance_3D} and \eqref{eq:moment_balance_3D}, where the force and moment densities (\textit{i.e.}, $\bp$ and $\bl$), can now be thought of as arising from the contact of the rod with the rigid cylinder. We assume that these force and moment densities can, at the most, have Dirac delta functions in their description, which immediately implies from \eqref{eq:force_balance_3D} and \eqref{eq:moment_balance_3D} that these delta functions must be balanced by jumps in the internal force $\bn$ and the internal moment $\bmo$ (and, therefore, $\bu$). It then follows from the kinematic relation $\bd'_i=\bu\times\bd_i$ that the directors (and consequently the tangent angle $\theta$) must always be continuous. We also assume that the force and moment densities are related by the expression $\bl = -\frac{\Delta}{2}\bd_1\times\bp$, which is to say that the only source of externally applied moment density loading in the rod is the couple exerted by the externally applied force density loading about the center of the cross-section. In the \textit{frictionless} case, which is the case of interest here, this relation implies that the moment density is identically zero, since the force density exerted by the rigid cylinder must be normal to the surface, and hence be parallel to $\bd_1$. This vanishing of the external moment density loading implies that the internal moment (and consequently the curvature $\kappa$) must be continuous at all points of contact, including at $s=s_1$. We also note that using the second Weirstrass-Erdmann corner condition for elastic rods~\cite{oreilly2017} (which states that the Hamiltonian function \eqref{eq:hamiltonian} must be continuous across the point of contact) and the continuity of internal moments, it can be deduced that the tension (or the $n_3$ component of the internal force) must be continuous across the point $s=s_1$. However, we do not require this condition to formulate the desired boundary value problem here.

Coming back to the problem formulation, we assume that the entire setup is symmetric under a reflection in the $x$-$z$ plane (shown by the red line in \figref{fig:elastica_schematics}). The problem of computing the opening angle $\phi$ can then be reduced to integrating Eqs.~[\ref{eq:theta}] to [\ref{eq:u2}], in the free region $s\in(0,s_1)$ with $p_1=p_3=l_2=0$, subject to the following boundary conditions,
\begin{alignat}{2}
y(0) &= y_0\, ,\quad &&y(s^-_1)=- \sin\alpha\, ,\label{eq:BC_Y_free}\\
z(0) &= z_0\, ,\quad &&z(s^-_1)=c_z + \cos\alpha\, ,\label{eq:BC_Z_free}\\
\theta(0)&=\frac{\pi}{2}\, ,\quad &&\theta(s^-_1)=\alpha\, , \label{eq:BC_theta_free}\\
\kappa(s_1^-) &= 1\, ,\label{eq:BC_u2_free}
\end{alignat}
where the condition in \eqref{eq:BC_u2_free} is deduced from the continuity of curvature, and $\alpha=\tfrac{L}{2}- s_1$ is the angle subtended by half of the contact region of the rod at the center of the rigid cylinder (see \figref{fig:elastica_schematics}). Once $\alpha$ has been computed, the opening angle can be evaluated as $\phi = \pi-2\alpha$.

The BVP mentioned above was solved using the function \texttt{bvp5c} of Matlab 2019b (MathWorks)  to obtain the theoretical curves for opening angles vs. the normalized force $f_z$ in Fig.~2(e) and Fig.~2(f) of the main article. As is clear from the comparison plots, these computations do not agree well with experiments nor FEM. The primary culprit for the significant mismatch between the measured local opening angles (from experiments and FEM) with the computations described above can be attributed to the approximation of the contact region as a circular arc, which has entered the analysis through the boundary conditions stated from \eqref{eq:BC_Y_free} to \eqref{eq:BC_Z_free}. It could have been envisioned that the large deformations of the rod reported in \secref{sec:largedeformation} could also play a role in this mismatch. However, in \secref{sect:ideal_clasp} and \secref{sect:friction}, we will provide evidence that the complex geometry of contact is dominant over other elasticity effects. 

It should also be pointed out that given the force and moment at the clamped end at $s=0$, the angle $\alpha$ can be computed by evaluating the integral in \eqref{eq:hamiltonian} at $s=0$ and $s=s_1$. This procedure would be a direct way of computing the opening angles if the tensions and moments, in place of positions and slopes, were prescribed at $s=0$.

\subsection{Frictionless ideal orthogonal clasp}
\label{sect:ideal_clasp}

In this subsection, we describe how to compute the equilibrium configurations of two interacting tubular volumes with uniform, circular, cross-sections of a small, but finite, diameter $D$ (cf.\ Fig.~1\textbf{d} of the main article). These equilibrium configurations then provide the curvilinear-diamond shape lines that are the idealized contact sets, which are plotted over the pressure maps in Fig.~3 of the main article.  We solve the appropriate equilibrium balance laws when the tubes are approximated as {\em strings}; \textit{i.e.}, we assume the constitutive relation that each tubular region can support no bending moment ($\bmo = \mathbf{0}$) across any of its cross-sections. We also make the approximations that the tubular string has undeformable cross-sections which remain perpendicular to its (deformable) centerline. In other words, we make the approximation that the string is unshearable. We further assume that the string is inextensible, so that the centerline of the string neither stretches nor contracts. With the assumptions of inextensibility and unshearability, the force acting across each cross-section of the tube is arbitrary, and becomes a basic unknown in the balance laws with no associated constitutive relation. All of these assumptions can be expected to be good approximations in the regime of small tube radius $D/2$ (compared to the minimum radius of curvature of the deformed equilibrium tube centerlines) and low, externally applied, force loadings (compared to a characteristic rigidity of the material appropriately scaled by the tube radius $D/2$). 

\begin{figure}[h!]
	\centering
	\includegraphics[width=0.75\textwidth]{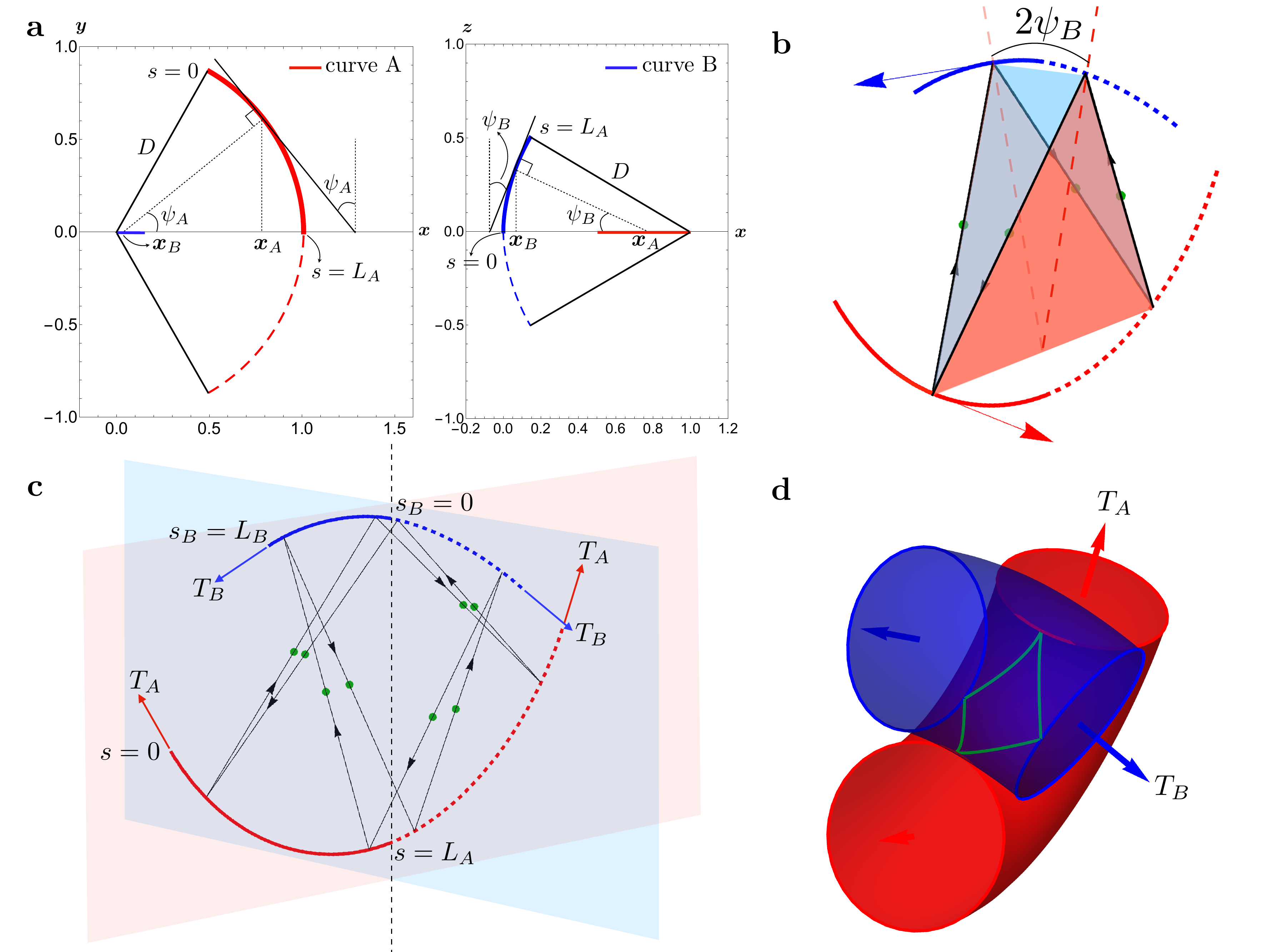}
	\caption{\textbf{Schematic figures of geometric theory for tubular rods} \textbf{a}, Two planar curves comprising mutually congruent halves (solid and dashed), lying in mutually orthogonal planes $x$-$y$ and $x$-$z$. The normal to the tangents of the two curves are depicted using dashed lines, which are also the projections of the corresponding edges of the equilateral quadrilateral onto the plane of the curve. \textbf{b}, An equilateral quadrilateral tetrahedron is depicted. Each such tetrahedron is fully characterized by the pair of angles $2\psi_\text{B}$ (on the blue curve) and $2\psi_\text{A}$ (on the red curve, not shown in the figure) between the faces of the EQT \textbf{c}, The curves are computed as the centerlines of tubular strings (cf.\ panel d) in equilibrium under applied tension end loads $T_\text{A} = 1$ (red) and $T_\text{B}=1.732$ (blue), and subject to the condition of frictionless, self-avoidance. The prescribed opening angles between end tangents are, respectively, $60^\circ$ and $120^\circ$ for the red and blue curves. 
   As the vertices of the one-parameter family of quadrilaterals trace the two curves, the (green) midpoints of the contact chords trace the 3D contact set between the two tubular regions shown in panel d. 
   \textbf{a}, By construction, tubular regions with uniform, circular cross-sections of diameter $D$, each perpendicular to its line of centers, form non-intersecting volumes that touch along the two-fold symmetric contact set shown in green. The perspective in panel d. has been modified slightly to make the contact set more visible.}
	\label{fig:schematics}
\end{figure}

In the absence of external loads distributed along such a tubular string, moment balance \eqref{eq:moment_balance_3D} shows that the only possible equilibria are the trivial ones with straight centerlines, and a purely tangential constant force, or tension, acting across each cross-section. We will therefore consider nontrivial tubular string equilibria with external distributed loading of two types. In \secref{sect:friction} we will consider one tubular string with a planar centerline wrapped around a rigid obstacle and interacting through both an appropriate pressure loading (which excludes interpenetration of the tube and obstacle), and an associated frictional interaction; in other words, a {\em capstan} problem. However, we first consider the case of two tubular strings, each with a planar centerline, interacting through avoidance of intersection of the two tubular volumes and a corresponding associated pressure field, but under the assumption of frictionless contact. With frictionless contact, the moment balance expressed in \eqref{eq:moment_balance_3D} again implies that the only internal force acting across each cross-section of each string is a constant tension, but now the pressure field $\bp$ in \eqref{eq:force_balance_3D} means that each centerline can be curved. The main point of interest in the equilibria of a pair of self-avoiding frictionless tubular strings is that their contact set is surprisingly complicated. In particular, we compute equilibria with the property that in the interior of the contact region between the strings, each point of each centerline is the endpoint of {\em two} minimal distance (\textit{i.e.}, length $D$) contact chords, with the two other ends being distinct points on the other centerline. In fact, the contact chords form a one-parameter family of equilateral quadrilateral tetrahedrons (EQT), whose corners trace out the centerlines of the interacting tubes, while the center points of the edges of the tetrahedron trace out the contact lines where the surfaces of the two tubes touch (see \figref{fig:schematics}\textbf{d}).

We construct equilibria for the semi-inverse problem in which the centerline of each tube is assumed to be planar, and, moreover, the two centerline planes are orthogonal one to the other. We then further assume that each centerline is symmetric under a reflection in the plane of the centerline of the other tube. Thus, each tube comprises two congruent halves, and the full configuration can be divided into four segments, made up of two co-planar congruent pairs, as shown in \figref{fig:schematics}. We note that the theory of orthogonal clasps was first described in detail by Starostin~\cite{Starostin2003}, and, thereafter, by Cantarella~\textit{et al.}~\cite{Cantarella2006}, both of whom were interested in the further special case of four-fold symmetry in which all four segments were assumed to be congruent. The derivation we describe here is a variant of the approach of Starostin, and certainly we recover his conclusions in the particular four-fold symmetric case. Cantarella~\textit{et al.} provide visualisations of the four-fold symmetric contact set, including the gap region between the apices.

With our assumed two-fold symmetry, we can solve for the equilibrium shapes of only two segments, which we index by $A$ and $B$, with each segment being one of the two congruent components in each of the two distinct tube centerlines, for example, the solid red and blue curves in \figref{fig:schematics} (a-c). The remainder of the equilibrium configuration can then be constructed by symmetry. We use subscripts $A$ and $B$ to denote comparable variables associated with the two segments. We assume curves $A$ and $B$ to be contained in mutually orthogonal planes $x$-$y$ and $x$-$z$ respectively (see \figref{fig:schematics}\textbf{a}). The coordinates of the two curves along the shared $x$ axis are denoted by $x_\text{A}$ and $x_\text{B}$, while no subscripts are used for the $y$ and $z$ components of curves $A$ and $B$, respectively.

The tensions in the two components are denoted by $T_\text{A}$ and $T_\text{B}$ (both are positive), representing the tangential component of the unknown force acting across each cross-section. Force balance in the tangential direction at each point is satisfied precisely when the two tensions $T_\text{A}$ and $T_\text{B}$ are both constant (but not necessarily equal). The assumed contact chords (all of equal lengths $D$; \textit{i.e.}, the tube diameter) define a one-to-one mapping between pairs of points on the $A$ and $B$ segments, so that all unknown variable functions can be parametrized by the arc-length coordinate $0\le s\le L_\text{A}$ in segment $A$, where the total length $L_\text{A}$ of the $A$ segment is itself an unknown. However, the two endpoints of the contact chords can evolve at different rates along their respective segments, which gives rise to the next basic unknown $\Gamma\equiv{\text{d}s_\text{B}}/{\text{d}s}$ that determines the differential relation between the arc-length $s$ in segment $A$ and the arc-length $0\le s_\text{B}(s)\le L_\text{B}$ in segment $B$. We chose that the arc-length $s$ in the $A$-segment increases from $s=0$ at the tip away from the plane of symmetry {\em toward} the plane of symmetry in segment $A$, which implies that the arc-length $s_\text{B}$ in the $B$-component increases from $s_\text{B}(L_\text{A})=0$ moving {\em away} from the plane of symmetry toward its tip. 

As mentioned previously, the corners of a one-parameter family of EQTs comprising the contact chords trace out the centerlines of the interacting tubes. We consider two such curves in the $x$-$y$ and $x$-$z$ planes (see \figref{fig:schematics}\textbf{a}), and parametrize their tangents as ${\text{d}x_\text{A}}/{\text{d}s}=\sin\psi_\text{A}$, ${\text{d}y}/{\text{d}s}=-\cos\psi_\text{A}$, ${\text{d}x_\text{B}}/{\text{d}s}=(\text{d}s_\text{B}/\text{d}s)(\text{d}x_\text{B}/\text{d}s_\text{B}) = \Gamma\sin\psi_\text{B}$, and ${\text{d}z}/{\text{d}s}=(\text{d}s_\text{B}/\text{d}s)(\text{d}z/\text{d}s_\text{B})=\Gamma\cos\psi_\text{B}$ by their respective tangent angles $0\le\psi_\text{A}\le\pi/2$ and $0\le\psi_\text{B}\le\pi/2$ as shown in \figref{fig:schematics}\textbf{a}. The angle between two congruent faces of an EQT,  joined by two shared corners lying on the same curve, is given by twice the tangent angle of the other curve at the corresponding point. For instance, as shown in \figref{fig:schematics}\textbf{b}, the angle between the congruent faces joined by an edge in the plane of the red curve (curve A) is given by twice the tangent angle of the blue curve (curve B) at the corresponding points.

Furthermore, we assume that at any given point on a curve, the tangent is perpendicular to the face of the EQT, whose corner traces the curve itself. With this assumption, and some geometrical considerations from \figref{fig:schematics}\textbf{a}, the difference between the $x$ coordinates of the two curves can be related as 
\begin{align}
x_\text{A} - x_\text{B} = y\cot\psi_\text{A}=z\cot\psi_\text{B}\, .\label{eq:geometric_relation}
\end{align}
We use the relation above to eliminate the tangent angles from the parametrization of the tangents stated previously and obtain the following set of ODEs describing the two curves traced by the corners of an EQT.
\begin{alignat}{2}
\frac{\text{d}x_\text{A}}{\text{d}s} &= \frac{y}{\sqrt{(x_\text{A}-x_\text{B})^2 + y^2}}\, ,\qquad &&\frac{\text{d}y}{\text{d}s} =-\frac{x_\text{A}-x_\text{B}}{\sqrt{(x_\text{A}-x_\text{B})^2+y^2}}\, ,\label{eq:curve_A}\\
\frac{\text{d}x_\text{B}}{\text{d}s}&=\Gamma\frac{z}{\sqrt{(x_\text{A}-x_\text{B})^2+z^2}}\,,\qquad&&\frac{\text{d}z}{\text{d}s}=\Gamma\frac{x_\text{A}-x_\text{B}}{\sqrt{(x_\text{A}-x_\text{B})^2+z^2}}\, .\label{eq:curve_B}
\end{alignat}
This set of ODEs admit an integral given by $y^2 + z^2 + (x_\text{A}-x_\text{B})^2 = D^2$, where the left side of the equality is the square of the distance between corresponding points on the two curves, and the constant $D$ prescribes the value of this distance, which is the edge length of the equilateral quadrilateral.

For any given function $\Gamma$, the above system of ODEs describes a valid configuration of two planar curves traced by the corners of an EQT. We, however, are only interested in configurations that also satisfy equilibrium conditions in \eqref{eq:force_balance_3D} and \eqref{eq:moment_balance_3D} for strings, and therefore, the corresponding $\Gamma$ function must come from these conditions. To compute $\Gamma$, we identify an integral of the system by noting that even though the internal force in individual tubes is not conserved, the force balance in \eqref{eq:force_balance_3D} for the four individual strands can be added and integrated to obtain $\sum_{i}\bn_i = \text{const.}$, where the index $i$ runs over four curves, namely, $A$ and $B$, and their congruent counterparts.
Furthermore, moment balance in \eqref{eq:moment_balance_3D} implies that, in the absence of external moment densities, the internal force in a string is purely tensile, i.e. $\bn = T\bd_3$. Substituting this expression into the integral, and projecting it onto the Cartesian $x$ axis, we obtain the following expression,
\begin{align}
2T_\text{A}\sin\psi_\text{A} + 2T_\text{B}\sin\psi_\text{B} = \text{const.}\, ,\label{eq:force_balance_clasp_integral}
\end{align}
where the value of the constant on the right side is determined by the boundary conditions. The tangent angles above can be eliminated in favour of the coordinates using \eqref{eq:geometric_relation}. Then, on differentiating the above equation, and using \eqref{eq:curve_A} and \eqref{eq:curve_B} to eliminate the derivatives of the coordinates, we obtain an explicit expression for $\Gamma$ purely in terms of the coordinates of the two curves. This expression is then substituted back in \eqref{eq:curve_A} and \eqref{eq:curve_B}, and the resulting system is subjected to the following boundary conditions,
\begin{align}
x_\text{A}(0) = D\sin\frac{\theta_\text{A}}{2}\, ,\quad y(0) = D\cos\frac{\theta_\text{A}}{2}\, ,\quad x_\text{B}(0) = 0\, ,\quad z(0) = 0\, ,\label{eq:initial_conditions}
\end{align}
\begin{align}
y(L_\text{A}) = 0\, ,\quad\psi_\text{A}(L_\text{A}) = 0\, ,\label{eq:stopping_criteria}
\end{align}
where $\theta_\text{A} = \pi-2\psi_\text{A}(0)$ and $L_\text{A}$ are the desired opening angle and the unknown total arc-length of segment $A$. The two conditions in \eqref{eq:stopping_criteria} are the stopping criteria for integration, which deliver the value of the unknown length $L_\text{A}$. The two conditions are, in fact, equivalent by construction (as can be seen from \eqref{eq:geometric_relation})  in the sense that the satisfaction of one guarantees the other.

The tension $T_\text{A}$ and the diameter $D$ of the tube may be chosen as the force and length scale to non-dimensionalize the system \eqref{eq:curve_A} to \eqref{eq:stopping_criteria}, which would then lead to $T_\text{B}/T_\text{A}$ as the only dimensionless unknown parameter remaining. The value of this parameter can be obtained by using the relation $T_\text{B}/T_\text{A} = \cos\frac{\theta_\text{A}}{2}/\cos\frac{\theta_\text{B}}{2}$, obtained by evaluating the integral \eqref{eq:force_balance_clasp_integral} at $s=0$ and $s=L_\text{A}$, where $\theta_\text{B}=\pi-2\psi_\text{B}(L_\text{A})$ is the desired opening angle of curve $B$. This expression for the tension ratios can be identified as the global force balance of the two tubes along the Cartesian $x$ axis. We note that the four-fold symmetric solutions of Starostin~\cite{Starostin2003} and Cantarella et al.~\cite{Cantarella2006} are obtained with the choice $T_\text{A}=T_\text{B}$.

We solved the set of ODEs \eqref{eq:curve_A} and \eqref{eq:curve_B}, in conjunction with \eqref{eq:initial_conditions}, as an initial value problem using  the \texttt{NDSolve} function of \texttt{Wolfram Mathematica, Version 12.1.0.0}, with the boundary condition in \eqref{eq:stopping_criteria} as the stopping criteria. We chose the values $T_\text{A}=1$ and $D=1$ to obtain dimensionless solutions from which any dimensionful solution can easily be reconstructed.

We remark that once the above system is solved numerically to obtain the centerline coordinates, several other quantities of interest, which will be required in the next subsection, can be readily computed. The curvatures $\kappa_\text{A}$ and  $\kappa_\text{B}$ of the two curves can be computed from the centerlines using a standard procedure, whereas the tangent angles are given by the relations in \eqref{eq:geometric_relation}, where $0\le\psi_\text{A}\le\pi/2$ and $0\le\psi_\text{B}\le\pi/2$. Of particular interest are the angles $\beta_\text{A}$ and $\beta_\text{B}$, which will be relevant in the next \secref{sect:friction}, between adjacent edges of an EQT. These can be readily computed as the smaller angle between the incoming and outgoing chords at any given point on a curve.
 Another point of interest in the system is the contact pressure (force per unit length) experienced by each curve due to mutual interaction, which can be deduced by considering the force balances of the two tubes individually. For instance, the force balance in the normal direction for curve $A$ can be deduced from \eqref{eq:force_balance_3D} as $-T_\text{A}\kappa_\text{A} + 2\Lambda_\text{A}\cos\frac{\beta_\text{A}}{2}=0$, from which the contact pressure $\Lambda_\text{A}$ is readily obtained. The contact pressure $\Lambda_\text{B}$ for tube $B$ can be obtained analogously.

\subsection{Frictional clasp}
\label{sect:friction}

\begin{figure}[ht!]
	\centering
	\includegraphics[width=0.75\textwidth]{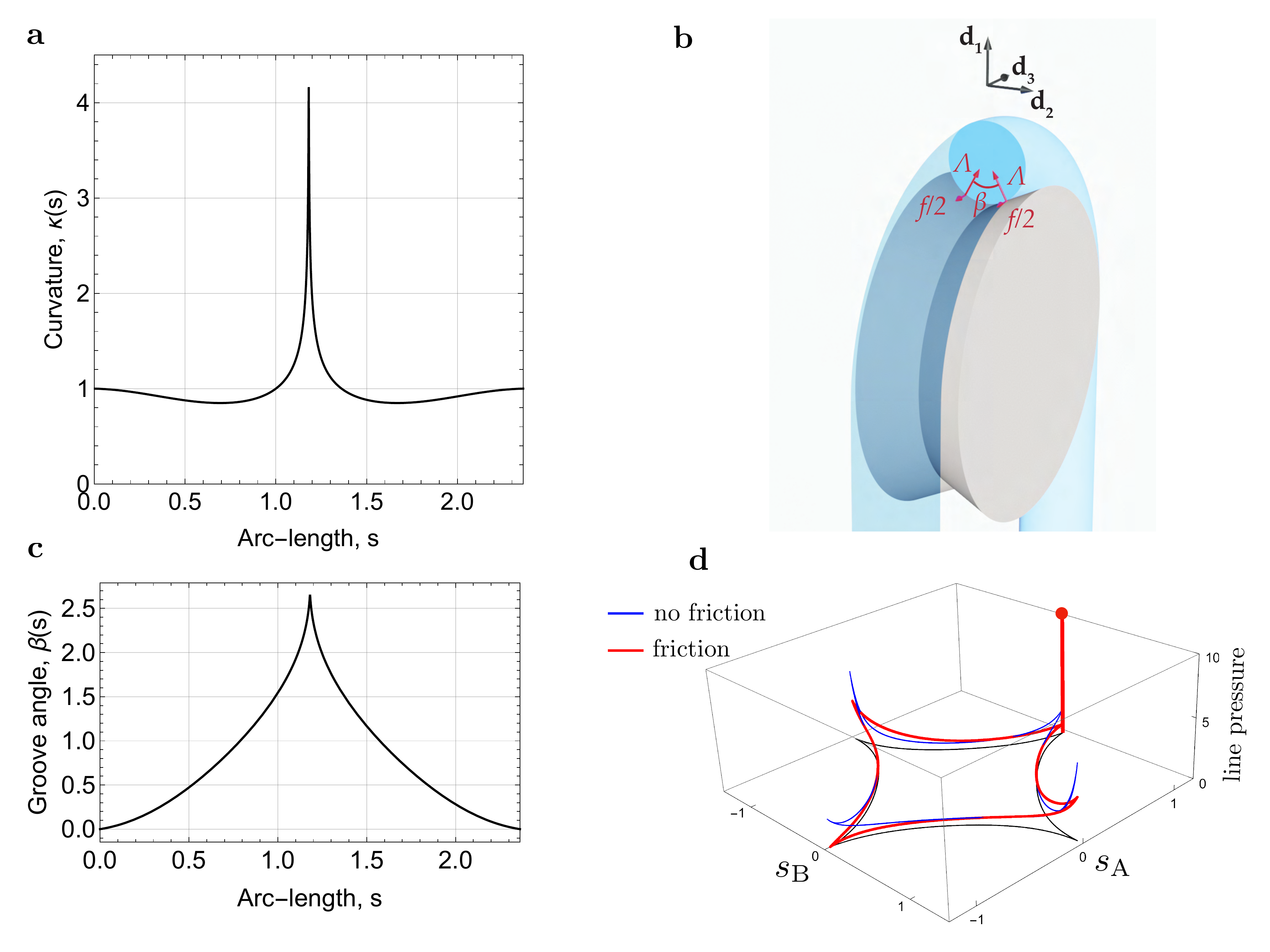}
	\caption{\textbf{a}, Curvature profile of the centerline of the flexible tube as a function of its arc-length. The opening angles of the flexible tube and the rigid capstan are $\theta_\text{A} = 46.5 \deg$ and $\theta_\text{B} = 27.0 \deg$ respectively. \textbf{b}, A flexible tube on a capstan with a V-shaped groove with varying angle $\beta(s)$ \textbf{c}, The groove angle $\beta(s)$ plotted as a function of the arc-length of the flexible tube. \textbf{d}, Line pressures for the frictionless case (in blue) and the frictional case (in red) plotted along the contact set. Two fold symmetry in the pressure profile observed in the frictionless case is broken when friction is introduced, where the pressure reduces to zero at the  touch-down point, while high values of pressure are still observed at the wings and a delta function (shown by the red vertical line hitting the bounding box) in pressure is observed at the lift off point.} 
	\label{fig:schematics2}
\end{figure}

The computations described in the previous section provided insight into the curvilinear diamond shape formed by the contact lines between the equilibria of two idealized, circular cross-section, tubular strings with planar centerlines. Furthermore, the analysis above also enabled us to compute equilibrium pressures $\Lambda_\text{A}$ and $\Lambda_\text{B}$ (in this case force per unit length in the centerlines of the $A$ and $B$ components, respectively) acting between the two tube surfaces along the contact lines. In the system of equilibrium equations for strings, the values of $\Lambda_\text{A}$ (and $\Lambda_\text{B}$) scale with $D/T_\text{A}$ (and $\Lambda_\text{B}$) where $D$ is the tube diameter and $T_\text{A}$ is the constant tension in one component. In \figref{fig:schematics2}\textbf{d}, 
we present a plot of the average of the two dimensionless pressures along the curvilinear contact set for the case $T_\text{B}/T_\text{A} = 1.058$. 
We find that the pressure distribution in the absence of friction has the same two-fold symmetry as the curvilinear contact set. The purpose of this section is to determine how the addition of friction into the model of contact can break this two-fold symmetry and lead to highly asymmetric pressure distributions.

To model the \textit{frictional clasp} experiment described in the main article, we will consider equilibria in which the tension increases as much as possible along one of the tubular centerlines. The change in tension in one component breaks the symmetry in the system so that it cannot be expected that there is an equilibrium configuration where the centerline of both of two deformable tubes remains planar. In turn, a computation of the contact set between tubular volumes with non-planar centerlines is too intricate to be described here, and beyond the scope of the present study. Instead, we provide qualitative insight into how asymmetric the pressure distribution can be in the presence of friction by making the further idealization that one of the tubular regions is replaced by a completely {\em rigid} tubular volume (but with the same centerline as computed for an equilibrium of a tubular string in the absence of friction). The other tubular volume will continue to be treated as a completely flexible string (with all the same assumptions as above) so that we have at hand a capstan problem, 
in which a deformable tubular string of finite diameter is wrapped around a rigid tubular body (or a curved capstan). It then remains to compute the equilibrium configuration of the deformable tube, and, in particular, the maximum possible growth in tension, given the available friction in the interaction with the surface of the rigid capstan. The great simplification in assuming that one of the tubular regions is now rigid is that, without loss of generality, we may assume that the contact set is already known, specifically in the projection onto the plane of the two arc lengths, it remains as the curvilinear diamond shape already computed previously in  \secref{sect:ideal_clasp}.

Both the experiment and FEM simulations reported in Fig.~4 of the main article consider the case of the interaction of two tubes with equal diameters $D$. In the theory for the equilibria of tubular strings, the parameter $D$ also appears through the initial conditions, but, there, it plays the role of the prescribed distance between the two tube centerlines. In the case of equal diameter tubes $D$, the distance between the two tube centerlines is indeed also $D= D/2 +  D/2$. But the theory for two touching strings works just as well for two uniform tubes with radii $R_\text{A}$ and $R_\text{B}$ with the interpretation $D = R_\text{A} + R_\text{B}$.  The governing equations for the equilibrium tube centerlines are unaltered with this generalized interpretation of the parameter $D$. The only difference is that the contact line between two tubes of unequal radii as embedded in 3D is no longer swept out by the central points of the contact chords, but rather is swept out by the points that divide the contact chord of length $D$ in the ratio $R_\text{A}:R_\text{B}$. We explicitly adopt this generalization in our treatment of our tubular capstan problem with the flexible tube of radius $R_\text{A} = \alpha D$ and the rigid tube of radius $R_\text{B} = (1-\alpha) D$, for any $0<\alpha<1$. The equal radii case corresponds to $\alpha=1/2$, and the introduction of the generalization allows us to recover results from classic capstan theory (equation 1 in the main article) in the limit $\alpha \to 0$. 

To further emphasize the connection between our theory and the literature on capstan problems, we now introduce a slightly modified notation. Specifically, we will have at hand a capstan problem where a belt (\textit{i.e.},\ our deformable tubular string of radius $R_\text{A}$) interacts with a rigid tubular capstan with circular cross-sections of radius $R_\text{B}$. The condition of contact between string and capstan then prescribes the curvature $\kappa(s)$ of the deformable tube centerline, along with a groove of prescribed (but variable) opening angle $\beta(s)$, cf.\  \figref{fig:schematics2}\textbf{b} and \textbf{c}. Both of these quantities we regard as having been computed via the analysis of the previous section with some prescribed $D$ and opening angles $\theta_\text{A}$ and $\theta_\text{B}$, and $\beta(s)=\beta_\text{A}$ with $0\le \beta < \pi$. The specific cases of $\kappa(s)$ and $\beta(s)$ shown in \figref{fig:schematics2}\textbf{a} and \textbf{c} will be considered as our numerical example (corresponding to $\theta_\text{A}=46.5 \deg $ and $\theta_\text{B}=27.0 \deg$). We note that both $\kappa(s)$ and $\beta(s)$ are continuous throughout, but that both of their first derivatives have a simple jump discontinuity at the central point. Now, we parametrize by arc length $s$ in the flexible tube with $0<s<L=2.362$ (\textit{i.e.},\ $s=0$ corresponds to the first point at which the tube touches the capstan), and $s=L$ to where the tube leaves the capstan. By the geometry of the pre-computed contact set, we have $\beta(0)=\beta(L)=0$. 

By construction, the centerline of the tubular string is planar, and we further assume that both the normal pressures and frictional interactions acting at the (non-planar) contact lines are symmetric at the two contact points arising in each normal cross-section to the tube for each value of $s$. Then, we introduce $P(s)\equiv2\Lambda(s) \cos\frac{\beta(s)}{2}\ge 0$ as the effective total normal pressure field at the centerline, and $f(s)$ as the total frictional force density, which has only a tangential component to the tube centerline (due to the assumption of symmetry of the two individual contact line force densities). We adopt the sign convention that $f>0$ corresponds to the total frictional force density being directed anti-parallel to the tangent direction with arc length $s$ increasing. Both $P$ and $f$ have units of force per unit length, and we adopt the scaling in which it is arc length along the centerline of the deformable tube. As we are now considering an obstacle problem comprising a flexible tube wrapped around a rigid capstan, it {\em cannot} be ruled out {\it a priori} that the surface interactions may involve concentrated Dirac delta function densities in $P$ or $f$. And indeed, we will see that a Dirac delta in the pressure $P$ will play an important role in our theory. However, a detailed consideration of moment and force balance for a tubular string of finite diameter reveals that the frictional density $f$ can be at most discontinuous, and correspondingly the tension $T(s)$ must always be continuous. Because the tubular string is in contact with a rigid surface, the internal force acting across each tube cross-section can now have a normal, or shear, component $N(s)$ in addition to the tangential tension $T(s)$. However, moment balance immediately implies that the shear force is directly related to the friction density $f(s)$ via the relation $N(s) = R_\text{A}\, f(s) \cos\frac{\beta(s)}{2}$. For a string with a planar centerline, as is assumed here, this identity is equivalent to moment balance. We use the identity to eliminate the frictional force density $f$ in favor of a shear force density $S\equiv N/R_\text{A} = f(s) \cos\frac{\beta(s)}{2}$ and we transfer the sign convention for $f > 0$ to $S > 0$.

The force balance and static friction laws can then be expressed as
\begin{align}
T' &= \left(\sec\frac{\beta}{2}-\alpha \kappa D\right)\, S\, ,\label{eq:tangential_force_balance}\\
\alpha D  S'&= \kappa\, T - P\, ,\label{eq:normal_force_balance}\\
S &\le \mu_{\rm s} \,  P \, , \label{eq:friction_inequality}
\end{align}
where the choice of the nondimensional constant $0< \alpha < 1$ sets the ratio of the two tube radii,  $1 \le \sec\frac{\beta(s)}{2} = 1/\cos\frac{\beta(s)}{2}$ is the secant function,  which is finite  for our range of values of the angles $\beta$, and $\mu_{\rm s}$ is the material static coefficient of friction.
As before, $T(0)$, which by continuity is the applied external load, can be taken as the force scale to be used in non-dimensionalization, so that, effectively, the above system can be solved subject to the initial condition $T(0)=1$. Then, the problem of interest is to determine the shear force density $S$ and pressure $P$ that leads to the maximum possible value of $T(L)$.

In the context of a tubular string interacting with a grooved capstan with prescribed groove opening angle function $\beta(s)$~\cite{Belofsky1976}, it is both sensible and of interest to first consider the limit $\alpha \to 0$ which corresponds to the standard capstan problem of a string of negligible diameter wrapped around a rigid body. With our choice of variables, $\alpha$ small is a regular perturbation in the tension balance \eqref{eq:tangential_force_balance}, a singular perturbation in the shear density balance \eqref{eq:normal_force_balance}, and the friction inequality \eqref{eq:friction_inequality} is independent of $\alpha$. 
As $\alpha \to 0$ the ODE in \eqref{eq:normal_force_balance}) reduces to an algebraic formula by which the pressure density $P$ can be eliminated, and we arrive at the simple system
\begin{align}
T' &= \sec\frac{\beta}{2} \, S\, ,\label{eq:tangential_force_balance_zero}\\
S&\le \mu_{\rm s}\,  \kappa \, T \, .\label{eq:friction_inequality_zero}
\end{align}
It is now evident that to arrive at the maximum growth in tension the shear density $S$ should be chosen to saturate the (static) friction inequalities \eqref{eq:friction_inequality_zero}. (This choice also corresponds to the formula for the shear, and therefore friction, density that arises in the case of sliding friction provided that static and sliding friction coefficients coincide $\mu = \mu_{\rm s}$.) 
Given that we are interested in the case of kinetic friction $\mu$, we arrive at the single differential equation and initial condition
\begin{equation}
T'= \mu\,  \kappa\, \sec\frac{\beta}{2} \, T\, , \qquad T(0)=1\, ,
\label{eq:max_tension_delta_zero}
\end{equation}
which is the classic capstan equation modified to include the effective increase in friction coefficient due to a groove in the capstan as expressed in the $\sec\frac{\beta}{2}\ge 1$ factor~\cite{Belofsky1976}. In the special case of constant friction coefficient $\mu$ (uniform material), and constant angle $\beta$ (uniform groove) its solution is the classic exponential growth in tension as function of turning angle around the capstan (independent of whether or not the curvature $\kappa(s) \ge 0$ is constant; \textit{i.e.}\ whether or not the capstan has a circular cross-section, cf.~\cite{maddockskeller1987}). 
We next show that this simple argument characterizing maximal possible tension growth can be generalized to a version applicable to the case $\alpha > 0$ of positive tube diameter that is of primary interest here.

We first remark that the geometric inequality $\left(1-\kappa \frac{D}{2}\right) \ge 0$ is the condition that a tube of diameter $D$ has no local self-intersection singularity when the curvature of its centerline is $\kappa(s)$. In fact, we expect the tubular string theory to be most accurate in the regime $\alpha\kappa D \ll 1$. In any case, with the additional term $\sec\frac{\beta}{2}\ge 1$, we conclude (all the more) that for all $0<\alpha< 1/2$ the coefficient on the right hand side of \eqref{eq:tangential_force_balance} is non-negative. Consequently, to achieve maximal tension growth, the shear density $S$  should be as large as possible. 

Turning to \eqref{eq:normal_force_balance} (and for the moment setting aside consideration of boundary conditions) in order to have $S$ as large as possible, we see that the term $-P\le 0$ must be chosen to be as large (\textit{i.e.}, as least negative) as possible. In particular, it can never be beneficial to have a delta function in the pressure $P$ (at $s=0$ or in the interior of the interval of definition of the ODE) as this would correspond to a jump {\em down} in the value of the shear density $S$. In the absence of any delta function in the pressure $P$, the solution $S(s)$ to \eqref{eq:normal_force_balance} is continuous, as is the pressure $P(s)$, which should be picked to saturate the friction inequality [\ref{eq:friction_inequality}]. (As before this choice of $P$ also corresponds to the law of sliding friction if static and dynamic coefficients of friction are assumed equal.) Thus we arrive at the system determining maximal tension growth for $\alpha>0$:
\begin{align}
T' &= \left(\sec\frac{\beta}{2}-\kappa \alpha D\right)\,S\, ,\label{eq:tangential_force_balance_sat}\\
\alpha D\,S' &=\kappa\, T - \frac{S}{\mu}\, .\label{eq:normal_force_balance_sat}
\end{align}

We have already noted the scaling initial condition $T(0)=1$, but for $\alpha > 0$, the linear system of \eqref{eq:tangential_force_balance_sat} and \eqref{eq:normal_force_balance_sat}) has two ODE, so we need a second initial condition for $S(0)$ in order to find a unique solution. This second initial condition can be obtained from the following argument. We have already observed that even in the presence of a Dirac delta function in the pressure $P(s)\ge 0$, the moment balance relation for a string implies that the tension $T(s)$ must be continuous. Moreover, for the equilibrium corresponding to maximum tension growth, we have also already observed that a delta function in the pressure cannot be beneficial. Consequently, on an equilibrium with maximal tension growth, all of the variables $T$, $N$, $f$, and $S$ must be continuous, and in particular, be continuous at $s=0$. But, by construction, there is no contact in the region $s=0^-$ so that necessarily $f(0^-)$ and by continuity $f(0)=0$. And moment balance relates $f$ and $N$, so that $N(0)=0$, which by definition implies that $S(0)=0$, which is our desired second initial condition. \eqref{eq:tangential_force_balance_sat} and \eqref{eq:normal_force_balance_sat}) can then be numerically integrated as an IVP subject to the initial conditions
\begin{equation}
T(0)=1\, , \qquad S(0)=0\, .
\label{eq:IC}
\end{equation}
The linear IVP solve for equations~\eqref{eq:tangential_force_balance_sat} and \eqref{eq:normal_force_balance_sat}) with initial conditions~[\ref{eq:IC}] can be continued up to any $L$ for which the contact geometry $\kappa(s)$ and $\beta(s)$ are known for $0\le s \le L$. For an initial input tension $T_0$ the corresponding maximum (dimensionful) value of tension possible at that arc length is $T_0\, T(s)$. With the interpretation that one of the tubular regions is a rigid body, the flexible tube will be in equilibrium if it lifts off from the capstan at any $L$, provided that the tension in the free region is $T(L)$ (and for $\alpha > 0$ moment balance requires that $T$ be continuous at lift-off). However, there remains a subtlety in the boundary condition at $s=L$ for the shear force density $S$, or equivalently the frictional force density $f$. Certainly, the IVP solve will deliver a non-zero, in fact positive, limiting value of $S(L^-)$. But, as there is no contact after lift-off, we know that $S(L^+)=f(L^+)=0$, so there must be discontinuity in $S$ at $s=L$. Such a discontinuity can be achieved by having a Dirac delta function in $P$ at $s=L$. This is possible while respecting $P \ge 0$ because the jump in $S$ is downward, and it is $-P$ that appears on the right-hand side of the ODE in~\eqref{eq:normal_force_balance}. Moreover, our argument that for solutions with maximal tension increase, delta functions in $P(s)$, with associated downward jumps in $S(s)$, cannot arise for $0\le s < L$,  does {\em not} apply at $s=L$, because the right hand side of \eqref{eq:normal_force_balance}) remains bounded, so that the maximal achievable value of $T(L)$ is unaffected by the discontinuity in $S$ at $s=L$. Thus we conclude that, for a maximally increasing tension equilibrium, a delta function in the pressure, {\em must} arise at the lift-off point $s=L$. This conclusion is compatible with the FEM simulations of fully elastic tubes, as observed in the 3D normalized pressure plot in Fig.~4(e) of the main article, in which we observed a rather high concentrated pressure distribution close to the lift-off point (cf.~\figref{fig:schematics2}\textbf{d}).

\section*{Bibliography}
\bibliography{references_SI}